\documentclass[a4paper,12pt]{article}
\usepackage[utf8]{inputenc}
\usepackage[T1]{fontenc}
\usepackage{textcomp}
\usepackage{lmodern}
\usepackage[margin=2.5cm]{geometry}
\usepackage{amsmath,amssymb}
\usepackage{enumitem}
\usepackage{xurl}
\usepackage[colorlinks=true,linkcolor=blue,citecolor=blue,urlcolor=blue]{hyperref}
\title{Fabless Quantum Chip Design and Commercial Production}
\author{
\begin{minipage}{0.94\textwidth}
\centering
\normalsize
Cai\textsuperscript{\dag},
Ling Qiao\textsuperscript{1,2,\dag},
Bin Yang\textsuperscript{1,2},
Fumin Luo\textsuperscript{1,2},
WeiGui Guo\textsuperscript{1,2},
GuoRong Zhang\textsuperscript{1,2},
XueFei Liu\textsuperscript{1,2},
Qinglang Guo\textsuperscript{1,2,*},
and Bin Wu\textsuperscript{*}
\\[0.6em]
\small
\textsuperscript{1}Yangtze Delta Industrial Innovation Center of Quantum Science and Technology, Suzhou, China, 215100
\\[0.25em]
\textsuperscript{2}China Academy of Electronics and Information Technology, No. 11 Shuangyuan Road, Shijingshan District, Beijing, China, 100041
\\[0.4em]
\textsuperscript{\dag}These authors contributed equally to this work.
\\[0.25em]
Emails: \href{mailto:qiaoling@tgqs.net}{qiaoling@tgqs.net}, \href{mailto:1763098000@qq.com}{1763098000@qq.com}
\\[0.25em]
\textsuperscript{*}Corresponding authors: Qinglang Guo, \href{mailto:gql1993@mail.ustc.edu.cn}{gql1993@mail.ustc.edu.cn}.
\end{minipage}
}
\date{June 16, 2026}

\begin{document}
\maketitle
\tableofcontents
\newpage

\section{Abstract}

Integrating quantum-chip design paradigms, superconducting quantum-hardware engineering, and the SPICE-Q multiphysics simulation framework, this paper proposes an architecture for ``fabless quantum-chip design and commercial production'' oriented toward the superconducting quantum-computing industry. The framework aims to build a scalable, reusable quantum-chip industrial ecosystem with engineering closed-loop capabilities through standardized process design kits (PDKs), parameterized device cells (PCells), SPICE-Q physical modeling and simulation languages, unified Q-EDA design flows, and quantum-IP (Q-IP) market mechanisms (Krantz et al., 2019; Kjaergaard et al., 2020; Blais et al., 2021; Levenson-Falk \& Shanto, 2025).

In this system, quantum-chip design firms do not simply outsource manufacturing. Rather, they complete design verification and tape-out preparation under the constraints of process-certified PDKs, traceable model cards, test structures, and foundry sign-off rules. The potential economic value of the fabless model lies in reducing repetitive device debugging, process exploration, and low-level layout development; however, its realized benefits depend on PDK maturity, foundry yield, cryogenic test throughput, model-prediction error, data-feedback mechanisms, and the boundaries of Q-IP licensing (Mead \& Conway, 1980; Macher et al., 2008; Cross et al., 2022; National Academies, 2024).

The central argument of this paper is that the quantum-chip industry can move from the currently dominant IDM (vertically integrated manufacturing) model toward a Fabless-Foundry collaborative ecosystem resembling that of the classical semiconductor industry only when quantum-hardware design is built upon standardized, verifiable, and reusable industrial software and process interfaces. This transition depends on five key pillars: process-certified PDKs, a PCell-based parameterized design system, SPICE-Q cross-physics simulation models, an end-to-end Q-EDA design toolchain, and a tradable and reusable Q-IP intellectual-property market (QPDK Documentation, 2025; EDA-Q Collaboration, 2025; GDSII-to-Wafer Collaboration, 2026; Shanto et al., 2024).

Related studies show that the success of the classical Fabless-Foundry model derives essentially from interface standardization between design and manufacturing, toolchain abstraction, and reusable IP mechanisms. If the quantum-chip industry is to achieve a comparable level of scalable expansion, it must likewise establish rigorous design rules, verifiable physical models, cross-tool metadata standards, cryogenic-test feedback mechanisms, and sign-off systems compatible across foundries (Macher et al., 2008; Mack, 2011; Wilkinson et al., 2016; National Academies, 2024).

\section{Introduction: Problem Definition and Core Argument}

This section defines the problem of ``fabless quantum-chip design and commercial production'' from three perspectives: systems engineering, device physics, and industrial organization. It also clarifies the research scope, core assumptions, and argumentative path of this paper. Here, fabless does not simply mean outsourcing manufacturing to a third-party factory. Instead, it refers to enabling quantum-chip designers to complete design, simulation, sign-off, and tape-out preparation without owning a complete cryogenic micro- and nanofabrication production line, on the basis of verifiable process interfaces, model interfaces, data interfaces, and sign-off interfaces.

Current superconducting quantum-computing systems still rely primarily on a vertically integrated (IDM) model: the same organization, or a tightly coupled research system, typically undertakes device design, micro- and nanofabrication, packaging and interconnection, cryogenic testing, system calibration, and cloud operation. This model facilitates rapid closed-loop iteration during early exploration, but it also tightly binds design knowledge, process experience, and test data within a single experimental system, resulting in long R\&D cycles, high trial-and-error costs, difficulty in cross-institutional reuse, and constraints on the expansion of industrial participants (Preskill, 2018; Krantz et al., 2019; Kjaergaard et al., 2020; National Academies, 2024).

The central argument of this paper is that, if the superconducting quantum-chip industry is to move from laboratory prototypes toward scalable commercial production, it must transform ``physical manufacturability'' into engineering interfaces that can be invoked, verified, and audited by external designers. Experience from the classical semiconductor Fabless-Foundry model shows that industrial specialization releases economies of scale only when design rules, model libraries, EDA tools, IP licensing, and manufacturing sign-off processes are sufficiently stable. The theory of modular design also indicates that open collaboration in complex systems presupposes clear design rules and replaceable module boundaries (Mead \& Conway, 1980; Baldwin \& Clark, 2000; Langlois \& Robertson, 1992; Macher et al., 2008). Accordingly, this paper defines the fabless quantum-chip model as an engineering institution jointly supported by PDKs, PCells, SPICE-Q, Q-EDA, Q-IP, cryogenic-test feedback, and data governance, rather than as a single commercial outsourcing arrangement.

\subsection{Research Question and Scope}

The core research question of this paper can be stated as follows: within superconducting quantum-chip systems, can a division-of-labor structure analogous to the Fabless-Foundry model of the classical semiconductor industry be constructed, such that quantum-chip design is partially decoupled from specific manufacturing processes without departing from physical realizability, and further develops into a replicable, auditable, and iterative commercial production flow? It must be emphasized here that quantum-chip design and manufacturing cannot be fully abstracted in the way mature CMOS digital chips can be. Josephson-junction discreteness, material-interface loss, frequency collisions, packaging modes, and cryogenic-test throughput all directly affect chip performance and yield (Koch et al., 2007; McRae et al., 2020; Levenson-Falk \& Shanto, 2025). Therefore, this paper discusses ``constrained decoupling,'' not unconditional process outsourcing.

Formally, the quantum-chip design problem can be abstracted as the following constrained optimization system:

\[
\min_{\theta \in \Theta} \; \mathcal{L}(\theta, \mathcal{H}_{\text{device}}, \mathcal{N})
\quad \text{s.t.} \quad
\theta \in \mathcal{P}_{\text{PDK}}, \;
\mathcal{V}_{\text{DRC/LVS/DFM}}(\theta)=1
\]

Here, $\theta$ denotes device, circuit, and layout design parameters; $\mathcal{H}_{\text{device}}$ is the effective Hamiltonian obtained by mapping geometric structures and material parameters; $\mathcal{N}$ denotes noise, loss, and process-statistical drift; $\mathcal{P}_{\text{PDK}}$ denotes the manufacturable parameter space defined by the process design kit; and $\mathcal{V}_{\text{DRC/LVS/DFM}}$ indicates whether design-rule, layout-versus-schematic, and design-for-manufacturability sign-off has passed. This formulation better reflects the engineering focus of recent Q-EDA and GDSII-to-wafer work than merely minimizing a device-loss function, namely the extension from design-file generation to manufacturing data packages and foundry-executable processes (EDA-Q Collaboration, 2025; GDSII-to-Wafer Collaboration, 2026).

The scope of this study is limited to superconducting quantum-computing platforms, with Transmons and their readout, coupling, and control structures as the primary objects, and with the ``design-simulation-fabrication-calibration-model feedback'' loop as the system boundary. This paper does not directly discuss ion-trap, photonic-quantum, or neutral-atom systems; however, methodologies such as PDK-ization, parameterized components, physical model cards, and sign-off flows may provide references for heterogeneous quantum-hardware platforms.

\subsection{Contributions of the Quantum Fabless Model}

The core contribution of the quantum fabless model is to gradually transform the quantum-chip industry from a ``laboratory-driven process-exploration model'' into a ``standardized-interface-driven engineering-design model.'' In this model, the PDK (Process Design Kit) defines the process layers, design rules, statistical corners, and model applicability ranges that a foundry can commit to; the PCell (Parameterized Cell) encapsulates common components such as Transmons, readout resonators, couplers, test structures, and subarrays as reusable objects; the SPICE-Q model connects geometry, electromagnetics, Josephson nonlinearity, noise, and Hamiltonian parameters; the Q-EDA workflow organizes these objects into a design flow that can be automatically checked and signed off; and Q-IP further consolidates design assets that can be traded, licensed, and reused across projects.

The role of this structure is similar to the ``design abstraction layer'' in classical semiconductors. It does not eliminate the physical complexity of quantum devices; rather, it converts that complexity into versioned, verifiable, and traceable interfaces. Mead and Conway's idea of VLSI design abstraction, Baldwin and Clark's theory of modular design rules, and studies of division of labor in the Fabless-Foundry industrial chain jointly show that external innovators can conduct parallel design and trial-and-error learning on the same platform only when underlying complexity can be encapsulated by stable interfaces (Mead \& Conway, 1980; Baldwin \& Clark, 2000; Langlois \& Robertson, 1992; Macher et al., 2008).

Furthermore, the economic advantage of the fabless model should not be understood as an unconditional reduction in cost. Rather, it should be understood as a reduction in repetitive R\&D costs and single-project process-exploration costs after PDK maturity, foundry yield, cryogenic test throughput, model-prediction error, and Q-IP reuse mechanisms have reached a certain level. This relationship can be expressed more cautiously as:

\[
\mathbb{E}[C_{\text{fabless}}] < \mathbb{E}[C_{\text{IDM}}], \quad
\mathbb{E}[T_{\text{iteration}}^{\text{fabless}}] < \mathbb{E}[T_{\text{iteration}}^{\text{IDM}}]
\]

where $C$ denotes R\&D and verification cost, and $T$ denotes the design-iteration cycle. The prerequisite for the above inequalities to hold is that manufacturing interfaces and model-prediction capabilities have reached a level trusted by external designers. If PDK rules are unstable, cryogenic test data cannot be fed back, or PCell models lack statistical confidence intervals, the fabless model may instead increase communication costs and verification risks (CMC Microsystems, 2023; National Academies, 2024; QPDK Documentation, 2025; GDSII-to-Wafer Collaboration, 2026).

\subsection{Definitions and Boundary Conditions}

This paper defines a \textbf{fabless quantum-chip design provider} as an enterprise or research institution that does not own a complete physical cryogenic manufacturing production line, but can complete quantum-circuit design, simulation verification, layout sign-off, and tape-out preparation on the basis of PDKs, PCell libraries, SPICE-Q models, and Q-EDA toolchains. Its core capability lies not in independently mastering a particular manufacturing process, but in understanding the process boundaries released by a foundry and completing system-level design optimization within those boundaries.

A \textbf{Quantum Foundry} is a facility that provides externally reusable superconducting quantum-chip fabrication capabilities, test structures, process data, and PDK interfaces. Its core output is not merely a single chip product, but a process platform comprising layer definitions, design rules, PCell libraries, model cards, test-chip data, manufacturing data packages, and sign-off flows. Recent work on superconducting PDKs, Q-EDA, and GDSII-to-wafer all indicates that, if a foundry is to serve external designers, it must transform internal experience into publishable, versionable, and verifiable engineering documentation, rather than merely providing one-off processing services (CMC Microsystems, 2023; QPDK Documentation, 2025; EDA-Q Collaboration, 2025; GDSII-to-Wafer Collaboration, 2026).

\textbf{SPICE-Q} is defined in this paper as an extended circuit-simulation language or intermediate representation for quantum circuits, used to describe in a unified manner the mapping relationships among classical control circuits, electromagnetic modes, Josephson nonlinearity, noise channels, and quantum Hamiltonian dynamics. Its basic idea can be written as:

\[
\mathcal{H}_{\text{SPICE-Q}} = \mathcal{H}_{\text{Josephson}} + \mathcal{H}_{\text{coupling}} + \mathcal{H}_{\text{noise}} + \mathcal{H}_{\text{package}}
\]

Here, $\mathcal{H}_{\text{package}}$ reminds the reader that packaging, interconnection, cavity modes, and the control-line environment are not external appendages, but physical factors that must be included within the modeling boundary of large-scale superconducting quantum-chip design (Blais et al., 2021; Levenson-Falk \& Shanto, 2025). Based on the above definitions, the discussion in this paper is limited to mK-scale cryogenic superconducting systems, and assumes that manufacturing processes can be modeled through statistical distributions, test structures, material-loss data, and cross-batch measurement results. It also assumes the existence of a traceable and versioned PDK release mechanism that can be invoked by external Q-EDA tools.

\subsection{Structure of the Paper}

The structure of this paper follows the engineering logic of ``progressively descending abstraction layers.'' Sections 2--3 first establish the industrial and physical problem definitions for fabless quantum chips, explaining which references can be drawn from classical fabless experience and in which respects that experience cannot be directly transferred. Sections 4--6 further discuss the physical foundations of superconducting quantum devices, PDK interfaces, PCell parameterized design, and the SPICE-Q modeling system. Sections 7--8 turn to Q-EDA workflows and the Q-IP ecosystem, analyzing how design automation and knowledge reuse form industrial infrastructure. Sections 9--10 discuss commercialization paths, foundry collaboration models, and mechanisms for restructuring the industrial organization.

This arrangement corresponds to the layered abstraction path in classical EDA systems, from ``device physics $\rightarrow$ standard cells/parameterized cells $\rightarrow$ design toolchains $\rightarrow$ industrial ecosystems.'' However, this paper places special emphasis on the additional issues faced by quantum chips, including Hamiltonian modeling, decoherence, cryogenic calibration, and manufacturing-data feedback (Hennessy \& Patterson, 2019; National Academies, 2024).

\subsection{Overview of the Literature Position}

This paper lies at the intersection of three research domains. The first body of literature comes from superconducting quantum-hardware engineering and focuses on foundational issues such as Transmon devices, readout systems, coupling structures, sources of decoherence, and material loss (Koch et al., 2007; Devoret \& Schoelkopf, 2013; Blais et al., 2021). The second body of literature comes from quantum-computing systems engineering and error correction, and focuses on scaling paths from physical qubits to logical qubits, architectural constraints, and resource overheads (Fowler et al., 2012; Gambetta et al., 2017; National Academies, 2024). The third body of literature comes from EDA, modular design, and theories of semiconductor industrial structure, explaining why the classical Fabless-Foundry model depends on stable design rules, standardized interfaces, specialized division of labor, and modular production networks (Mead \& Conway, 1980; Baldwin \& Clark, 2000; Langlois \& Robertson, 1992; Sturgeon, 2002; Macher et al., 2008).

Early quantum-hardware literature focused mainly on qubit physics, small-scale system implementation, and resource estimation for error correction. Recent work has begun to discuss systematically superconducting quantum-chip design automation, GDSII-to-wafer manufacturing-data conversion, QPDK, KLayout/KQCircuits, Qiskit Metal, and open hardware toolchains (Minev et al., 2021; KQCircuits Documentation, 2024; QPDK Documentation, 2025; EDA-Q Collaboration, 2025; GDSII-to-Wafer Collaboration, 2026; Levenson-Falk \& Shanto, 2025). The contribution of this paper is not to replace these engineering literatures, but to organize them within the Fabless-Foundry industrial division-of-labor framework, propose a unified industrialization path of SPICE-Q + PDK + PCell + Q-IP + Q-EDA, and further explain that this path has commercial significance only when models are trustworthy, interfaces are stable, test data can be fed back, and intellectual property can be reused.

\section{Industrial Background: Classical Fabless and the Chinese Quantum-Chip Ecosystem}

\subsection{Why Classical Fabless Is Effective}

The fundamental reason why the classical fabless model has achieved decisive success in the semiconductor industry is that it realized the ``encapsulation of physical complexity'' and the ``stabilization of the design abstraction layer.'' In this system, manufacturing complexity is encapsulated at the foundry layer. Fabless companies do not need to directly manage wafer fabs, lithography equipment, yield ramp-up, or process maintenance; instead, they search the design space under the constraints of certified PDKs, standard-cell libraries, IP cores, and EDA toolchains. In other words, fabless does not simply outsource manufacturing; it converts manufacturing capability into design interfaces that are computable, verifiable, and subject to sign-off (Mead \& Conway, 1980; Macher et al., 2008; Mack, 2011).

Formally, the design flow can be expressed as:

\[
\theta^\ast = \arg\min_{\theta \in \mathcal{P}_{\text{CMOS}}} \mathcal{L}(\theta)
\]

where $\mathcal{P}_{\text{CMOS}}$ is the manufacturable design space defined by the process design kit (PDK). Fabless companies can iterate rapidly within this space not because they are detached from manufacturing physics, but because manufacturing physics has already been abstracted into design rules, device models, layout-layer definitions, and sign-off flows.

The effective operation of the classical fabless system depends on the joint action of three types of mechanisms. First, GDSII/OASIS, DRC/LVS, SPICE models, and PDK version management form stable interfaces between design and manufacturing. Second, foundries amortize high capital expenditure through wafer-scale manufacturing and cross-customer process reuse. Third, mature EDA toolchains automate logic synthesis, placement and routing, timing analysis, power optimization, and final sign-off, enabling design innovation to occur without rebuilding manufacturing infrastructure. Modular design theory and research on modular production networks further show that, when interfaces are sufficiently clear and module boundaries can be governed, industrial specialization promotes parallel innovation, specialized investment, and external economies of scale (Baldwin \& Clark, 2000; Langlois \& Robertson, 1992; Sturgeon, 2002).

This system enables design innovation to evolve, to some extent, independently of the manufacturing capabilities of any single firm, thereby forming a positive feedback loop of ``expanded design demand--foundry capacity and yield optimization--maturation of tool and IP ecosystems--entry of more design actors'' (Macher et al., 2008).

\subsection{Limits of Direct Analogy}

Although the fabless model succeeded in classical semiconductors, it cannot be mechanically transferred to quantum chips. Most key behaviors in classical CMOS design can be encapsulated in engineering terms through transistor models, timing models, and statistical corners. By contrast, the performance of superconducting quantum devices depends simultaneously on Josephson-junction parameters, material-interface loss, the electromagnetic environment, packaging modes, frequency allocation, control crosstalk, and cryogenic calibration processes. In other words, the ``manufacturing outcome'' in a quantum chip not only determines whether a device conducts, but also directly changes the Hamiltonian, decoherence channels, and system-level error budget (Krantz et al., 2019; Kjaergaard et al., 2020; Levenson-Falk \& Shanto, 2025).

Its dynamics can be summarized as an open quantum-system evolution problem:

\[
\frac{d\rho}{dt} = -\frac{i}{\hbar} [\mathcal{H}, \rho] + \mathcal{L}_{\text{noise}}(\rho)
\]

Here, $\mathcal{H}$ is not a fixed object determined by layout alone, but is jointly determined by geometric structure, material parameters, coupling networks, packaging boundary conditions, and control waveforms. $\mathcal{L}_{\text{noise}}$ includes multiple mechanisms, such as dielectric loss, TLS defects, Purcell loss, crosstalk, and nonideal readout (Blais et al., 2021; McRae et al., 2020; Murray, 2021). Therefore, the statistical drift of quantum fabrication processes, device-frequency discreteness, and cryogenic-test feedback are more difficult to encapsulate once and for all as stable models than in mature CMOS.

Another limitation is that quantum chips have not yet formed a mature abstraction layer analogous to the standard-cell library of digital CMOS. Transmons, couplers, readout resonators, and control lines can be parameterized, but these PCells cannot be treated merely as geometric templates; they must also carry Hamiltonian models, noise parameters, process-statistical distributions, and applicable sign-off conditions. For this reason, the fabless quantum model must introduce additional abstraction layers such as SPICE-Q, PCells, Q-IP, Q-EDA, and cryogenic-test data feedback before the industrial division-of-labor logic of classical fabless can be translated into an executable engineering flow for quantum hardware (EDA-Q Collaboration, 2025; QPDK Documentation, 2025; GDSII-to-Wafer Collaboration, 2026).

\subsection{Classical Semiconductor Chip Providers}

After decoupling chip design from manufacturing, the classical semiconductor industry underwent significant structural expansion and eventually formed an industrial division-of-labor system composed of fabless companies, foundries, EDA tool vendors, IP providers, and packaging and testing firms. The essence of this transformation was not merely organizational optimization, but the conversion of chip design from a ``physical manufacturing problem'' into a ``constrained optimization and systems-design problem'' through the establishment of standardized interfaces and design abstraction layers (Mead \& Conway, 1980; Hennessy \& Patterson, 2019).

In this system, fabless companies are responsible for chip-architecture design, logic optimization, system-level performance tuning, and market-application definition; foundries are responsible for nanometer-scale manufacturing-process implementation and yield optimization; and EDA/IP enterprises provide software and design assets reusable across projects. The core advantage of this division-of-labor structure is that it encapsulates complex physical-manufacturing uncertainty within the process layer and provides stable interfaces upward through process design kits (PDKs, Process Design Kits). Research on modular production networks shows that this type of division of labor can exist because interfirm interfaces are highly codified, allowing trading parties to collaborate on innovation without sharing all internal knowledge (Sturgeon, 2002; Macher et al., 2008).

Formally, the classical chip-design process can be abstracted as:

\[
\theta^\ast = \arg\min_{\theta \in \mathcal{P}_{\text{CMOS}}} \Big( \mathcal{L}_{\text{perf}}(\theta) + \lambda \mathcal{C}_{\text{manufacturing}}(\theta) \Big)
\]

where $\mathcal{P}_{\text{CMOS}}$ denotes the manufacturable design space defined by the PDK, $\mathcal{L}_{\text{perf}}$ denotes the performance loss function, and $\mathcal{C}_{\text{manufacturing}}$ denotes the manufacturing-constraint cost function. The key significance of this abstraction is that design companies do not need to master transistor-level manufacturing recipes or production-line details; they need only optimize within the rule-constrained space, thereby substantially improving design-iteration speed and innovation efficiency.

The rise of the Fabless-Foundry model depends on the long-term co-maturation of three types of infrastructure. PDKs define design rules, layout-consistency checks, device models, and process corners, enabling design behavior to be signed off within a given manufacturing node. EDA toolchains automate logic synthesis, placement and routing, timing analysis, power optimization, and physical verification, transforming manual-experience-driven design flows into algorithm-driven engineering flows. Foundries amortize capital expenditure through wafer-scale mass production, statistical process control, and multi-customer platform reuse. Together, the three constitute the ``interface--tool--scale'' loop of the classical semiconductor industry.

From the perspective of system evolution, this model has produced an ``accelerating effect on design innovation.'' Fabless enterprises can rapidly iterate architectural designs independently of manufacturing nodes, while foundries continuously optimize process stability through statistical process control (SPC). This dual-wheel driving mechanism forms a typical positive feedback loop:

\[
\text{Design Innovation} \rightarrow \text{Higher Volume} \rightarrow \text{Process Refinement} \rightarrow \text{Lower Cost} \rightarrow \text{More Design Innovation}
\]

Historical studies show that this structure drove the continuous evolution of the integrated-circuit industry from micrometer-scale to nanometer-scale process nodes, and enabled the computing industry to expand within a highly specialized global value chain (Macher et al., 2008; Mack, 2011). It should be noted, however, that the success of this system depends on a crucial premise: device behavior must be capable of stable statistical modeling. Although CMOS transistors exhibit randomness at nanometer scales, their statistical distributions can be incorporated into SPICE models, process corners, and PDK parameter systems, providing a foundational guarantee for EDA automation. Therefore, the essence of the classical fabless system is not ``separation between design and manufacturing,'' but rather ``the conversion of physical complexity into a computable design space through standardized models.'' This also provides a direct methodological reference for the subsequent construction of the quantum fabless model.

\subsection{Current Status of Chinese Quantum-Chip Providers}

The current superconducting quantum-computing industry as a whole still operates mainly under the integrated device manufacturer (IDM, Integrated Device Manufacturer) model or a quasi-IDM model. That is, the design, fabrication, packaging, cryogenic testing, system calibration, and cloud operation of quantum chips are often completed by a single organization or a tightly coupled research-engineering system. International representative systems such as IBM Quantum and Google Quantum AI achieve full-stack control from device design to quantum cloud services through vertical integration. This route is reasonable during the early stage of hardware exploration, because device parameters, manufacturing drift, packaging conditions, and control systems require rapid closed-loop iteration (Arute et al., 2019; Kjaergaard et al., 2020).

The advantage of this IDM structure lies in its ability to rapidly iterate the closed loop between device design and experimental verification, especially in coherence-time optimization, readout-fidelity improvement, multi-qubit gate calibration, and system-error diagnosis. However, its systemic drawbacks are equally significant: capital expenditure is high, because cryogenic experimental facilities, nanofabrication lines, high-precision microwave measurement systems, and system-software teams must be maintained simultaneously; design knowledge and process experience are highly internalized, making reuse by external teams difficult; and the lack of standardized interfaces also makes it difficult for cross-regional collaboration, third-party tool development, and Q-IP transactions to scale. McKinsey's quantum-technology monitoring report also emphasizes that the quantum industry is moving from scientific validation toward commercial pilots, but value chains, supply chains, and application ecosystems are still in the formation stage, and the coordination capability among hardware, software, component, and application actors will directly affect the pace of commercialization (McKinsey \& Company, 2025).

From a systems-engineering perspective, the IDM model can be represented as a highly coupled closed-loop system:

\[
\mathcal{S}_{\text{IDM}} = \{ \mathcal{D}_{\text{design}} \leftrightarrow \mathcal{M}_{\text{fabrication}} \leftrightarrow \mathcal{T}_{\text{test}} \leftrightarrow \mathcal{C}_{\text{control}} \}
\]

where all subsystems are iteratively optimized within the same organization or tight alliance. Although this increases the speed of local convergence, it reduces the scalability of the overall ecosystem. For the quantum-computing industry, which requires large numbers of physical qubits, stable processes, and long-term application development, a single actor bearing full-stack R\&D over the long term will encounter bottlenecks in cost, talent, and data reuse.

In the development landscape of Chinese quantum chips, a relatively clear ``multi-center distributed IDM/quasi-IDM structure'' has currently taken shape. Public industrial materials show that Hefei has formed a superconducting quantum-computing cluster around the University of Science and Technology of China, Origin Quantum, and the National Quantum Information Science Infrastructure; Beijing has gathered universities, national laboratory platforms, and enterprises related to quantum software, photonic quantum technologies, and ion traps; Suzhou is more characterized by micro- and nanofabrication, semiconductor supporting industries, and industrial-park hosting capabilities; and Shenzhen has formed an application-driven ecosystem relying on the electronic-information industry, cloud services, and application enterprises (ECNS, 2024; ChinaQuantum, 2025; Entangled Future, 2026; The Quantum Insider, 2026). These regions are not simply clearly defined nodes in an industrial chain, but are at a stage of intersecting evolution among research capability, engineering capability, and application demand.

\textbf{Beijing} is more oriented toward universities, research institutions, quantum software, and algorithm ecosystems, with strengths in theoretical research, system software, platform prototypes, and interdisciplinary resources, while its engineering manufacturing capabilities are relatively dispersed. \textbf{Hefei}, relying on national-level experimental platforms, the USTC system, and actors such as Origin Quantum, has formed a relatively complete ``laboratory--engineering'' transition system, with representative strengths in superconducting quantum-computing prototypes, system integration, and quantum cloud services, and is gradually expanding toward engineering manufacturing (ECNS, 2024). \textbf{Suzhou} has a foundation in micro- and nanofabrication, semiconductor industrial support, and application-scenario coordination, and is a potential candidate region for a quantum-chip process node; however, if it is to assume the role of a quantum foundry, it still needs to form externally releasable standardized PDKs, PCell libraries, statistical models, test-chip data, and sign-off flows. \textbf{Shenzhen} is more oriented toward application-driven industrialization exploration, has advantages in the electronic-information industrial ecosystem, and participates in the construction of quantum cloud, educational hardware, and application systems through actors such as Huawei Cloud and SpinQ; however, cryogenic superconducting-device fabrication and fundamental-physics verification still rely more heavily on cross-regional collaboration or support from external platforms (Entangled Future, 2026; The Quantum Insider, 2026).

This multi-center structure helps form a multi-path exploration mechanism in the early stage, but in the absence of unified process standards (PDKs) and design abstraction layers (Q-EDA), it also creates systemic friction. Qubit parameters, layout rules, test structures, and data formats from different regions and teams are often difficult to migrate, resulting in repeated development of similar devices. The absence of unified Q-IP encapsulation methods for devices, couplers, readout structures, and control modules also makes it difficult for design assets to be traded and reused across projects. As a result, the threshold for industrial participation remains high, and small and medium-sized fabless quantum-chip design companies struggle to enter the market in the way classical semiconductor design companies do on the basis of stable PDKs and third-party EDA tools.

From the perspective of system evolution, China's quantum-chip ecosystem remains in a ``pre-standardization stage'': industrial development is driven mainly by major research projects, leading enterprises, and regional platforms, rather than jointly by open PDKs, interoperable Q-EDA, and Q-IP markets. It should be pointed out cautiously that this does not mean China lacks manufacturing capability or physical implementation capability. On the contrary, the core bottleneck is closer to ``distributed capabilities but insufficient interfaces.'' Only after cross-institutionally shared design abstraction layers, process-data standards, and cryogenic-test feedback mechanisms are established can the multi-center ecosystem move from parallel exploration toward reusable industrial coordination.

\subsection{Expanding Markets and Applications Through Fabless Quantum-Chip Design Providers}

To reach the scalable gate counts, numbers of logical qubits, and long-term stable operation required for fault-tolerant quantum computation (Fault-Tolerant Quantum Computation, FTQC), quantum-hardware systems need to evolve from the current vertically integrated (IDM) model dominated by experimental drive toward a standardized-interface-driven fabless ecosystem. Existing studies show that error-correction schemes such as the surface code require substantial physical-qubit redundancy, with resource overhead approximately expressible as:

\[
N_{\text{physical}} \sim \mathcal{O}(d^2 \cdot N_{\text{logical}})
\]

where $d$ denotes the error-correction code distance (code distance), and $N_{\text{logical}}$ denotes the number of logical qubits. This relationship is not itself exponential growth; however, when the target logical error rate is reduced, the number of logical qubits increases, and readout and wiring constraints are incorporated into system design, engineering resource requirements expand rapidly, and the difficulty of a single organization sustaining a full-stack IDM R\&D model over the long term rises accordingly (Fowler et al., 2012; Gambetta et al., 2017; Litinski, 2019; National Academies, 2024).

The fabless quantum ecosystem proposed in this paper can be understood as a combination of three types of infrastructure, rather than as a stack of isolated tools. The Q-EDA toolchain is responsible for automated mapping from quantum-circuit descriptions to physical layouts. Its core functions include quantum-circuit compilation, topology optimization, error-budget analysis, frequency allocation, crosstalk checking, and SPICE-Q-level multiphysics simulation. Standardized PDKs define the design-space boundaries of superconducting quantum chips, including Josephson-junction parameter distributions, coupling-strength ranges, resonator-frequency windows, material-loss statistics, and process-error models. Their objective is to convert manufacturing uncertainty into a computable and sign-off-ready set of design constraints:

\[
\mathcal{P}_{\text{PDK}} = \{ \theta \mid \theta_{\min} \leq \theta \leq \theta_{\max}, \; \mathbb{E}[\delta \theta] < \epsilon \}
\]

The Q-IP core market is responsible for consolidating verified physical-device IP, such as Transmon cells; circuit-level IP, such as readout resonators and tunable couplers; and logic-level IP, such as surface-code patches or subarray templates, into licensable, reusable, and composable design assets. The significance of this mechanism is not merely to reduce the cost of a single project, but more importantly to enable design knowledge to be reused across teams, projects, and application scenarios, thereby forming a long-tail innovation structure similar to the IP economy of classical semiconductors (Mead \& Conway, 1980; Weste \& Harris, 2011).

On top of this structure, fabless companies can transform the quantum-chip design process into a ``module-combination optimization problem,'' whose system objective can be expressed as:

\[
\theta^\ast = \arg\min_{\theta \in \mathcal{P}_{\text{PDK}}} \Big( \mathcal{L}_{\text{logical}}(\theta) + \lambda \mathcal{R}_{\text{noise}}(\theta) \Big)
\]

where $\mathcal{L}_{\text{logical}}$ denotes logical-computation error loss, and $\mathcal{R}_{\text{noise}}$ denotes the noise-accumulation risk function. Compared with the current experimental model of tuning parameters device by device, module-combination optimization can transfer part of the basic device trial-and-error process to the foundry-PDK layer for accumulation, allowing fabless teams to focus more on architecture, application mapping, error budgets, and system-level tradeoffs.

In the Chinese industrial context, if regions such as Suzhou, which possess micro- and nanofabrication capabilities and semiconductor-support foundations, can gradually form stable foundry nodes, fabless quantum-design companies will be able to conduct design, simulation, and tape-out verification under unified PDK constraints. This structure is expected to expand device trial and error from within a single laboratory into a multi-actor parallel design ecosystem, and to continuously optimize PDK parameter-distribution models through process-feedback data, thereby forming a Design--Fabrication--Calibration Loop. It must be emphasized that this transition will not occur automatically in the short term. It depends on whether foundry openness, PDK version stability, cryogenic-test throughput, Q-IP licensing systems, data-security rules, and application-market demand mature synchronously (CMC Microsystems, 2023; QPDK Documentation, 2025; GDSII-to-Wafer Collaboration, 2026; McKinsey \& Company, 2025).

Ultimately, the fabless model is expected to gradually form a cluster of small and medium-sized quantum-chip design companies in China, analogous to the long-tail fabless structure in the classical semiconductor industry. More precisely, however, the near-term significance of this goal is not to immediately replicate the mature scale of classical semiconductors, but first to establish engineering interfaces that external designers can trust, enabling more application-oriented teams to participate in quantum-hardware innovation without repeatedly constructing underlying manufacturing capabilities.

\section{Technical Foundations of Superconducting Quantum Chips}

Starting from the device-physics foundations of superconducting quantum computing, this section systematically summarizes Transmon qubits, readout resonators, coupling structures, material-induced decoherence mechanisms, and the mapping from physical qubits to logical error-correction structures. For Fabless quantum chip design, these topics are not merely independent physical background; rather, they constitute the physical sources of subsequent PDK, PCell, SPICE-Q, and Q-EDA signoff rules: every reusable design unit must simultaneously satisfy Hamiltonian parameters, noise models, layout constraints, and cryogenic test verifiability.

Related studies show that the central challenge in superconducting quantum systems lies in jointly optimizing the trade-offs among coherence time, gate fidelity, readout speed, frequency assignability, crosstalk suppression, material loss, and the packaging environment (Koch et al., 2007; Devoret \& Schoelkopf, 2013; Kjaergaard et al., 2020; Blais et al., 2021; Levenson-Falk \& Shanto, 2025). Therefore, the aim of this section is not to exhaust the theory of superconducting quantum circuits, but to extract the physical constraints that most strongly affect Fabless design abstractions.

\subsection{Transmon Physics and Design Variables}

The Transmon qubit is one of the most widely adopted implementations in superconducting quantum computing today. In essence, it is a weakly nonlinear oscillator composed of a Josephson junction and a capacitive structure. Compared with the earlier Cooper-pair box, the Transmon substantially reduces sensitivity to charge noise by increasing the ratio $E_J/E_C$, but at the cost of reduced anharmonicity; consequently, engineering trade-offs must be made among addressability, gate speed, frequency crowding, and leakage errors (Koch et al., 2007; Krantz et al., 2019; Kjaergaard et al., 2020). Its Hamiltonian can be written as:

\[
\mathcal{H}_{\text{transmon}} = 4E_C (n - n_g)^2 - E_J \cos(\phi)
\]

where $E_C$ is the charging energy, $E_J$ is the Josephson energy, $n_g$ is the background charge offset, and $\phi$ is the superconducting phase difference. In the Transmon limit $E_J/E_C \gg 1$, the dispersive sensitivity of the energy levels to charge offset is reduced, thereby improving coherence time; however, because the level spacings are approximately equidistant, the anharmonicity is typically of the same order as $E_C$ and cannot be suppressed arbitrarily.

From the perspective of chip design, the key variables of a Transmon include the Josephson junction area and oxide thickness, capacitor-pad geometry, target operating frequency, anharmonicity, coupling capacitance, coupling to packaging modes, and parasitic coupling to control/readout lines. These variables are not optimized independently: increasing the coupling strength may accelerate gate operations or readout, but it may also aggravate Purcell loss and crosstalk; adjusting the capacitor geometry can reduce surface participation, but it will change chip area, frequency distribution, and packaging modes. Recent studies on high-coherence Transmon materials also indicate that the material system and surface/interface quality directly affect $T_1$ and $T_2$. Accordingly, a Transmon PCell cannot contain only geometric parameters; it must also be associated with materials, processes, and loss models (McRae et al., 2020; Murray, 2021; Place et al., 2021).

Thus, the system is essentially a ``parameterized quantum nonlinear-oscillator design problem'' and may be viewed as a core PCell primitive in the Fabless design space. For external Fabless designers, a qualified Transmon PCell should expose the target frequency, anharmonicity, tunable parameter ranges, manufacturing statistical distributions, and model-validity boundaries, rather than merely providing a drawable layout template.

\subsection{Readout Resonators and Purcell Constraints}

State readout of superconducting qubits is usually realized through a circuit quantum electrodynamics (cQED) system: the qubit is dispersively coupled to a readout resonator, the qubit state changes the resonator response, and the measurement is then completed through feedlines, amplifiers, and room-temperature electronics. The readout chain must simultaneously pursue a high signal-to-noise ratio, short measurement time, low crosstalk, and weak back-action. It is therefore not a peripheral circuit appended to the qubit, but a core design object that determines system scalability (Blais et al., 2021).

Coupling between the readout resonator and the qubit gives rise to the Purcell effect, namely radiative decay of the qubit through the resonator and external feedline, thereby limiting the $T_1$ relaxation time. The Purcell limit can be approximated as:

\[
T_1^{\text{Purcell}} \approx \frac{\Delta^2}{g^2 \kappa}
\]

where $\Delta$ is the detuning between the qubit and the resonator, $g$ is the coupling strength, and $\kappa$ is the resonator loss rate or external coupling linewidth. This expression shows that increasing readout speed usually requires a larger $g$ or $\kappa$, but doing so compresses the Purcell-limited lifetime; increasing detuning can protect the qubit, but may reduce readout signal strength and measurement speed (Blais et al., 2021).

In engineering practice, this tension is often mitigated through Purcell filters, frequency-selective coupling, readout multiplexing, and optimization of the amplification chain. Reed et al.'s work on Purcell filtering demonstrates that a suitable filter can suppress the spontaneous-emission channel while maintaining strong readout coupling; this idea subsequently became an important constraint in the design of superconducting readout chains (Reed et al., 2010). For a Fabless system, a readout PCell should not merely describe the geometric dimensions of a CPW resonator; it should also include impedance continuity, the target frequency window, the Purcell limit, feedline coupling, multiplexing spacing, packaging modes, and measurability metrics. In the SPICE-Q model, this problem corresponds to a ``readout subcircuit constrained-optimization problem'' and is also a physical constraint that must be explicitly checked during Q-EDA signoff.

\subsection{Couplers, Frequency Crowding, and Layout}

In multi-qubit superconducting chips, interactions among qubits are realized through capacitances, inductances, bus resonators, or tunable couplers. The corresponding coupling Hamiltonian can usually be simplified as:

\[
\mathcal{H}_{\text{coupling}} = g_{ij} (a_i^\dagger a_j + a_i a_j^\dagger)
\]

where $g_{ij}$ denotes the effective coupling strength between qubits $i$ and $j$. Coupling in an actual chip is not determined solely by an ideal parameter, but is jointly affected by coupler geometry, the electromagnetic environment, residual coupling, packaging modes, and control waveforms. As the system size increases, frequency crowding becomes a major limiting factor: unintended proximity or hybridization may occur among different qubits, readout resonators, and coupling modes, leading to crosstalk, leakage, increased gate errors, or reduced usable chip yield.

Fixed-frequency Transmon processors are particularly affected by fabrication variability in Josephson junctions, and deviations between target and actual frequencies can significantly reduce usable chip yield. Frequency assignment has therefore been formalized as mixed-integer programming, linear programming, or other constrained-optimization problems in order to maximize the probability of avoiding frequency collisions and the usable chip yield under manufacturing statistical fluctuations (Murray et al., 2021; McDonald et al., 2022; Shi et al., 2024). This problem is essentially a high-dimensional constrained layout-optimization problem. It is structurally similar to placement and routing in classical EDA, but the objective function must explicitly incorporate the quantum spectrum, crosstalk, and error budgets.

Layout design must not only ensure that the coupling topology satisfies the logical gate structure, but also simultaneously control frequency spacing, microwave-routing crosstalk, readout multiplexing windows, parasitic coupling of control lines, ground continuity, packaging cavity modes, and manufacturing repeatability. For Q-EDA, this means that a layout is not merely the result of geometric drawing, but a design object jointly constrained by electromagnetic fields, Hamiltonians, noise models, and manufacturing rules. Thus, quantum layout design is essentially a ``geometric optimization problem under electromagnetic-field constraints'' and is also one of the key physical constraints that Q-EDA must handle in addition to those in classical EDA (EDA-Q Collaboration, 2025; GDSII-to-Wafer Collaboration, 2026; Levenson-Falk \& Shanto, 2025).

\subsection{Materials, Interfaces, and Decoherence}

The performance bottleneck of superconducting qubits arises primarily from decoherence mechanisms. Common physical sources include two-level-system (TLS) defects, dielectric loss at metal--dielectric and substrate--vacuum interfaces, surface charge noise, flux noise, thermally excited quasiparticles, and additional losses introduced by the packaging and radiation environment. For large-scale chips, these noise sources are not merely lifetime issues for individual devices; they also affect the credibility of PDK models through batch-to-batch fluctuations, spatial nonuniformity, and incomplete test data (Mueller et al., 2019; McRae et al., 2020; Murray, 2021).

The decoherence process is commonly described by the Lindblad master equation:

\[
\frac{d\rho}{dt} = -\frac{i}{\hbar}[\mathcal{H}, \rho] + \sum_k \mathcal{L}_k(\rho)
\]

where $\mathcal{L}_k$ denotes different noise channels (Breuer \& Petruccione, 2002). In engineering modeling, this expression cannot remain at the formal level; it must be translated into parameters that are measurable, fitable, and usable in simulation, such as $T_1$, $T_2$, frequency-noise spectra, resonator internal quality factors, dielectric loss tangents, and quasiparticle-related loss rates.

Materials-engineering optimization plays a critical role in improving $T_1$ and $T_2$ times, including metal-film selection, interface cleaning, control of etching residues, oxide-layer quality, surface treatment, and packaging-contamination control. Superconducting microwave resonators are often used as proxy test structures for material loss and interfacial TLS loss, and the corresponding measurement results can be fed back into the noise models and statistical-corner definitions in the PDK (McRae et al., 2020; Murray, 2021). Place et al.'s work on tantalum-based Transmons further shows that changes in the material platform can significantly improve the coherence time of two-dimensional Transmons. This reinforces a conclusion that is highly important for the Fabless model: a PDK cannot publish only geometric design rules; it must also publish the material stack, surface treatments, loss statistics, and applicable process windows (Place et al., 2021). In the Fabless model, this layer corresponds to the ``statistical noise-model layer'' in the PDK and is also the core basis for determining whether the same PCell can be reused across fabrication batches.

\subsection{From Physical Qubits to Logical Patches}

Because a single physical qubit is limited by decoherence, control errors, and measurement errors, it cannot directly support long-duration fault-tolerant computation. Logical qubits must therefore be constructed through quantum error-correction codes. The surface code is one of the most frequently discussed fault-tolerant routes in superconducting quantum computing, because its local nearest-neighbor coupling structure is well matched to two-dimensional superconducting qubit arrays from an engineering perspective (Fowler et al., 2012; Gambetta et al., 2017; Litinski, 2019).

The logical error rate can be approximated as:

\[
p_L \sim \left(\frac{p}{p_{\text{th}}}\right)^{(d+1)/2}
\]

where $p$ is the physical error rate, $p_{\text{th}}$ is the threshold, and $d$ is the code distance (Fowler et al., 2012). The applicability of this expression presupposes that the physical error rate is below the threshold and that the error model, syndrome measurement, and decoding workflow satisfy the corresponding assumptions. Practical engineering systems must also address leakage errors, correlated errors, measurement crosstalk, real-time feedback latency, and routing congestion. Experiments by Barends et al. show that superconducting quantum-gate fidelities can approach the threshold requirements of the surface code, but moving from a small number of high-fidelity gates to scalable logical patches still requires massively parallel readout, stable calibration, and system-level error management (Barends et al., 2014; National Academies, 2024).

This structure maps physical devices into a ``logical patch.'' From a Fabless perspective, this means that the PCell abstraction cannot remain forever at the level of a single Transmon, readout resonator, or coupler. As the system matures, PCells and Q-IP need to advance to the level of logical units that include data qubits, measurement qubits, syndrome-extraction circuits, readout multiplexing, and decoding interfaces. A PDK likewise should not provide only device-level parameters; it should gradually provide logical-level error budgets, supported topological constraints, synchronous-measurement rules, and boundaries for cryogenic/room-temperature control interfaces.

Therefore, the transition from physical qubits to logical patches constitutes the most critical abstraction leap in a Fabless quantum chip system. It connects underlying materials, devices, and layout problems to system architecture, Q-IP reuse, and resource estimation for fault-tolerant computing, and it also determines whether the quantum Fabless ecosystem can progress from ``device-design reuse'' to ``logical-module reuse.''

\section{Quantum PDKs as Core Products}

In the Fabless quantum chip design paradigm, the Process Design Kit (PDK) constitutes the core infrastructure layer of the entire ecosystem. A PDK is not only a collection of process parameters, but also a unified abstraction interface connecting physical fabrication, layout design, device models, cryogenic testing, and logical architecture. The experience of the classical semiconductor industry shows that PDKs can support the Fabless ecosystem because they encapsulate the internal process capabilities of the Foundry into design rules, device models, layer definitions, standard-cell/IP interfaces, and signoff flows. The experience of the open SkyWater SKY130 PDK also indicates that PDK release, version tagging, tool adaptation, and test-silicon validation jointly determine whether external designers can trust the process interface (Mead \& Conway, 1980; Macher et al., 2008; SkyWater PDK Authors, 2020).

On this basis, superconducting quantum PDKs become further complicated. At a minimum, they must include layer definitions, material stacks and process windows, PCell libraries, Josephson junction/resonator/coupler models, DRC/LVS/DFM rules, Hamiltonian and noise models, statistical corners, test-chip structures, cryogenic measurement data feedback, and manufacturing data-package generation mechanisms. Recent work such as QPDK, KQCircuits, Qiskit Metal/Quantum Metal, EDA-Q, and GDSII-to-wafer has already shown the prototype of this technology stack from different perspectives: code-based PCell generation, GDSII layout output, electromagnetic/circuit model coupling, KLayout technology files, design-rule checking, test-chip examples, and data-conversion workflows from \texttt{tape-out} to \texttt{fab-out} (Minev et al., 2021; KQCircuits Documentation, 2024; QPDK Documentation, 2025; EDA-Q Collaboration, 2025; GDSII-to-Wafer Collaboration, 2026).

Unlike a classical CMOS PDK, a quantum PDK must simultaneously describe geometric manufacturability, electromagnetic behavior, quantum dynamics, material loss, and decoherence noise processes. Thus, the PDK in a Fabless quantum chip system is essentially a ``system-level constraint definition space across physical layers'':

\[
\mathcal{P}_{\text{QPDK}} = \{ \mathcal{L}, \mathcal{D}, \mathcal{C}_{\text{PCell}}, \mathcal{M}, \mathcal{S}, \mathcal{N}, \mathcal{T}, \mathcal{G} \}
\]

where $\mathcal{L}$ denotes layer definitions and the material/process stack, $\mathcal{D}$ denotes the design-rule set (DRC/LVS/DFM), $\mathcal{C}_{\text{PCell}}$ denotes the parameterized component library, $\mathcal{M}$ denotes physical model cards and Hamiltonian interfaces, $\mathcal{S}$ denotes statistical distribution models, $\mathcal{N}$ denotes noise and loss models, $\mathcal{T}$ denotes test structures and measurement data, and $\mathcal{G}$ denotes release governance and signoff interfaces.

The core dependency relationship of the Fabless system can be summarized as:

\[
\text{QPDK} \Rightarrow \{\text{PCell}, \text{SPICE-Q}, \text{Q-DRC/LVS/DFM}, \text{Model Cards}, \text{Test Chips}\}
\]

Here, PCell and SPICE-Q respectively undertake the roles of structural abstraction and physical simulation. Q-DRC/LVS/DFM is responsible for guaranteeing the manufacturability and executability boundaries of the design space, while model cards and test-chip data maintain the trust loop between design prediction and real cryogenic experiments.

\subsection{PDK Content and Version Management}

The version-management problem of a quantum PDK is essentially a dual-constraint problem of ``physical-system evolution and design compatibility.'' Because superconducting quantum devices exhibit significant process drift, material batch differences, Josephson junction variability, and cryogenic test fluctuations, fabrication in different batches can change the distributions of frequency, coupling, loss, and readout parameters. A PDK therefore cannot be regarded as a static document; it should instead be treated as a versioned engineering product that is continuously updated with process statistics and test-chip data.

This versioned relationship can be expressed as:

\[
\mathcal{P}^{(t)} = \{ \mathcal{L}^{(t)}, \mathcal{D}^{(t)}, \mathcal{M}^{(t)}, \mathcal{S}^{(t)}, \mathcal{N}^{(t)}, \mathcal{T}^{(t)} \}
\]

where each version describes not only geometric rules and process parameters, but also traceable experimental calibration records, test-chip batches, model-fitting errors, cryogenic measurement conditions, and scopes of applicability. For Fabless designers, a version number means more than a ``file update''; it indicates whether a particular set of PCells, model cards, and signoff rules can still be used for a new design project.

Version updates must satisfy a stability constraint:

\[
\|\mathcal{P}^{(t+1)} - \mathcal{P}^{(t)}\|_{\text{KL}} < \epsilon
\]

This expression can be understood as follows: the key statistical distributions in the new PDK version must not exhibit undeclared severe drift relative to the previous version; otherwise, an old design may fail at the frequency-collision, readout-mismatch, or decoherence boundary even if it passes geometric DRC. In actual governance, this constraint should be implemented through release notes, migration guides, deprecation policies, regression-test layouts, and cryogenic benchmark data, rather than relying solely on version numbering.

Existing studies indicate that frequency drift in superconducting quantum devices, Josephson junction variability, TLS defect distributions, and material-interface loss are key factors limiting large-scale uniformity. Therefore, the PDK versioning system must be tightly coupled with the experimental feedback mechanism (Koch et al., 2007; McRae et al., 2020; McDonald et al., 2022; Place et al., 2021; Levenson-Falk \& Shanto, 2025). Tools such as QPDK and KQCircuits show that code-based PCells and KLayout technology files can significantly improve version management and regression-testing capabilities; however, in a quantum PDK, what truly determines version credibility remains whether the deviation between model predictions and cryogenic measurements is continuously recorded and converged (KQCircuits Documentation, 2024; QPDK Documentation, 2025).

\subsection{Design Rules for Quantum Manufacturability}

Quantum Design Rule Checking (Q-DRC) extends the DRC system in classical EDA, but its constraints derive not only from geometric structures, but also from quantum electromagnetic coupling, cryogenic physics, material loss, and system-level error budgets. Classical DRC mainly answers whether a layout can be manufactured, whereas Q-DRC must also answer whether the fabricated quantum device is still likely to fall within usable Hamiltonian and noise boundaries. Therefore, Q-DRC should constitute a signoff flow together with LVS, DFM, electromagnetic simulation, frequency assignment, and SPICE-Q model checking.

In a superconducting quantum PDK, geometric rules remain fundamental, including Josephson junction dimensions, current density, interlayer alignment, CPW linewidth/spacing, air bridges or crossover structures, ground continuity, and chip-edge keep-out margins. However, these rules alone are insufficient to guarantee correct quantum functionality. The PDK must also define microwave impedance matching, readout resonator frequency windows, qubit frequency spacing, upper bounds on non-target coupling, Purcell-loss boundaries, avoidance of packaging cavity modes, thermal-noise paths, quasiparticle injection paths, and cryogenic routing constraints. Recent GDSII-to-wafer work further emphasizes that wafer-level quantum chip fabrication must incorporate layer mapping, mask-data preparation, process-stack mapping, and manufacturing file-package integrity into \texttt{fab-out} checks, rather than remaining at traditional \texttt{tape-out} file delivery (GDSII-to-Wafer Collaboration, 2026).

These constraints can be uniformly represented as a feasible design space:

\[
\mathcal{F}_{\text{Q}} = \{ \theta \mid g_{ij} < g_{\max}, \; |\omega_i - \omega_j| > \Delta_{\min}, \; Q_{\text{loss}} > Q_{\text{threshold}}, \; \Gamma_{\text{Purcell}} < \Gamma_{\max} \}
\]

where $g_{ij}$ denotes residual or target coupling strength, $\omega_i$ denotes device frequency, $Q_{\text{loss}}$ denotes a loss-related quality factor, and $\Gamma_{\text{Purcell}}$ denotes the radiative loss rate through the readout environment. The core role of Q-DRC is to translate these continuous physical constraints into discrete rules that can be automatically checked by Q-EDA, thereby reducing the risk of omissions in manual layout review (Blais et al., 2021; Reed et al., 2010). Unlike classical DRC, superconducting quantum chips must explicitly incorporate constraints such as frequency collisions, crosstalk, Purcell loss, packaging modes, and cryogenic thermal paths into the signoff flow, a point that has been systematically discussed in recent Q-EDA and GDSII-to-wafer workflows (EDA-Q Collaboration, 2025; GDSII-to-Wafer Collaboration, 2026).

\subsection{Model Cards and Statistical Corners}

Model Cards are used to describe the statistical behavior of quantum devices under different process conditions. Their goal is to transform experimental measurement results into computable, auditable, and reproducible probabilistic models and embed them in the PDK system. The concept of a ``model card'' was originally developed for transparent reporting of machine-learning models, emphasizing that model releases must specify intended use, performance boundaries, evaluation data, and usage limitations. When transferred to quantum PDKs, this idea is equally applicable to the release governance of device models, noise models, and statistical corners (Mitchell et al., 2019). For superconducting quantum chips, a model card should provide not only nominal parameters, but also data sources, measurement temperature ranges, sample sizes, fitting methods, confidence intervals, applicable layout ranges, failure modes, and version dependencies. Otherwise, designers may easily misuse empirical parameters from a particular laboratory, batch, or test structure as cross-Foundry general models.

Quantum device parameters usually exhibit non-ideal distributions:

\[
\theta \sim \mathcal{N}(\mu, \sigma^2) + \mathcal{T}_{\text{TLS}} + \mathcal{J}_{\text{JJ}} + \mathcal{P}_{\text{package}}
\]

where $\mathcal{T}_{\text{TLS}}$ denotes non-Gaussian tail effects introduced by two-level-system defects, $\mathcal{J}_{\text{JJ}}$ denotes Josephson junction fabrication variability, and $\mathcal{P}_{\text{package}}$ denotes systematic shifts caused by packaging and environmental modes. This expression is not a strict single statistical model; rather, it serves as a reminder that a PDK must simultaneously handle Gaussian batch drift, non-Gaussian defect tails, and system-level coupling errors.

``Statistical Corners'' define worst-case or representative boundaries to support robust simulation and signoff. For superconducting quantum chips, statistical corners should at least cover the lowest $T_1/T_2$ lifetime boundary, maximum frequency drift, weakest/strongest coupling strength, readout resonator Q-factor distribution, strongest crosstalk case, TLS-loss tails, Purcell-loss upper bounds, frequency-collision probabilities, and packaging-mode coupling. These dimensions correspond respectively to the existing research foundations of materials testing, frequency assignment, electromagnetic simulation, and cryogenic system testing (McRae et al., 2020; Murray et al., 2021; Place et al., 2021; Levenson-Falk \& Shanto, 2025).

These statistical corners are key inputs for SPICE-Q robust simulation, enabling Fabless design to perform worst-case analysis and yield-aware design and thereby improve fabrication success rates. S-parameter models, CPW parameter models, Josephson junction models, and test-chip examples in QPDK indicate that model libraries can extend from layout objects to composable circuit/electromagnetic models. For Fabless signoff, however, the more critical requirement is that each model card specify ``within which processes, temperature ranges, frequencies, and layout ranges the model is valid,'' preferably in the form of machine-readable metadata bound to the PDK version (QPDK Documentation, 2025; Wilkinson et al., 2016).

\subsection{Test Structures and Data Feedback}

A quantum PDK must include dedicated test structures for extracting key physical parameters and feeding them back into process-model updates. Similar to process control monitors (PCMs) and test chips in classical semiconductors, the role of quantum test structures is not to validate a particular product function, but to measure the stability, repeatability, and model error of the process platform itself. Typical test structures should cover single-qubit $T_1/T_2$ measurement arrays, Josephson junction critical-current and frequency-monitoring structures, CPW/readout resonator Q-factor structures, coupling-strength calibration structures, TLS noise-spectrum measurement structures, Purcell/readout-chain test structures, packaging-mode test structures, and cross-wafer positional distribution monitoring structures.

The data-feedback mechanism can be expressed as:

\[
\mathcal{P}^{(t+1)} = \mathcal{P}^{(t)} + \eta \cdot (\mathcal{D}_{\text{measured}} - \mathcal{D}_{\text{predicted}})
\]

This expression indicates that the PDK should be updated according to the deviation between measured data and model predictions, but the actual process requires stricter governance: the Foundry should distinguish single anomalies, batch drift, tool drift, and systematic model mismatch, and map the update results to PCell parameter ranges, model-card confidence intervals, statistical corners, and design-rule changes. For external Fabless designers, data feedback cannot remain only in internal Foundry reports; it should take the form of publishable summaries, version notes, and revalidation requirements. According to the FAIR data principles, key test data and their derived models should, as far as possible, have metadata structures that are findable, accessible, interoperable, and reusable, so that different Q-EDA tools, PDK versions, and Foundry workflows can trace the origin and applicability boundaries of the same parameter (Wilkinson et al., 2016).

This closed-loop system constitutes the learning core of the Fabless-Foundry ecosystem, enabling process models to improve continuously with manufacturing data (Kjaergaard et al., 2020). Material-loss testing, Josephson junction frequency trimming, resonator Q-factor measurements, and cross-batch test-chip data are especially suitable as model-feedback entry points and can be used to update PDK statistical corners and PCell parameter distributions (McRae et al., 2020; McDonald et al., 2022; CMC Microsystems, 2023). QPDK test-chip examples and the GDSII-to-wafer workflow also show that a mature quantum PDK should not provide only individual device cells, but should include standardized test layouts, measurement metadata, and manufacturing data packages for the continuous calibration of the process platform (QPDK Documentation, 2025; GDSII-to-Wafer Collaboration, 2026).

\subsection{Release Governance and Trust Mechanisms}

Because a quantum PDK directly determines chip-design correctness, physical manufacturability, and the predictability of cryogenic experiments, its release must be supported by rigorous governance and trust mechanisms. The experience of classical open PDKs shows that whether an external ecosystem can form depends not only on whether files are made public, but also on the clarity of documentation completeness, toolchain adaptation, version tagging, validation status, licensing terms, and test-silicon feedback (SkyWater PDK Authors, 2020). For quantum PDKs, this point is even more important because design rules, model cards, and cryogenic measurement data are more strongly coupled.

A trustworthy PDK system should at least satisfy traceability, verifiability, process consistency, cross-tool compatibility, access-permission governance, and data-security governance. Traceability requires that every PCell, model card, and design rule be traceable to process versions, test structures, and measurement batches. Verifiability requires that external designers be able to reproduce experimental boundaries through public or controlled test layouts, regression cases, and signoff scripts. Process consistency requires the Foundry to continuously publish cross-batch statistical summaries. Cross-tool compatibility requires the PDK to be callable by Q-EDA, KLayout/KQCircuits, GDSII/OASIS conversion tools, and SPICE-Q simulators. Access-permission and data-security governance are used to distinguish public design rules, controlled process parameters, and commercially confidential data. If FAIR principles are used to organize PDK metadata and test data, PDK release can be upgraded from ``document download'' to an ``engineering interface that can be automatically discovered, parsed, verified, and reused by toolchains'' (Wilkinson et al., 2016).

The trustworthiness function can be defined as:

\[
\mathcal{T}(\mathcal{P}) = P(\text{fabrication consistency} \mid \text{design constraints}, \text{model validity}, \text{test feedback})
\]

When $\mathcal{T}(\mathcal{P}) \rightarrow 1$, the Fabless ecosystem acquires the conditions for scaled expansion. This function should not be understood as the success rate of a single tape-out, but rather as the conditional probability that, within the given PDK rules and the applicable scope of the models, external designers can repeatedly obtain predictable fabrication results.

Therefore, a PDK is not only a technical product, but also a ``trust infrastructure'' connecting design, manufacturing, and verification. Its governance mechanism determines the upper bound and scalability of the entire quantum Fabless industry. For external Fabless designers, a trustworthy PDK should also provide versioned release notes, design-rule change logs, model scopes of applicability, lists of validated PCells, test-chip data summaries, signoff scripts, failure-case explanations, and fab-out manufacturing file specifications. Together, these contents determine whether a design can move from a one-time \texttt{tape-out} to a repeatable \texttt{fab-out} manufacturing flow (CMC Microsystems, 2023; QPDK Documentation, 2025; GDSII-to-Wafer Collaboration, 2026).

\section{PCell Libraries for Physical and Logical Components}

In the classical semiconductor industry, standard cell libraries and Parameterized Cells (PCells) jointly support large-scale chip design. Digital-circuit designers can directly invoke validated standard cells such as NAND, NOR, and Flip-Flop cells; analog and RF-circuit designers rely more heavily on PCells with adjustable dimensions, regenerable layouts, and associated models to explore the design space (Mead \& Conway, 1980; Weste \& Harris, 2011). Similarly, the Fabless quantum chip ecosystem needs to establish standardized, reusable, and verifiable quantum component libraries, but their abstraction objects are not merely geometric layouts; rather, they are design units jointly constrained by ``geometry--electromagnetics--Hamiltonian--noise--process statistics.''

Parameterized components are an important part of a quantum PDK. In essence, they are reusable design objects that simultaneously contain geometric structures, physical parameters, simulation models, manufacturing rules, and test-data boundaries. Unlike classical digital standard cells, quantum PCells not only describe layout structures, but must also contain the corresponding Hamiltonian models, noise parameters, manufacturing statistical information, and applicable signoff conditions (Kjaergaard et al., 2020; Levenson-Falk \& Shanto, 2025). Tools such as Qiskit Metal/Quantum Metal, KQCircuits, QPDK, and EDA-Q have already demonstrated feasible routes for generating layouts of Transmons, CPW resonators, couplers, readout lines, and test chips from code. SQuADDS further demonstrates a database-oriented route that supports device-design reuse using pre-simulation and experimentally validated data (Minev et al., 2021; KQCircuits Documentation, 2024; Shanto et al., 2024; QPDK Documentation, 2025; EDA-Q Collaboration, 2025).

In engineering implementation, a PCell should expose a limited set of high-level parameters with clear physical meaning, such as target frequency, anharmonicity, coupling capacitance, readout resonator frequency, layout occupancy boundary, material/process version, and model scope of applicability. Internally, it should encapsulate geometric generation rules, electromagnetic simulation interfaces, SPICE-Q models, DRC/LVS/DFM constraints, and test-chip calibration records. As a result, what designers invoke is not an arbitrary layout-library object, but a manufacturable design unit constrained by the PDK, validated by models, and calibrated with experimental data.

The Fabless design flow can be abstractly represented as:

\[
\text{PCell Parameters}
\rightarrow
\text{SPICE-Q Simulation}
\rightarrow
\text{Layout Generation}
\rightarrow
\text{DRC/LVS Verification}
\rightarrow
\text{Tape-out/Fab-out}
\]

In a mature setting, designers need not directly handle the low-level process details of every Josephson junction, CPW cross-section, or readout resonator, but they still need to understand the physical meaning, scope of applicability, and PDK-version boundaries of PCell parameters. This design-abstraction approach has engineering significance similar to that of standard cells and analog PCells in classical EDA. However, because quantum devices are more sensitive to frequency, loss, crosstalk, packaging, and cryogenic calibration, their verification burden is substantially heavier than that of classical digital standard cells.

Furthermore, as fault-tolerant quantum computing develops, the abstraction level of PCells will gradually expand from physical qubits to logical qubits. For example, a PDK may directly provide a surface-code logical-qubit PCell with distance 3 (Distance-3). What chip designers obtain will no longer be a single Transmon structure, but a validated logical error-correction module that internally contains multiple data qubits and measurement qubits, and that automatically carries the corresponding SPICE-Q model, readout schedule, statistical parameters, and resource-overhead description.

Thus, a PCell is not only a design component, but also an encapsulation form for quantum hardware knowledge. Its role is analogous to that of IP cores and standard cell libraries in the classical industry. It determines the abstraction level at which Fabless designers can reuse hardware knowledge, and also determines whether a PDK can be upgraded from a collection of documents to a genuinely callable, verifiable, and tradable engineering platform.

\subsection{Physical-Qubit PCells}

Physical-qubit PCells are the foundational components of the entire quantum PCell library. In the current field of superconducting quantum computing, the most widely adopted structure is the Transmon qubit. Its design objective is to retain sufficient anharmonicity while reducing sensitivity to charge noise, so as to enable addressable control and low-leakage gate operations (Koch et al., 2007; Krantz et al., 2019).

The basic Hamiltonian of a Transmon is:

\[
\mathcal{H}=4E_C(n-n_g)^2-E_J\cos\phi
\]

where $E_J$ is the Josephson energy and $E_C$ is the charging energy. A physical-qubit PCell should connect these Hamiltonian parameters with the actual layout and process variables, including the target frequency $\omega_q$, anharmonicity $\alpha$, capacitor-pad geometric parameters, Josephson junction dimensions, coupling capacitance, target $T_1/T_2$ metrics, material/interface loss parameters, SPICE-Q model files, GDS/OASIS layout generation rules, and PDK-version dependencies.

The high-level parameters modified by designers can be summarized as:

\[
\theta=\{\omega_q,\alpha,Q,g,\mathcal{B}_{\text{layout}},\mathcal{V}_{\text{PDK}}\}
\]

where $Q$ denotes the loss or quality-factor target, $g$ denotes the target coupling strength, $\mathcal{B}_{\text{layout}}$ denotes layout boundaries and connection constraints, and $\mathcal{V}_{\text{PDK}}$ denotes the applicable PDK version. Layout candidates generated by the PCell must enter electromagnetic simulation, parameter extraction, Hamiltonian fitting, noise evaluation, and PDK rule-checking flows. Design databases such as SQuADDS show that a searchable mapping can be established between pre-simulated device geometries and target Hamiltonian parameters, thereby providing physical PCells with a reuse mechanism of ``initial design candidates + validation data'' (Shanto et al., 2024).

This abstraction allows Fabless enterprises to reuse validated Transmon design templates, but it does not imply complete elimination of device-level verification. Target frequency, anharmonicity, coupling strength, material loss, and packaging coupling still need to be jointly calibrated through simulation models and test-chip data (Koch et al., 2007; Minev et al., 2021; Place et al., 2021; Levenson-Falk \& Shanto, 2025).

\subsection{Readout and Control PCells}

Readout and control structures are important components of quantum processors, and their complexity is usually no less than that of the qubits themselves (Blais et al., 2021). In a Fabless design flow, readout and control PCells cannot be regarded as peripheral routing templates; they should instead be treated as system-level components jointly coupled to qubit lifetime, readout fidelity, multiplexing capability, packaging modes, and control crosstalk.

Typical readout PCells include CPW readout resonators, Purcell filters, feedline structures, readout multiplexing structures, parametric readout/amplification interfaces, and packaging transition structures. The dispersive readout relation can be approximated as:

\[
\chi \approx \frac{g^2}{\Delta}
\]

where $g$ is the coupling strength between the qubit and the readout resonator, and $\Delta$ is the detuning. This expression reminds designers that enhancing the readout signal usually affects Purcell loss and qubit lifetime; therefore, a readout PCell must simultaneously carry the target readout frequency, linewidth, coupling Q, multiplexing spacing, Purcell limit, and packaging-mode constraints (Blais et al., 2021; Reed et al., 2010).

Control PCells include XY control lines, flux-bias lines, Z-control structures, microwave packaging interfaces, ground and shielding structures, and connection boundaries with room-temperature/cryogenic control electronics. These structures can be generated parametrically, but they must be jointly verified under constraints such as impedance continuity, crosstalk, thermal load, cryogenic filtering, parasitic coupling, and readout fidelity. Therefore, readout and control PCells should simultaneously be associated with electromagnetic simulation, noise models, packaging models, and layout signoff rules, rather than being reused merely as metal shapes in GDS (Blais et al., 2021; Levenson-Falk \& Shanto, 2025).

\subsection{Coupler and Subarray PCells}

When the system scale expands to tens or even hundreds of qubits, the coupling network gradually becomes a key factor determining performance. Couplers determine not only two-qubit gate speed, but also residual ZZ coupling, crosstalk, frequency crowding, leakage errors, and calibration complexity. The effective coupling Hamiltonian between two qubits can be simplified as:

\[
\mathcal{H}_{int}=g(a_i^\dagger a_j+a_i a_j^\dagger)
\]

where $g$ is the effective coupling strength jointly determined by the coupling structure, frequency detuning, control waveform, and parasitic paths. Therefore, the PDK needs to provide standardized coupler PCells, including fixed capacitive couplers, tunable couplers, gmon structures, bus resonator structures, and coupling-network templates applicable to specific gate schemes. Geller et al.'s analysis of gmon/Xmon tunable couplers and Yan et al.'s study of tunable-coupling schemes for high-fidelity two-qubit gates both show that coupler PCells must simultaneously expose model parameters such as coupling range, off-state residual coupling, nonlinear corrections, control channels, and parasitic ZZ terms (Geller et al., 2015; Yan et al., 2018).

Furthermore, multiple qubits, couplers, readout resonators, and control lines can form higher-level subarray PCells, such as $2\times2$ qubit modules, $4\times4$ tiles, surface-code unit arrays, and readout clusters. This hierarchical design approach is similar to macro-block design in modern digital chips (Hennessy \& Patterson, 2019), but quantum subarrays must also explicitly handle frequency assignment, residual coupling, crosstalk, readout multiplexing, packaging boundary conditions, and cross-module calibration. Application-driven architectural design frameworks such as MQT-DASQA also show that future Q-EDA needs to integrate application requirements, connection topology, frequency planning, and physical layout into a single automated flow (Kunasaikaran et al., 2024).

Accordingly, subarray PCells should be regarded as modules jointly constrained by ``geometry--electromagnetics--Hamiltonian--noise--calibration,'' rather than as simple layout macros. Only when a subarray PCell is accompanied by verifiable models, frequency-assignment rules, readout multiplexing rules, and test-chip data does it acquire cross-project reuse and Q-IP transaction value (Murray et al., 2021; Levenson-Falk \& Shanto, 2025).

\subsection{Logical-Qubit PCells}

With the development of fault-tolerant quantum computing, the fundamental unit of future quantum chip design will gradually expand from physical qubits to logical qubits. The ``logical PCell'' here is not a logical gate at the abstract algorithmic level, but a hardware module that contains data qubits, measurement qubits, syndrome extraction, readout multiplexing, decoding interfaces, and boundary-connection rules.

According to surface-code theory, a logical qubit with distance $d$ typically requires:

\[
N_{physical}\approx 2d^2
\]

physical qubits in order of magnitude (Fowler et al., 2012). This estimate depends on the specific patch definition, boundary conditions, and whether readout/ancillary structures are included. It is therefore more suitable as an order-of-magnitude judgment for area and resources than as a fixed unit-area formula. When invoking a logical PCell, designers need not redesign the entire underlying error-correction layout, but they still need to understand the constraint relationships among code distance, measurement cycle, logical error rate, area overhead, control-line resources, readout parallelism, and real-time decoding latency.

Accordingly, a mature PDK may gradually provide logical PCells such as distance-3 logical qubit, distance-5 logical qubit, logical memory block, logical entanglement block, and lattice-surgery boundary. These logical PCells need to carry assumptions about physical PCell quality, supported gate/measurement schedules, boundary-connection rules, scalable routing methods, and logical-error-rate models. Without such metadata, a logical PCell is merely a high-level diagram and cannot become an engineering component callable by Fabless designers.

This elevation of the abstraction layer is a key condition for the scalable development of the quantum Fabless ecosystem. Surface codes, lattice surgery, and modular surface-code architectures provide the theoretical basis for logical PCells, but their engineering realization still depends on physical PCell quality, readout parallelism, control-electronics bandwidth, and real-time decoding capability (Fowler et al., 2012; Litinski, 2019; Nickerson et al., 2016; Gambetta et al., 2017; QuIRC Collaboration, 2023).

\subsection{PCell Qualification Metrics}

To ensure that PCells can be used as industry-grade design components, a unified certification and evaluation system must be established. A publishable PCell should at least reach thresholds in manufacturing consistency, SPICE-Q model accuracy, layout-rule compatibility, tunable parameter range, cross-version stability, test-chip validation, documentation completeness, and tape-out/manufacturing data signoff pass rate. For quantum PCells, additional evaluation is required for cryogenic-measurement repeatability, noise-model coverage, frequency-collision risk, packaging sensitivity, and cross-tool interoperability.

Its comprehensive evaluation function can be defined as:

\[
Q_{\text{PCell}} = w_1M + w_2A + w_3C + w_4S + w_5T + w_6I
\]

where $M$ is the manufacturing-consistency score, $A$ is model accuracy, $C$ is design-rule compatibility, $S$ is long-term stability, $T$ is the sufficiency of test-chip validation, and $I$ is interoperability across Q-EDA tools and PDK versions. The weights $w_i$ should not be fixed, but should be adjusted according to the PCell type: physical-qubit PCells place greater emphasis on material loss and frequency prediction; readout PCells place greater emphasis on Purcell limits and measurement fidelity; and logical PCells place greater emphasis on resource estimation, boundary connections, and decoding interfaces.

Only after $Q_{\text{PCell}}$ reaches the prescribed threshold may the PCell enter the formal PDK release and become a standard quantum component callable by Fabless designers. Before entering the formal library, it should also complete regression layout generation, DRC/LVS/DFM checking, SPICE-Q simulation, model-card review, test-chip comparison, and version-compatibility checking. SQuADDS-style pre-simulation databases, QPDK test-chip examples, and GDSII-to-wafer manufacturing data packages jointly show that PCell quality evaluation should cover the complete chain from parameter input to fab-out delivery, rather than considering only whether a single layout generation succeeds (Shanto et al., 2024; QPDK Documentation, 2025; GDSII-to-Wafer Collaboration, 2026).

Therefore, a PCell library is not merely a collection of geometric structures, but a unified encapsulation of quantum hardware knowledge, process experience, and simulation models. Its maturity can be assessed through model error, process coverage, version stability, test-chip pass rate, documentation/model-card completeness, and cross-tool interoperability, and it will directly affect the development speed and scale of the Fabless quantum chip industry (EDA-Q Collaboration, 2025; GDSII-to-Wafer Collaboration, 2026).
\section{SPICE-Q Models and Multiphysics Simulation}

In the classical semiconductor industry, SPICE (Simulation Program with Integrated Circuit Emphasis) models constitute the foundation of the EDA ecosystem. Before tape-out, designers can predict circuit behavior on the basis of device models and circuit equations, thereby transforming chip development from ``trial-and-error driven'' to ``model driven'' (Nagel, 1975; Weste \& Harris, 2011). For the quantum-chip industry, an analogous standardized simulation system is likewise a necessary condition for establishing a Fabless ecosystem, but it must cover multiple layers, including classical electromagnetics, Josephson nonlinearity, quantum Hamiltonians, open-system noise, and cryogenic measurement feedback (Nigg et al., 2012; Blais et al., 2021; Levenson-Falk \& Shanto, 2025).

This paper proposes SPICE-Q (SPICE for Quantum Integrated Circuits) as a unified intermediate representation and multiphysics simulation framework for quantum-chip design and verification. It should be emphasized that SPICE-Q should not be understood as a single piece of software or a one-off solver, but rather as a model chain linking PCells, PDKs, EM solving, EPR/black-box quantization, Hamiltonian construction, noise model cards, and cryogenic data calibration. Existing superconducting quantum-device design tools typically map geometric layouts to Hamiltonian parameters through electromagnetic simulation, lumped-oscillator models, energy participation ratio (EPR), or black-box quantization methods, providing a technical basis from which the interface design of SPICE-Q can draw (Nigg et al., 2012; Minev et al., 2021; Shanto et al., 2024).

SPICE-Q files and simulators should constitute an important component of a quantum PDK. Every released PCell should correspond to a Foundry-validated SPICE-Q model, statistical corners, and model applicability boundaries; when invoking a PCell, designers obtain not only layout-generation capability but also a simulation entry point consistent with the process version. The flow may be summarized as follows:

\[
\text{PCell}
\rightarrow
\text{SPICE-Q Model}
\rightarrow
\text{Multi-physics Simulation}
\rightarrow
\text{Sign-off/Fab-out}
\]

Through this mechanism, designers can evaluate, before tape-out, metrics such as qubit frequencies, $T_1/T_2$ lifetimes, readout performance, gate-error budgets, crosstalk levels, and logical-error-rate trends. Here, ``prediction'' should be understood as an engineering estimate with confidence intervals and applicability conditions, rather than as a guarantee that the performance of an actual chip will exactly match the simulation results; its credibility must rely on continuous calibration through test structures, cross-batch data, and cryogenic measurement feedback (McRae et al., 2020; CMC Microsystems, 2023; GDSII-to-Wafer Collaboration, 2026).

Through the coupling of SPICE-Q and PCells, a PDK can be upgraded from a static process document into a simulatable, verifiable, and sign-off-capable quantum-chip design interface. For China's quantum industry, if potential Foundry nodes such as Suzhou can continuously release validated SPICE-Q models, test-chip data, and versioned PDKs, Fabless design enterprises nationwide will obtain a more unified physical-simulation standard, thereby reducing duplicated low-level device-calibration work. This model is structurally similar to the industrial arrangement in which classical Foundries provide PDKs and SPICE models to Fabless companies, but quantum systems must additionally handle cryogenic measurement throughput, material-loss statistics, and the credibility of Hamiltonian models (Macher et al., 2008; National Academies, 2024).

Therefore, the core value of SPICE-Q is not merely to provide simulation capability, but to establish a credible predictive bridge among design, manufacturing, and cryogenic verification, enabling Fabless design to enter the \texttt{fab-out} flow under explicit physical boundaries and statistical confidence.

\subsection{The SPICE-Q Netlist Concept}

The SPICE-Q netlist inherits the hierarchical design concept of classical SPICE, but its extension is not merely to add ``quantum component names''. Rather, it enables the same design object to simultaneously carry classical ports, electromagnetic modes, Josephson nonlinearities, Hamiltonian parameters, noise channels, measurement structures, and PDK version information. In classical SPICE, a device is usually represented by a current-voltage relation or an equivalent circuit model, for example $I=f(V)$; in superconducting quantum systems, however, device behavior must also be jointly determined by the effective Hamiltonian and open-system dynamics:

\[
i\hbar\frac{\partial}{\partial t}|\psi\rangle=\mathcal{H}|\psi\rangle
\]

Therefore, the basic descriptive object in SPICE-Q should not be simply replaced by a ``quantum state'', but should instead be traceable quantum degrees of freedom, microwave ports, and coupling relations. A typical SPICE-Q netlist needs to simultaneously express qubit nodes, resonator nodes, Josephson-junction elements, capacitor/inductor elements, microwave control ports, noise models, measurement structures, and package/interconnect boundary conditions. These objects should remain consistent with layout coordinates, PCell parameters, EM simulation results, EPR/black-box quantization parameters, and cryogenic calibration data (Nigg et al., 2012; Minev et al., 2021; Blais et al., 2021).

From an engineering perspective, SPICE-Q is in effect an intermediate representation for quantum-chip design, analogous in status to the SPICE Netlist in classical EDA, but it must also be bound to layout geometry, effective Hamiltonians, noise channels, calibration data, and PDK versions. In other words, a SPICE-Q netlist cannot be merely a text representation of physical equations; it should become a structured data object shared by Q-EDA during simulation, verification, and sign-off, and should maintain parameter traceability throughout the \texttt{layout -> EM -> Hamiltonian -> noise -> sign-off} chain (Levenson-Falk \& Shanto, 2025; EDA-Q Collaboration, 2025).

\subsection{Hamiltonian Extraction Flow}

Automatically generating an effective Hamiltonian from circuit structure is one of the core functions of SPICE-Q, but this process cannot be reduced to directly applying a fixed formula to a layout. For superconducting quantum circuits, a common route is first to generate capacitances, inductances, impedances, port parameters, and electromagnetic modes from the layout and material stack, and then to extract low-dimensional Hamiltonian parameters through the circuit Lagrangian, black-box quantization, or EPR methods (Nigg et al., 2012; Minev et al., 2021; Shanto et al., 2024).

For a superconducting circuit described by node flux variables $\phi_i$, the starting point can be written as the circuit Lagrangian:

\[
\mathcal{L}=T-V,
\quad
T=\frac{1}{2}\dot{\boldsymbol{\phi}}^T C\dot{\boldsymbol{\phi}},
\quad
V=-\sum_j E_{J,j}\cos\phi_j+V_{L/C}(\boldsymbol{\phi})
\]

Here, $C$ denotes the capacitance matrix, and $V_{L/C}$ denotes the potential-energy terms introduced by linear inductances, capacitances, external magnetic fluxes, or boundary conditions. After a Legendre transformation, the system Hamiltonian can be obtained. For an idealized multi-Josephson-junction circuit, it can be written in a similar form:

\[
\mathcal{H}=4\mathbf{n}^T E_C\mathbf{n}-\sum_j E_{J,j}\cos\phi_j+\mathcal{H}_{\mathrm{lin}}
\]

Here, $E_C$ usually comes from the inverse of the capacitance matrix, rather than from a simple substitution of a single isolated capacitance; $\mathcal{H}_{\mathrm{lin}}$ contains the contributions of resonators, coupling lines, package modes, or equivalent linear networks. Subsequently, through modal expansion and quantization, zero-point fluctuations, participation ratios, and mode couplings are converted into quantum-operator representations, such as $\phi_i=\phi_{\text{zpf}}(a_i+a_i^\dagger)$.

In an engineering model, the effective Hamiltonian ultimately obtained can be summarized as:

\[
\mathcal{H}_{eff}=\sum_i \omega_i a_i^\dagger a_i+\sum_{ij}g_{ij}(a_i^\dagger a_j+a_j^\dagger a_i)+\mathcal{H}_{\mathrm{anh}}+\mathcal{H}_{\mathrm{drive}}+\mathcal{H}_{\mathrm{noise}}
\]

Here, $\mathcal{H}_{\mathrm{anh}}$ denotes transmon anharmonicity and higher-order nonlinear terms, $\mathcal{H}_{\mathrm{drive}}$ denotes control drives, and $\mathcal{H}_{\mathrm{noise}}$ denotes noise channels that require further mapping in open-system solving. This expression better satisfies practical Q-EDA sign-off requirements than retaining only linear coupling terms, because gate errors, leakage, readout backaction, and residual ZZ coupling often arise from these higher-order or environmental terms (Blais et al., 2021; Levenson-Falk \& Shanto, 2025).

This flow has a theoretical foundation consistent with the Black-box Quantization method (Nigg et al., 2012), and it is also consistent with the engineering route in Qiskit Metal, SQuADDS, and EPR/pyEPR-like work, where quantum-device parameters are extracted through finite-element simulation and participation-ratio analysis (Minev et al., 2021; Shanto et al., 2024). For Fabless design, the key is not that every designer re-derive the Hamiltonian, but that PDK and SPICE-Q models clearly specify parameter sources, approximation conditions, truncation dimensions, and calibration errors relative to cryogenic measurements.

\subsection{Noise and Loss Model Cards}

An important distinction between quantum chips and classical chips is that noise not only affects simulation accuracy or timing margins, but also directly changes quantum-state evolution, readout results, and logical error rates. Therefore, SPICE-Q must include standardized noise model cards and make them part of PDK releases, rather than adding empirical parameters temporarily at the end of simulation.

For superconducting quantum chips, the major sources of noise and loss include TLS defects (Two-Level Systems), charge noise, flux noise, Purcell loss, quasiparticle loss, thermal-photon noise, package-mode coupling, and additional noise introduced by readout/control chains. At the single-device level, these noise channels manifest as $T_1$, $T_2$, frequency drift, and readout error; at the system level, they affect gate fidelity, leakage error, crosstalk, and error-correction threshold margins (Mueller et al., 2019; McRae et al., 2020; Murray, 2021; Levenson-Falk \& Shanto, 2025).

Open-quantum-system evolution can be expressed as:

\[
\frac{d\rho}{dt}=-\frac{i}{\hbar}[\mathcal{H},\rho]+\sum_k\mathcal{L}_k(\rho)
\]

where the Lindblad dissipator can be written as:

\[
\mathcal{L}_k(\rho)=L_k\rho L_k^\dagger-\frac{1}{2}\{L_k^\dagger L_k,\rho\}
\]

This expression provides the standard form for open-system modeling, but an engineering model card cannot remain at the symbolic level. It must also provide, for each channel, the physical source, parameter-extraction method, measurement temperature range, sample size, confidence interval, applicable layout range, and version dependencies (Breuer \& Petruccione, 2002; Mitchell et al., 2019).

These model cards should derive from Foundry statistical test data, material-loss measurements, device-level calibration, and system-level error analysis, and should be released as part of the PDK. For external designers, the model cards should also specify which parameters come from test structures, which come from product chips, and which are only simulation priors, in order to avoid misusing laboratory-specific models as cross-process general models. Organizing measurement metadata according to the FAIR data principles helps different Q-EDA tools and PDK versions trace the sources and applicability boundaries of noise parameters (Wilkinson et al., 2016; CMC Microsystems, 2023).

\subsection{Calibration with Cryogenic Data}

Any SPICE-Q simulation model must be consistent with real cryogenic experimental results, but this consistency should be understood as model calibration in a statistical sense, rather than as exact agreement between a single sample and a single simulation. Therefore, SPICE-Q needs to establish a Continuous Calibration Framework that continuously writes test structures, product chips, and cross-batch measurement data back into PDK model cards, statistical corners, and PCell parameter ranges (McRae et al., 2020; CMC Microsystems, 2023; QPDK Documentation, 2025).

For a set of key design parameters:

\[
\theta=\{\omega_q,T_1,T_2,g,\chi,Q_i,Q_c\}
\]

the error between simulation results and experimental results can be defined as a weighted residual:

\[
L(\theta)=\sum_i w_i\left|\theta_{sim,i}-\theta_{exp,i}\right|^2
\]

where the weights $w_i$ should reflect the impact of different metrics on design sign-off. For example, frequency errors directly affect frequency collisions and gate scheduling, $T_1/T_2$ errors affect noise models and logical-error-rate estimates, and readout-parameter errors affect measurement time, Purcell limitations, and multiplexing capability.

The PDK publisher may update model parameters on the basis of cryogenic test data, but the update process should not simply be written as unconstrained gradient descent. A more reasonable approach is to distinguish among model recalibration, process-drift correction, anomalous-batch investigation, and interface-incompatible changes: the first two can be handled through in-version parameter updates and adjustments to model-card confidence intervals, whereas the latter two may require releasing new major versions of the PDK/SPICE-Q and completing PCell verification again. Formally, this can be written as:

\[
\mathcal{M}^{(t+1)}=\mathrm{Update}\left(\mathcal{M}^{(t)},\mathcal{D}_{\mathrm{cryo}},\mathcal{D}_{\mathrm{fab}},\mathcal{R}_{\mathrm{validation}}\right)
\]

where $\mathcal{M}$ denotes the set of model cards, $\mathcal{D}_{\mathrm{cryo}}$ denotes cryogenic measurement data, $\mathcal{D}_{\mathrm{fab}}$ denotes process data, and $\mathcal{R}_{\mathrm{validation}}$ denotes regression-validation results. This formulation is more consistent with Foundry release governance requirements than a single gradient update.

Therefore, SPICE-Q is not a static file, but a dynamic model system that evolves continuously with manufacturing data. The ultimate goal of cryogenic data calibration is not to make the model ``fit historical data'', but to increase the probability that external Fabless designers will obtain predictable manufacturing outcomes in new projects. A mature calibration mechanism should provide measurement metadata, sample sizes, fitting residuals, version-change notes, and regression-test layouts, making the model-update process auditable, reproducible, and callable by Q-EDA tools (Wilkinson et al., 2016; GDSII-to-Wafer Collaboration, 2026).

\subsection{Simulation Acceptance Criteria}

To ensure that Fabless designs can enter an executable \texttt{fab-out} flow, unified Simulation Sign-off Criteria must be established. Here, ``acceptance'' should not be understood as a guarantee that a chip will achieve its target performance after tape-out, but rather as indicating that, within the coverage of a given PDK version, model cards, statistical corners, and test data, the design has completed sufficiently thorough risk assessment and sign-off checks.

Typical acceptance metrics include frequency-prediction error, $T_1/T_2$ prediction error, gate-fidelity estimation error, readout-fidelity estimation error, crosstalk-prediction error, Purcell limitations, frequency-collision probability, package-mode risk, and logical-error-rate trends. These metrics need to have thresholds separately defined for device-level, subarray-level, and logical-level designs; a single numerical value cannot cover all PCell and Q-IP types.

A composite acceptance metric can be defined as:

\[
Q_{sim}=w_1E_f+w_2E_{T1}+w_3E_{T2}+w_4E_g+w_5E_{xo}+w_6E_{ro}
\]

where $E_f$ is frequency error, $E_{T1}$ and $E_{T2}$ denote lifetime and decoherence prediction errors, respectively, $E_g$ denotes gate-error-budget deviation, $E_{xo}$ denotes crosstalk/residual-coupling-related deviation, and $E_{ro}$ denotes readout-chain-related error. The weights $w_i$ should be determined by PCell type, target application, and Foundry sign-off strategy, rather than treated as fixed constants. Only when $Q_{sim}<Q_{threshold}$ and all key physical constraints are satisfied can a design be considered to meet the simulation sign-off requirements under the current PDK version.

From the perspective of the Fabless ecosystem, the significance of SPICE-Q is not merely to improve the accuracy of a single simulation, but to establish Foundry-consistent physical prediction capability at the design stage, enabling designers to complete auditable quantum-chip designs without owning complete laboratories and manufacturing lines. SPICE-Q and PCells together constitute the two core pillars of a quantum PDK; their maturity should be evaluated through simulation-experiment deviation, cross-batch stability, sign-off coverage, model-card completeness, and interoperability with Q-EDA tools, rather than solely on the basis of a single simulation result (Minev et al., 2021; Shanto et al., 2024; GDSII-to-Wafer Collaboration, 2026).

\section{Q-EDA Flow and Standard Interoperability}

In the classical semiconductor industry, Electronic Design Automation (EDA) tools are the core bridge connecting designers with manufacturing processes. From RTL design, logic synthesis, and placement and routing to final sign-off, EDA tools define the working mode of the entire Fabless industry (Mack, 2011). For the quantum-chip industry, PDKs, PCells, and SPICE-Q alone are still insufficient to support a scalable design ecosystem; a compatible Quantum Electronic Design Automation (Q-EDA) system must also be constructed.

The goal of Q-EDA is not simply to transplant classical EDA, but to establish a unified design environment among quantum devices, quantum circuits, quantum error correction, control systems, and manufacturing processes. Its core role is to automatically invoke PCells, SPICE-Q models, statistical corners, and design rules in the PDK during the design stage, thereby enabling conversion from system requirements to manufacturable GDSII/OASIS layouts and fab-out manufacturing data packages. Recent work on EDA-Q, GDSII-to-wafer, KQCircuits, and QPDK has already demonstrated prototypes of the Q-EDA technology stack from the perspectives of automated topology design, layout generation, process mapping, DRC/LVS/DFM, and mask-data preparation, respectively (KQCircuits Documentation, 2024; QPDK Documentation, 2025; EDA-Q Collaboration, 2025; GDSII-to-Wafer Collaboration, 2026). In particular, the GDSII-to-wafer work points out that Q-EDA for wafer-scale superconducting quantum-chip manufacturing cannot merely complete \texttt{tape-out} in the traditional sense, but must also generate manufacturing data packages executable by the process line, extending delivery from design files to the \texttt{fab-out} flow (GDSII-to-Wafer Collaboration, 2026).

For Fabless quantum-chip enterprises, Q-EDA in effect constitutes the foundational software platform for industrial scaling. Without a unified Q-EDA system, each enterprise would need to develop its own design tools, simulation environment, layout-conversion scripts, and manufacturing-submission flow, ultimately leading to duplicated investment of industrial resources and ecosystem fragmentation. OpenQASM 3, the QIR/LLVM intermediate representation, and IEEE quantum-terminology standardization efforts respectively show, from the perspectives of logical-circuit description, compiler interoperability, and terminological consistency, that the quantum software and hardware ecosystem is forming multilayer interface standards; however, these standards still need to be connected with quantum PDKs, SPICE-Q, and manufacturing sign-off flows in order to serve real chip production (Cross et al., 2022; OpenQASM Working Group, 2022; QIR Alliance, 2021; IEEE P7130 Working Group, 2021).

Therefore, in the quantum Fabless system:

\[
\text{PDK} \rightarrow \text{Q-EDA} \rightarrow \text{Sign-off/Fab-out} \rightarrow \text{Foundry}
\]

constitutes the core workflow of the entire industrial chain. The maturity of Q-EDA is reflected not only in whether it can draw layouts, but in whether it can organize logical descriptions, physical models, parameterized layouts, process constraints, sign-off rules, and manufacturing data packages into a traceable, auditable, and repeatably executable engineering flow.

\subsection{End-to-End Q-EDA Workflow}

Quantum-chip design involves multiple levels of abstraction, extending from application requirements, quantum circuits, and error-correction structures all the way down to underlying superconducting circuits, layouts, packaging, and manufacturing data. Therefore, Q-EDA needs to provide design capabilities covering the entire flow, rather than merely supplying auxiliary scripts at a single stage.

A typical workflow can be expressed as:

\[
\text{Application}
\rightarrow
\text{Quantum Circuit/IR}
\rightarrow
\text{Logical Layout}
\rightarrow
\text{Physical Layout}
\rightarrow
\text{Sign-off}
\rightarrow
\text{Fab-out}
\]

In this flow, application requirements are first converted into quantum circuits or an intermediate representation, and then enter the stages of error-correction encoding, logical layout, and hardware-topology mapping; in the physical implementation stage, the PCells, SPICE-Q models, and process rules in the PDK are invoked to generate verifiable physical layouts; in the sign-off stage, SPICE-Q performance verification, noise and yield prediction, DRC/LVS/DFM checks, frequency-collision checks, and crosstalk checks are completed; finally, GDSII/OASIS, mask data, layer mappings, process-stack mappings, and fab-out manufacturing file packages are output. Such a flow is consistent with the wafer-scale manufacturing-data conversion emphasized by GDSII-to-wafer, and is mutually complementary with the automated layout/verification workflows demonstrated by EDA-Q and QPDK (QPDK Documentation, 2025; EDA-Q Collaboration, 2025; GDSII-to-Wafer Collaboration, 2026).

In a future mature ecosystem, designers will not necessarily need to directly handle every low-level Josephson-junction structure, but may instead build systems by invoking validated physical PCells, subarray PCells, or logical PCells. For example, a distance-3 surface-code logical qubit can be inserted into the system architecture as a design unit:

\[
\text{Logical PCell} \rightarrow \text{Automatic Physical Expansion}
\]

However, this abstraction should not be understood as completely hiding physical details. A logical PCell must still carry code distance, boundary type, measurement cycle, readout multiplexing, decoder interface, physical-error-rate assumptions, and layout resource overhead. This abstraction is similar to the standard-cell design concept in classical EDA (Mead \& Conway, 1980), but quantum Q-EDA must also cover steps such as Hamiltonian extraction, noise propagation, frequency-collision checking, package-mode checking, and cryogenic manufacturing-data conversion, which either do not exist or do not occupy a dominant position in classical EDA (Fowler et al., 2012; Gambetta et al., 2017; EDA-Q Collaboration, 2025; GDSII-to-Wafer Collaboration, 2026).

\subsection{Data Formats and Metadata}

Whether the quantum Fabless ecosystem can form an open market depends to a large extent on the degree of standardization of data formats and metadata. One reason the classical EDA industry has been able to form a global collaboration network is the widespread adoption of formats and sign-off flows such as Verilog, SPICE, LEF/DEF, and GDSII/OASIS (Macher et al., 2008; Mack, 2011). On the quantum side, OpenQASM 3, QIR/LLVM, GDSII outputs from Qiskit Metal/KQCircuits, and the parameterized component system of QPDK have already covered portions of the standardization needs for logical-circuit description, intermediate representation, geometric layout generation, and process-constraint expression, respectively (Cross et al., 2022; OpenQASM Working Group, 2022; QIR Alliance, 2021; Minev et al., 2021; KQCircuits Documentation, 2024; QPDK Documentation, 2025).

Similarly, quantum EDA needs to establish a layered data-format system:

\[
\mathcal{D}=\{QASM/QIR, SPICE\text{-}Q, PCell, GDSII/OASIS, Metadata\}
\]

Here, QASM or QIR describes logical quantum programs and compiler intermediate representations, SPICE-Q describes physical device models and Hamiltonian/noise parameters, PCell describes parameterized structures and their PDK constraints, GDSII/OASIS describes the final manufacturing layout, and Metadata records cross-layer parameter sources, versions, units, applicability boundaries, and sign-off status. This system cannot merely solve the problem of whether a file can be opened; it must also solve the problems of whether semantics are consistent and whether parameters are traceable.

Quantum-specific metadata should at least include target frequencies, $T_1/T_2$ lifetimes, gate fidelities, coupling strengths, readout frequencies, material/process stacks, test-chip batches, manufacturing version numbers, PDK version numbers, model-card versions, temperature ranges, and measurement conditions. Organizing such metadata according to the FAIR data principles can improve the findability, accessibility, interoperability, and reusability of design objects across different Q-EDA tools, PDK versions, and Foundry flows (Wilkinson et al., 2016).

It should be cautiously noted that a unified metadata system cannot guarantee ``lossless migration'' of designs across different Q-EDA tools and different Foundries. Because quantum devices are highly sensitive to process, materials, packaging, and cryogenic testing, cross-Foundry migration usually still requires revalidation; the value of standardization lies in significantly reducing information loss, repeated modeling, and revalidation costs, rather than eliminating all physical differences.

\subsection{Compiler and Hardware Co-Design}

Traditional quantum software compilers usually focus on quantum-gate decomposition, circuit depth, mapping, and scheduling, whereas Fabless quantum-chip design is more concerned with co-design between the compiler and hardware. In real quantum chips, compilation results are affected not only by the logical gate set, but also by physical topology, coupling graphs, frequency planning, crosstalk, readout parallelism, control-electronics bandwidth, and error-correction-cycle constraints. Therefore, Q-EDA needs to connect compiler outputs with chip physical models, rather than treating quantum circuits only as abstract gate sequences.

System performance can be expressed as:

\[
P=f(A,H,E,C)
\]

where $A$ denotes the algorithmic structure, $H$ denotes the hardware architecture, $E$ denotes the error model, and $C$ denotes control and compilation constraints. OpenQASM 3's support for timing, pulses, and real-time classical control, together with QIR's focus on multi-language and multi-backend compilation interfaces, indicates that quantum compilation is moving from ``gate-level lists'' toward ``hardware-constraint-aware intermediate representations'' (Cross et al., 2022; QIR Alliance, 2021).

In the era of fault-tolerant quantum computing, the resource requirements of logical circuits often far exceed physical chip capabilities; therefore, compilers must simultaneously consider physical-topology limitations, frequency planning, coupling-network structures, logical-qubit layout, measurement scheduling, error-correction overhead, and decoding latency. For example, the logical error rate in the surface code approximately satisfies:

\[
P_L \approx A\left(\frac{p}{p_{th}}\right)^{\frac{d+1}{2}}
\]

where $P_L$ is the logical error rate, $p$ is the physical error rate, $p_{th}$ is the fault-tolerance threshold, and $d$ is the code distance. This formula shows that when choosing logical layouts, code distances, gate schedules, and measurement cycles, the compiler must jointly optimize with the error rates, readout capabilities, and topological structures of the physical chip (Fowler et al., 2012; Litinski, 2019; Gambetta et al., 2017).

Therefore, compiler selection and chip-architecture design are in fact a joint optimization problem. The task of Q-EDA at this layer is to connect logical programs, intermediate representations, error-correction layouts, physical topologies, and SPICE-Q error models into an iterative design-space search, rather than treating compiler and hardware design as two independent stages.

\subsection{Design Rule Checking and Sign-off}

Design Rule Check (DRC) is a key component of the Q-EDA system, but quantum-chip sign-off cannot remain at the level of traditional geometric DRC. In classical chips, DRC mainly focuses on line widths, spacing, layer overlaps, and manufacturing rules; in superconducting quantum chips, however, whether a design is usable also depends on frequency collisions, crosstalk, Purcell loss, package modes, thermal paths, control-line congestion, cryogenic wiring, readout multiplexing, and model applicability ranges.

The feasible design space can be defined as:

\[
\Omega=\{\theta\mid C_i(\theta)\le 0\}
\]

where $C_i$ denotes manufacturing, physical, noise, control, and system-performance constraints. Unlike classical DRC, the $C_i$ in quantum sign-off are not necessarily all geometric inequalities; some constraints come from SPICE-Q simulation results, PDK statistical corners, test-chip model cards, and error-correction resource estimates. For example, frequency-collision checking requires statistical manufacturing drift, Purcell-loss checking requires a readout-environment model, and logical-error-rate evaluation requires joint inputs of physical error rate, code distance, and measurement cycle.

The sign-off stage must not only complete traditional DRC/LVS/DFM verification, but also complete SPICE-Q model-consistency verification, noise and decoherence simulation, yield prediction, logical-error-rate evaluation, PDK-version compatibility checks, model-card applicability-range checks, and fab-out manufacturing-file integrity checks. Only designs satisfying all constraints are allowed to enter tape-out. For wafer-scale quantum-chip manufacturing, sign-off should further confirm GDSII/OASIS layer mappings, process-stack mappings, mask-data preparation, wafer layout, insertion of test structures, and manufacturing-file package completeness; otherwise, a design may pass front-end layout checks but still fail to be stably executed by the Foundry (QPDK Documentation, 2025; EDA-Q Collaboration, 2025; GDSII-to-Wafer Collaboration, 2026).

Therefore, Q-EDA sign-off should be understood as a joint verification flow of ``geometric manufacturability + physical realizability + model credibility + manufacturing executability''. Its output should not be merely a pass/fail flag, but should also include causes of constraint failure, model versions, statistical corners, revalidation requirements, and the fab-out data package that can be submitted to the Foundry.

\subsection{Tool-Vendor Ecosystem}

One important reason for the success of the Fabless model is the formation of an ecosystem of independent EDA tool vendors. For example, companies such as Cadence, Synopsys, and Mentor Graphics provide unified design tools for chip enterprises worldwide (Mack, 2011). The quantum industry may similarly form such a structure in the future, but its path to maturity will be slower, because Q-EDA tools must simultaneously understand quantum-device physics, cryogenic measurement, manufacturing statistics, control electronics, and error-correction architectures.

Industrial participants in a mature ecosystem include Foundries, PDK suppliers, Q-EDA tool vendors, Q-IP suppliers, Fabless design enterprises, cloud verification platforms, cryogenic testing service providers, and control-electronics suppliers. Their relationships can be expressed as:

\[
\text{Foundry} \leftrightarrow \text{PDK} \leftrightarrow \text{Q-EDA} \leftrightarrow \text{Fabless/Q-IP}
\]

In a mature ecosystem, Q-EDA tool companies need neither own quantum-chip fabs nor develop complete quantum processors; instead, they focus on design automation, model management, layout generation, sign-off, simulation orchestration, and data interoperability itself. Work on QIR Alliance, OpenQASM, KQCircuits, QPDK, EDA-Q, and GDSII-to-wafer respectively covers compilation interfaces, logical descriptions, parameterized layouts, PDKs, automated flows, and manufacturing-file package conversion, showing that the tool-vendor ecosystem needs to form around open interfaces and verifiable data links, rather than around a single closed software stack (Cross et al., 2022; QIR Alliance, 2021; KQCircuits Documentation, 2024; QPDK Documentation, 2025; EDA-Q Collaboration, 2025; GDSII-to-Wafer Collaboration, 2026).

This specialized division of labor is expected to lower barriers to entry and promote the gradual evolution of the quantum-chip industry from the current model dominated by a small number of large research institutions into an open innovation network jointly composed of Fabless design enterprises, tool vendors, and Foundries (Cross et al., 2022). This process must be premised on open formats, verifiable PDKs, stable PCell interfaces, auditable sign-off flows, data-permission governance, and model-version governance, and cannot rely solely on the internal flow of a single tool or a single factory.

From an industrial perspective, Q-EDA is not merely a software tool, but a unified execution platform connecting PDKs, PCells, SPICE-Q, Q-IP, and manufacturing processes. Its maturity will largely determine the scale and pace of innovation that the future quantum Fabless ecosystem can achieve.
\section{Fabless Quantum Chip Design and Commercial Production}

The core of fabless quantum chip design and commercial production does not lie in improving the performance of an individual qubit, but in establishing an industrial mode of organization that is reproducible, scalable, and collaborative. In the classical semiconductor industry, the success of the fabless model did not stem from design firms possessing more advanced transistors, but from the emergence of unified PDKs, EDA tools, standard-cell libraries, and manufacturing interfaces across the industrial chain (Macher et al., 2008). Similarly, for the quantum computing industry, the viability of the fabless model does not depend merely on obtaining longer coherence times, but on building an engineering infrastructure composed of standardized PDKs, PCells, SPICE-Q, and Q-EDA.

The core of a fabless quantum chip system is the PDK, but a PDK should not be reduced to two components, PCells and SPICE-Q. More comprehensively, a quantum PDK should include layer definitions, design rules, PCell libraries, SPICE-Q/Hamiltonian models, noise and materials models, statistical corners, test structures, version governance, and sign-off interfaces. This section further explains the importance of standardization and softwareization, and how they determine whether the quantum fabless industry can form a scaled commercial ecosystem.

From the perspective of industrial structure, the quantum fabless system can be represented as follows:

\[
\text{Foundry}
\leftrightarrow
\text{PDK}
\leftrightarrow
\text{Q-EDA}
\leftrightarrow
\text{Fabless}
\leftrightarrow
\text{Application Market}
\]

Here, the PDK constitutes the industrial standard, Q-EDA constitutes the design platform, and fabless firms develop market-oriented quantum chip products on the basis of these standardized infrastructures.

\subsection{The Importance of Standardization}

To make the fabless ecosystem scalable, standardized data formats, model interfaces, and manufacturing sign-off flows are essential. This paper advocates the use of open or interoperable standards to represent quantum circuits, physical models, parameterized layouts, and manufacturing data, such as OpenQASM, SPICE-Q, PCell interfaces, and GDSII/OASIS formats with quantum-specific metadata. Standardization does not guarantee that a design can be migrated "seamlessly" to any foundry, but it can significantly reduce information loss and the cost of revalidation across Q-EDA tools, PDK versions, and foundries (OpenQASM Working Group, 2022; GDSII-to-Wafer Collaboration, 2026).

The development history of the classical semiconductor industry shows that standardization is a prerequisite for scaled industrial expansion. After the 1980s, as industrial standards such as SPICE, Verilog, VHDL, LEF/DEF, and GDSII gradually matured, chip design firms began to free themselves from dependence on a single manufacturer, thereby forming a highly specialized division of labor among design firms, EDA tool vendors, IP suppliers, and foundries (Mead \& Conway, 1980). This standardized division-of-labor model ultimately gave rise to the modern fabless industry and significantly lowered the entry barrier for chip design.

The quantum computing industry is still at a developmental stage similar to that of the semiconductor industry in the 1970s. Most quantum chip design flows still rely on laboratory-internal toolchains, while different institutions use different data formats, parameter-naming conventions, and layout-description methods. As a result, the same quantum device design often cannot be directly transferred to other teams or manufacturing platforms, leading to extensive duplicated development work (Cross et al., 2022). Recent work on Q-EDA and quantum PDKs has begun to formulate this fragmentation problem in concrete terms, including GDSII-to-wafer manufacturing data conversion, PDK-driven DRC/LVS, metadata consistency, and multi-tool interoperability (QPDK Documentation, 2025; EDA-Q Collaboration, 2025; GDSII-to-Wafer Collaboration, 2026).

For a quantum fabless system, standardization is first reflected in the unification of design data formats. The future quantum chip industry needs to establish a unified data structure covering the logic layer, circuit layer, device layer, and manufacturing layer:

\[
\mathcal{D}
=
\{
QASM,
SPICE\text{-}Q,
PCell,
GDSII
\}
\]

Here, QASM describes logical quantum circuits and compilation-layer instructions, SPICE-Q describes physical device models, Hamiltonian interfaces, and noise channels, PCell describes parameterized quantum structures and their PDK constraints, and GDSII/OASIS describes the final manufacturing layout and mask data entry. It should be noted that these four types of data are not mutually independent file collections; rather, they should be kept consistent through metadata, version numbers, and parameter mappings. Otherwise, semantic discontinuities will still arise among logic-layer objectives, physical models, and manufacturing layouts (Cross et al., 2022; QIR Alliance, 2021; GDSII-to-Wafer Collaboration, 2026).

Together, the above data structures constitute the information chain for quantum chip design:

\[
\text{Algorithm}
\rightarrow
\text{Circuit}
\rightarrow
\text{PCell}
\rightarrow
\text{Layout}
\rightarrow
\text{Fabrication}
\]

If any one link lacks a unified standard, design information will be lost or distorted during transmission, thereby increasing engineering risk and verification cost.

Standardization concerns not only file formats, but also the definition of device models. In current research on superconducting quantum computing, different teams often adopt different naming conventions for transmon parameters, coupler parameters, and readout resonator parameters. For example, parameters denoting the same qubit frequency may be written as $f_q$, $\omega_q$, $F_{01}$, or $\nu_q$. For a single research group, such differences have limited impact; for an industrial-scale ecosystem, however, they substantially increase the cost of tool interoperability.

Therefore, a quantum PDK should establish a unified parameter interface. For example, a standardized transmon PCell can be defined as:

\[
Qubit(
\omega_q,
\alpha,
T_1,
T_2,
C_c
)
\]

Here, $\omega_q$ denotes the qubit frequency, $\alpha$ denotes the anharmonicity, $T_1$ and $T_2$ denote the energy relaxation time and dephasing time, respectively, and $C_c$ denotes the coupling capacitance or an equivalent coupling parameter. An actual PDK should also provide parameter units, statistical distributions, measurement methods, temperature regimes, layout applicability ranges, and version dependencies. Otherwise, the same symbol may still represent different physical meanings across different foundries or tools.

Through a unified interface, devices provided by different foundries can be more readily integrated into the same Q-EDA environment for design and verification. Such integration, however, does not mean that designs can migrate unconditionally across processes; each reuse across foundries or PDK versions still requires rechecking frequency distributions, loss models, packaging boundaries, and sign-off rules.

The second dimension of standardization is model standardization. In the classical semiconductor industry, SPICE model cards enable designers to predict transistor behavior before tape-out. Similarly, in a quantum fabless system, SPICE-Q models need to become a unified description language shared across tools, institutions, and foundries. Their core task is to convert device physical parameters into computable, verifiable, and sign-off-ready engineering models (Krantz et al., 2019).

For superconducting qubits, the basic Hamiltonian can be written as:

\[
H
=
4E_C \hat{n}^2
-
E_J \cos \hat{\phi}
\]

Here, $E_C$ denotes the charging energy, $E_J$ denotes the Josephson energy, $\hat{n}$ denotes the Cooper-pair number operator, and $\hat{\phi}$ denotes the phase operator. For a transmon, $E_C$ is usually determined by the total capacitance, while $E_J$ is related to the critical current of the Josephson junction; both must therefore be linked to geometry, electromagnetic extraction, and process statistics (Koch et al., 2007; Nigg et al., 2012; Blais et al., 2021).

Future SPICE-Q standards should be able to express the above physical models in a unified manner and directly associate them with PCells in the PDK through parameter interfaces, thereby improving the consistency between simulation results and actual manufacturing results. The term "consistency" here should be understood as predictability within clearly specified statistical confidence intervals and model applicability ranges, rather than as an absolute guarantee of single-chip performance.

Going further, future logical qubits should also be standardized. As the surface code gradually becomes the mainstream approach to fault-tolerant quantum computing, a logical qubit is no longer merely a single physical device, but a functional module composed of tens or even hundreds of physical qubits (Fowler et al., 2012). For fabless designers, a logical qubit should appear in a form analogous to a standard cell in classical EDA:

\[
LogicalQubit(d)
\]

Here, $d$ denotes the surface-code distance. When logical designers invoke a verified logical PCell, they can focus on the main performance metrics of the logical qubit at a higher level of abstraction, but they still need to understand its underlying PDK version, physical error-rate assumptions, readout parallelism, control-line resources, and manufacturing applicability boundaries. A logical PCell should at minimum declare the logical error rate $P_L$, logical lifetime $T_L$, and occupied chip area $Area$, and should specify the surface-code cycle time, decoding latency, number of measurement rounds, and physical error-rate range corresponding to these metrics. Google Quantum AI's surface-code scaling experiments and below-threshold experiments show that the key to reusing a logical module is not the abstract symbol itself, but whether the logical error rate exhibits a verifiable suppression trend as the code distance increases (Google Quantum AI, 2023; Google Quantum AI and Collaborators, 2025).

According to surface-code theory, the logical error rate approximately satisfies:

\[
P_L
\approx
A
\left(
\frac{p}{p_{th}}
\right)^{\frac{d+1}{2}}
\]

Here, $p$ denotes the physical error rate, $p_{th}$ denotes the fault-tolerance threshold, $d$ denotes the code distance, and $A$ is a constant related to the implementation structure, noise model, and decoding flow. This expression has predictive meaning only when the physical error rate is below the threshold, the error model and decoding assumptions hold, and correlated errors and leakage errors are controlled (Fowler et al., 2012; Gambetta et al., 2017). Therefore, after logical qubits are standardized, their design and verification process can partly draw on the reuse logic of classical standard cells, but it cannot be directly equated with a classical digital cell library; every reuse across PDKs or foundries still requires revalidation of the logical error-rate model, readout scheduling, and real-time decoding interface.

Standardization also directly affects the efficiency of industrial-chain collaboration. Suppose that a quantum chip design project involves four types of participants: design firms, EDA suppliers, IP suppliers, and foundries. The project complexity approximately satisfies:

\[
Complexity
\propto
N^2
\]

Here, $N$ denotes the number of participating entities. This expression is not a strict law of complexity, but indicates that in the absence of unified interfaces, the costs of bilateral adaptation, format conversion, responsibility confirmation, and revalidation among participants increase rapidly with the number of entities. When unified standards exist, participants can interact more through standard interfaces, model cards, and sign-off flows, and the coordination cost can approximately converge toward a platform-type structure:

\[
Complexity
\approx
O(N)+C_{governance}
\]

Here, $C_{governance}$ denotes the institutional costs arising from version governance, licensing boundaries, data permissions, and revalidation rules. Standardization can therefore reduce friction in industrial collaboration, but it does not automatically compress all cross-organizational costs into linear or negligible costs (Baldwin \& Clark, 2000; Sturgeon, 2002; Wilkinson et al., 2016).

For the fabless model, the ultimate value of standardization is not technical consistency itself, but the capacity for market-scale expansion. Once PDK, SPICE-Q, PCell, and Q-EDA form unified standards, a Q-IP module developed by one design team can be reused by many firms, a process platform developed by one foundry can serve an entire industry, and one EDA tool can cover multiple manufacturing nodes. This knowledge-reuse effect will create network externalities, thereby pushing the industry into a stage of positive-feedback growth. Similar to parameter transfer and circuit optimization in quantum algorithms, reusable modules, model cards, and standard interfaces in hardware design can also reduce repeated search costs, but their effectiveness must be confirmed through manufacturing data and cross-project verification (Shaydulin et al., 2023; QPDK Documentation, 2025).

From an economic perspective, the core benefit brought by standardization can be expressed as:

\[
Value
=
N \times R
\]

Here, $N$ denotes the number of ecosystem participants, and $R$ denotes the reuse rate of design resources. This equation is more appropriate as a conceptual benefit function than as a formula for predicting market size; whether $R$ increases depends on PDK stability, the pass rate of Q-IP revalidation, the quality of test-data feedback, and tool interoperability, while whether $N$ expands also depends on foundry capacity, cryogenic testing throughput, and real application demand. In other words, standardization can increase reuse potential, but that potential becomes industrial scale only when models are trustworthy and contractual responsibilities are clear.

Therefore, for the future quantum computing industry, standardization is not an auxiliary task, but a foundational condition determining whether the fabless model can be established. Only by building unified PDK interfaces, unified SPICE-Q model descriptions, unified PCell specifications, and unified Q-EDA data-exchange standards can quantum chip design achieve cross-organizational collaboration and scaled commercial production in the same way as today's software development and classical integrated-circuit design (Cross et al., 2022; Shaydulin et al., 2023).

\subsection{Softwareization (Technology and Engineering)}

Softwareization is one of the core prerequisites for scaled expansion in the quantum fabless industry, but it does not mean "full automation" or the complete elimination of expert judgment. More precisely, softwareization is the transformation of repeatable human experience, experimental calibration results, and process rules into auditable, versionable, and reusable models and tools.

At present, superconducting quantum chip design still depends heavily on the long-term experimental experience accumulated by research teams. From qubit parameter selection, frequency planning, and coupling-structure optimization to packaging design, cryogenic testing, and calibration flows, many key decisions still rely on the experiential judgment of senior researchers. This human-centric development mode is reasonable to a certain extent in the research stage, but it will become an important bottleneck limiting industrial expansion at the industrialization stage (Krantz et al., 2019; Levenson-Falk \& Shanto, 2025).

From an engineering perspective, experience is essentially knowledge that has not yet been formalized. When design capability depends on individual experience, its diffusion speed is constrained by the training cycle for talent; when experience is transformed into algorithms, model cards, and software tools, its reuse cost decreases, but does not approach zero. The transfer of quantum chip experience is still constrained by PDK versions, materials stacks, cryogenic testing conditions, packaging environments, and foundry data permissions. Therefore, the essence of the fabless model is not merely the separation of design and manufacturing, but the softwareization, versioning, and auditability of knowledge and experience within controlled boundaries (Mitchell et al., 2019; Wilkinson et al., 2016).

This process can be represented as:

\[
\text{Experience}
\rightarrow
\text{Model}
\rightarrow
\text{Algorithm}
\rightarrow
\text{Software}
\]

That is:

\[
\text{Human Experience}
\rightarrow
\text{Mathematical Model}
\rightarrow
\text{Automatic Optimization Algorithm}
\rightarrow
\text{Engineering Software}
\]

In the development of the classical semiconductor industry, EDA tools gradually replaced large amounts of manual design work precisely through this process (Mack, 2011). Similarly, in the quantum chip industry, Q-EDA, SPICE-Q, and parameterized PCells will assume analogous roles.

For current superconducting quantum chip development, a typical design flow often simultaneously includes frequency planning, coupling-network design, readout-system design, packaging design, electromagnetic simulation, process verification, tape-out sign-off, cryogenic testing, and parameter calibration. These steps are highly coupled: frequency planning affects coupling networks and readout multiplexing; packaging structures alter electromagnetic modes and crosstalk paths; and cryogenic test results feed back to revise PDK model cards and statistical corners. Therefore, softwareization is not the simple arrangement of a process into automated scripts, but the transformation of these coupling relationships into traceable models, constraints, and regression-verification flows.

In the traditional research mode, many steps in the above flow require repeated manual adjustment and trial and error. Under a softwareized system, these processes should gradually be assisted by automated tools:

\[
\text{Design}
\rightarrow
\text{Simulation}
\rightarrow
\text{Optimization}
\rightarrow
\text{Verification}
\]

thereby forming a closed-loop design flow.

For superconducting quantum chips, frequency planning is one of the most typical objects of softwareization. As the number of qubits increases, frequency collisions rapidly become a major factor limiting system-scale expansion; related work has formulated this as a problem of frequency allocation and yield optimization under manufacturing-discreteness constraints (Murray et al., 2021; McDonald et al., 2022; Shi et al., 2024).

The frequency-planning problem can be expressed as:

\[
\min_{\omega_i}
\sum_{i\neq j}
P_{collision}
(\omega_i,\omega_j)
\]

Here, $\omega_i$ denotes the frequency of the $i$-th qubit, and $P_{collision}$ denotes the probability of a frequency conflict under given manufacturing statistics, coupling strengths, anharmonicity, and gate-operation bandwidth. This objective function should also incorporate constraints on readout resonator frequencies, coupler modes, packaging modes, and test pass rates, rather than optimizing qubit frequencies in isolation (Murray et al., 2021; Shi et al., 2024).

In traditional development flows, this process often relies on manual adjustment by experienced designers. In a Q-EDA environment, candidate solutions can be generated by automatic optimizers, but they must ultimately still pass SPICE-Q simulation, PDK sign-off, and cryogenic data validation.

Similarly, quantum chip layout design can also be formulated as a constrained optimization problem:

\[
\mathbf{x}^{*}
=
\arg\min_{\mathbf{x}}
L(\mathbf{x})
\]

subject to:

\[
\mathbf{x}
\in
\Omega_{DRC}
\cap
\Omega_{Performance}
\]

Here, $\Omega_{DRC}$ denotes the manufacturing constraint space, $\Omega_{Performance}$ denotes the performance constraint space, and $L(\mathbf{x})$ denotes an optimization objective function integrating factors such as frequency collisions, crosstalk, loss, packaging modes, and area. The optimization problem here is usually not a one-shot convex optimization problem, but a search for candidate solutions under joint constraints from multiple objectives, multiple physical domains, and multiple statistical corners.

The second important direction of softwareization is automated verification. In current quantum chip development, design verification usually requires the coordinated use of multiple software tools, including electromagnetic simulation, circuit simulation, noise analysis, logical simulation, and process checking. The goal of a mature fabless system is not to have a single tool replace all expert judgment, but to integrate these verification steps into a unified, traceable, and regression-capable verification platform:

\[
\text{Verification}
=
f(
EM,
Circuit,
Noise,
Logic
)
\]

Here, $EM$ denotes electromagnetic simulation, $Circuit$ denotes circuit simulation, $Noise$ denotes the noise model, and $Logic$ denotes logic-level verification. For fault-tolerant roadmaps, decoding latency, syndrome measurement cycles, and logical error-rate models should also be included; for NISQ/ASQIC roadmaps, variational-algorithm training stability, measurement overhead, and error-mitigation cost should also be included (Cerezo et al., 2021; McClean et al., 2018).

The third dimension of softwareization is the mechanism of knowledge accumulation.

In traditional research modes, a large amount of experience exists in the form of papers, experimental records, or the personal experience of researchers. When researchers leave a team, part of this knowledge is often lost. A softwareized system can instead solidify experience into PDKs, SPICE-Q models, and Q-EDA tools.

Its evolution can be represented as:

\[
Knowledge_{n+1}
=
Knowledge_n
+
Data_n
+
Feedback_n
\]

Here, $Knowledge_n$ denotes the current knowledge base, $Data_n$ denotes tape-out and cryogenic testing data, and $Feedback_n$ denotes user feedback, failure records, and customer benchmark results. As the number of designs increases, the knowledge base of the entire ecosystem has the opportunity to expand continuously, but only if the data have comparable measurement conditions, clear access permissions, stable metadata, and traceable versions; otherwise, more data will merely amplify model conflicts and governance costs (Wilkinson et al., 2016).

Furthermore, as machine-learning methods gradually enter quantum device design, parameter transfer, and circuit optimization, future Q-EDA systems can assist design-space search on the basis of historical manufacturing data and simulation data. Such methods, however, should be positioned as design assistance and candidate-solution generation, rather than as final substitutes for PDK verification, cryogenic testing, and foundry sign-off (Shaydulin et al., 2023; Levenson-Falk \& Shanto, 2025):

\[
\theta_{t+1}
=
\theta_t
+
\eta \nabla J
\]

where:

\begin{itemize}
\item $\theta$ denotes the design parameters;
\item $J$ denotes the performance objective function;
\item $\eta$ denotes the learning rate.
\end{itemize}

If the development cost of a quantum chip company is represented as:

\[
C
=
C_{Research}
+
C_{Engineering}
+
C_{Manufacturing}
\]

then softwareization can significantly reduce:

\[
C_{Research}
\rightarrow
\alpha C_{Research}
\]

where:

\[
0 < \alpha < 1
\]

denotes the cost-reduction ratio brought about by knowledge reuse.

Therefore, the essence of softwareization is not simply writing more software, but gradually transforming quantum chip development from a research activity that depends entirely on individual experience into an engineering activity jointly supported by standard toolchains, model cards, sign-off scripts, and regression tests. Only when key flows such as frequency planning, layout generation, simulation verification, process sign-off, and performance optimization can be tool-assisted, version-tracked, and continuously calibrated against cryogenic data can the quantum chip industry draw on the development path of the classical fabless industry and form an open innovation ecosystem composed of foundries, Q-EDA vendors, Q-IP suppliers, and numerous fabless design firms (Cross et al., 2022; Shaydulin et al., 2023; Wilkinson et al., 2016).

\section{Q-IP: Device, Circuit, and Logical Intellectual Property}

The commercialization of superconducting quantum computing requires a structural transition from the integrated device manufacturer (IDM) model to a fabless--foundry ecosystem. Devoret and Schoelkopf's industrial outlook for superconducting quantum information circuits indicates that, as system scale expands from a small number of qubits to hundreds or even thousands of qubits, design methods must shift from "one-off experimental optimization" to "repeatable, composable, and verifiable" engineering flows (Devoret \& Schoelkopf, 2013; Krantz et al., 2019). Q-IP is precisely the knowledge-encapsulation layer in this flow: it abstracts reusable design units in quantum hardware into "intellectual property modules," enabling different teams to invoke, combine, and revalidate these modules under shared PDK and Q-EDA interfaces, rather than repeating all underlying device development in every project.

Within this framework, Q-IP (Quantum Intellectual Property) is no longer merely a design document or experimental structure, but a class of engineering assets that are verifiable, parameterizable, licensable, tape-out-ready, and reusable under PDK constraints. Its form is similar to IP cores in classical semiconductors, but its complexity is greater: classical IP mainly encapsulates logical functions, interface protocols, or layout macro cells, whereas Q-IP must simultaneously encapsulate microwave device physics, Hamiltonian models, noise parameters, control/readout interfaces, layout rules, cryogenic test data, and logical error-correction structures (Fowler et al., 2012; Keating \& Bricaud, 2002). From a layered perspective, Q-IP typically corresponds to three types of composable modules: the physical device layer (transmons, readout resonators, test structures), the circuit layer (tunable couplers, readout chains, subarrays), and the logical layer (surface-code patches, lattice-surgery interfaces). Together with PCell libraries and SPICE-Q models, these modules constitute "tradable design assets" in fabless quantum chip design.

\subsection{IP in Classical Circuits}

In classical integrated-circuit design, an IP core (Intellectual Property Core) is a functional module that has been predesigned, verified, and packaged according to standards, such as a processor core, memory controller, PCIe interface module, SerDes, or analog/mixed-signal macro cell. Such IP reduces SoC (System-on-Chip) design complexity and improves the predictability and scalability of chip development through the engineering paradigm of "design once, reuse many times" (Weste \& Harris, 2011; Mack, 2011). Classical IP reuse methodology emphasizes that a module acquires genuine cross-project reuse value only after its interfaces, functions, timing, verification, documentation, versions, and integration constraints have all been explicitly encapsulated (Keating \& Bricaud, 2002).

Its core mechanism is to divide complex systems into composable modules, allowing system-level design to move from "transistor-level design" to "module-level assembly." Formally, a classical IP core can be abstracted as:

\[
IP_{classical} = \{Interface, Function, Timing, Verification, Documentation, Version\}
\]

Here, Interface denotes standard input-output protocols, Function denotes the functional behavior of the module, Timing denotes timing constraints, Verification denotes preverified models and test benchmarks, and Documentation and Version ensure that integration teams can understand the IP's applicability, change history, and responsibility boundaries. This structure enables IP to be reused across design teams and projects, but it does not mean that it can be migrated to any process node without cost. In their SoC reuse methodology, Keating and Bricaud further distinguish hard IP (Hard IP, usually delivered as GDSII or a layout macro bound to a specific process node) from soft IP (Soft IP, delivered as RTL or synthesizable HDL and requiring resynthesis, place-and-route, and timing sign-off at a new node); the two differ systematically in licensing scope, verification responsibility, and migration cost (Keating \& Bricaud, 2002). This distinction is equally important for quantum fabless: layout-bound physical Q-IP is closer to hard IP, while circuit/logical Q-IP that carries Hamiltonian models and control interfaces and can migrate after parameter recalibration is closer to soft IP, but both must be accompanied by cryogenic verification data and model cards.

From the perspective of system complexity, the value of the classical IP mechanism does not lie in strictly reducing arbitrary exponential complexity to linear complexity, but in significantly reducing repeated design, repeated verification, and integration risk through interface stabilization and preverification. More cautiously, IP reuse transforms system development from "designing every module from scratch" into a flow of "integrating, configuring, and verifying around already verified modules":

\[
C_{reuse} \approx C_{integration}+C_{verification}+C_{adaptation}
\]

Here, $C_{integration}$, $C_{verification}$, and $C_{adaptation}$ denote interface integration, system-level verification, and process/project adaptation costs, respectively. This cost is usually significantly lower than redesigning all modules, but it is not automatically zero. It is precisely this reusable, licensable, and verifiable modular method that has supported the expansion of classical SoC design and the fabless industry (Mead \& Conway, 1980; Keating \& Bricaud, 2002).

In the quantum fabless system, this idea is further extended into Q-IP (Quantum Intellectual Property). Unlike classical IP, Q-IP concerns not only logical functions, but must also encapsulate the continuous-variable behavior of physical quantum systems, such as Hamiltonian dynamics, decoherence mechanisms, cryogenic noise models, control/readout interfaces, and manufacturing statistical boundaries. Therefore, the abstraction levels of quantum IP are more complex, and its structure can be represented as:

\[
QIP = \{H_{phys}, C_{ctrl}, L_{logic}, M_{model}, D_{data}, V_{license}\}
\]

Here, $H_{phys}$ denotes the physical Hamiltonian model, $C_{ctrl}$ denotes the control and readout circuits, $L_{logic}$ denotes logical and system-level behavioral abstractions, $M_{model}$ denotes simulatable models, $D_{data}$ denotes cryogenic verification and test-chip data, and $V_{license}$ denotes licensing, versioning, and responsibility boundaries.

Under this framework, the quantum fabless model relies on Q-IP cores as foundational building blocks, enabling designers to invoke qubits, couplers, readout chains, subarrays, and error-correction modules without solving all underlying physical implementation problems from scratch in every project. A typical superconducting quantum system can be decomposed into physical-layer Q-IP (transmon qubits, readout resonators, Purcell filters), circuit-layer Q-IP (tunable couplers, microwave control chains, amplifier modules, subarrays), and logic-layer Q-IP (syndrome extraction, surface-code patches, lattice-surgery boundaries). This layered structure enables quantum system design to move gradually from "device-level experimental tuning" to "module-level system integration," but each level must retain traceable physical models and verification data.

Therefore, the classical IP system provides an important structural reference for the quantum fabless model, while Q-IP extends and reconstructs this paradigm under quantum-physical constraints. By uniformly encapsulating physical devices, circuit modules, and logical structures into reusable IP units, the quantum chip industry may form a scaled design and ecosystem division of labor similar to that of the classical semiconductor industry. Its viability, however, presupposes the simultaneous maturation of PDKs, SPICE-Q, test data, licensing contracts, and responsibility boundaries (Weste \& Harris, 2011; Krantz et al., 2019; Cross et al., 2022).

\subsection{Definition and Scope of Q-IP}

On the basis of the preceding discussion of classical IP reuse mechanisms, this section further formalizes the definition, layered scope, and reuse conditions of Q-IP. The problem Q-IP seeks to solve is not "whether reusable layouts exist," but how to decompose the "design--manufacturing--testing--calibration" knowledge originally embedded in the IDM closed loop into module interfaces that can be externally invoked, audited, and licensed within specific PDK, foundry, and Q-EDA flows (Krantz et al., 2019; Cross et al., 2022).

From a systems-engineering perspective, the definition of Q-IP should not be limited to a single device or circuit module, but should be formalized as a unified engineering abstraction spanning physical, control, model, verification, and logical layers. Its essence is to encapsulate reusable structures in quantum hardware so that they can be invoked, verified, licensed, and reused within specific PDKs, specific foundries, and specific Q-EDA flows. Therefore, Q-IP can be defined as:

\[
QIP = \{P_{phys}, C_{circuit}, L_{logic}, M_{model}, D_{verify}, R_{license}\}
\]

Here, $P_{phys}$ denotes physical device-layer structures (such as transmons, couplers, and readout resonators), $C_{circuit}$ denotes circuit-level interconnect and control structures, $L_{logic}$ denotes logical computation and error-correction abstractions, $M_{model}$ denotes model descriptions usable for simulation and verification (such as SPICE-Q or equivalent Hamiltonian models), $D_{verify}$ denotes test-chip, cryogenic measurement, and model-card data, and $R_{license}$ denotes the licensing scope, version, responsibilities, and revalidation requirements.

In this framework, Q-IP is not only a design asset, but also the core intermediate representation (Intermediate Representation, IR) connecting PDKs and Q-EDA systems. Its role is similar to that of standard-cell libraries and IP cores in classical semiconductors, but its complexity is significantly higher because it must simultaneously satisfy quantum coherence, cryogenic noise constraints, sensitivity to manufacturing process variations, and error-correction architecture constraints.

From the perspective of industrial structure, introducing a Q-IP mechanism can gradually transform the quantum chip design flow from "experiment-driven" to "platform-driven." In the traditional IDM model, design, manufacturing, and verification are highly coupled, causing each generation of chips to require repeated full-stack optimization. In the fabless--foundry system, Q-IP functions as a reusable module, allowing different design firms to carry out differentiated system design on the basis of the same physical platform, thereby forming a layered ecosystem structure similar to that of the classical semiconductor industry (Cross et al., 2022).

The scaled value of Q-IP should not be simplistically described as directly reducing exponential complexity to linear complexity. More accurately, Q-IP reduces repeated search costs and integration uncertainty through stable interfaces, preverified models, and cryogenic data boundaries. Its reuse benefit can be summarized as:

\[
C_{QIP} = C_{integration}+C_{revalidation}+C_{licensing}
\]

Only when $C_{QIP}$ is lower than the cost of developing a comparable device, circuit, or logical module from scratch does Q-IP reuse have economic significance. This condition depends on PDK maturity, model accuracy, foundry process stability, cryogenic testing capacity, and licensing price; it is not automatically guaranteed by modular form itself.

In addition, the Q-IP system must also satisfy cross-platform portability (Portability), verifiability (Verifiability), and responsibility definability (Accountability). Portability requires that the same Q-IP module maintain its functional boundaries under different PDK constraints through parameter recalibration and re-sign-off. Verifiability requires every Q-IP to be accompanied by repeatable simulation models, test-chip data, and experimental verification records, and to follow the provenance, applicability conditions, and version metadata required by model cards and FAIR data principles (Mitchell et al., 2019; Wilkinson et al., 2016). Responsibility definability requires clear specification of which responsibilities for models, processes, and system integration are borne respectively by IP suppliers, fabless designers, and foundries (Gambetta et al., 2017). The SQuADDS database proposed by Shanto et al. shows that cross-team reuse has engineering credibility only when Q-IP is accompanied by independently regression-validated parameter--performance mappings, rather than relying solely on a single experimental result (Shanto et al., 2024). From the perspective of logical abstraction, surface-code patches, lattice surgery, and modular quantum networks have already provided the theoretical basis for encapsulating underlying physical devices into logical blocks (Fowler et al., 2012; Nickerson et al., 2016; Litinski, 2019; QuIRC Collaboration, 2023); terminology standards such as IEEE P7130 provide a shared language for Q-IP interface naming and metadata interoperability (IEEE P7130 Working Group, 2021).

Therefore, Q-IP is not merely a technical abstraction, but also a foundational institutional unit of the fabless quantum ecosystem. Its definitional scope covers the full-stack quantum computing chain from device physics to logical algorithms, and it is the key bridge connecting PDKs, Q-EDA, and foundries. Through standardized Q-IP interfaces, the quantum chip industry can gradually evolve from a "vertically integrated R\&D system" toward a "horizontally divided industrial system" (Krantz et al., 2019; Cross et al., 2022).

\subsection{Physical Device Q-IP}

Physical Q-IP constitutes the lowest-level reusable unit in the quantum fabless system. Its core objects include optimized qubit topologies (such as transmon designs), high-fidelity readout resonators, Purcell filters, coupling capacitances, test structures, and structures for characterizing material/interface losses. This level corresponds directly to genuinely manufacturable superconducting circuits, and therefore must simultaneously satisfy electromagnetic simulation constraints, material-loss constraints, PDK geometric rules, and cryogenic decoherence constraints (Koch et al., 2007; Krantz et al., 2019; McRae et al., 2020).

From a modeling perspective, a transmon qubit can be described by a reduced Josephson circuit:

\[
H = 4E_C \hat{n}^2 - E_J \cos(\hat{\phi})
\]

Here, $E_C$ is the charging energy, and $E_J$ is the Josephson energy. The core task of physical Q-IP is to optimize $E_J/E_C$, capacitance geometry, target frequency, anharmonicity, and material participation ratio within the PDK constraint space, so as to balance charge-noise suppression, controllability of nonlinearity, chip area, and manufacturing repeatability. The work by Place et al. on high-coherence tantalum-based transmons shows that the reusability of physical device Q-IP depends not only on layout shape, but also on the material platform, interface processing, and cryogenic loss statistics (Place et al., 2021).

At this level, readout resonators are used to realize dispersive readout of quantum states, and their effective Hamiltonian can be written as:

\[
H_{disp} = \chi a^\dagger a \sigma_z
\]

Here, $\chi$ denotes the dispersive coupling strength. Physical Q-IP must reduce decoherence channels introduced by the measurement chain while ensuring measurement fidelity. Purcell filters are used to suppress spontaneous-emission loss through the readout channel, and their effective decay rate can be expressed as:

\[
\Gamma_{eff} = \Gamma_0 \cdot F_{Purcell}
\]

Here, $F_{Purcell}$ is the frequency-selective suppression factor. Reed et al.'s work on Purcell filters shows that readout speed, coupling strength, and qubit lifetime must be co-designed; therefore, such structures should be released as physical Q-IP with model cards and verification data, rather than as isolated layout templates (Reed et al., 2010).

In addition to functional devices, physical Q-IP should also include test structures and material characterization structures, whose role is similar to that of PCM (Process Control Monitor) in classical semiconductors: structures such as resonator $Q$ factors, TLS loss, quasiparticle generation rates, and Josephson-junction critical currents continuously calibrate PDK statistical corners and model cards (McRae et al., 2020; Murray, 2021). The black-box superconducting circuit quantization method proposed by Nigg et al. provides a standardized path for automatically extracting effective Hamiltonians from layouts and junction parameters, so that the $H_{phys}$ of physical Q-IP need not rely entirely on manual derivation, but can be automatically generated in SPICE-Q/Q-EDA flows and compared with cryogenic data (Nigg et al., 2012). Therefore, a qualified physical Q-IP deliverable usually includes a parameterized PCell, Hamiltonian/noise models, test-structure layouts, and verification records under the target PDK.

\subsection{Circuit-Level Q-IP}

Circuit-level Q-IP lies above physical devices, and its core function is to realize controllable coupling, low-noise readout, microwave control, and mesoscale subarray integration among multiple qubits. Typical objects include fixed/tunable couplers, gmon or bus resonator structures, Purcell-protected readout chains, Josephson parametric amplifiers, traveling-wave parametric amplifiers (TWPA/JTWPA), and standard subarray structures (Blais et al., 2021).

Tunable couplers usually realize coupling-strength modulation through external magnetic-flux or voltage control, and their general form can be expressed as:

\[
g = g(\Phi_{ext})
\]

Here, $\Phi_{ext}$ is the external control flux. This mechanism allows dynamic adjustment of interactions between qubits during computation, thereby reducing crosstalk and improving gate fidelity. The analysis of gmon/Xmon tunable couplers by Geller et al. and the study by Yan et al. on tunable-coupling schemes for high-fidelity two-qubit gates show that coupler Q-IP must simultaneously provide the coupling range, residual coupling in the off state, parasitic ZZ, control channels, calibration curves, and applicable frequency windows, rather than only a layout cell (Geller et al., 2015; Yan et al., 2018).

Parametric amplifiers are used in circuit-level Q-IP to realize microwave signal amplification close to the quantum limit, and are important components of high-fidelity readout chains. Josephson parametric amplifiers (JPAs) are suitable for low-noise narrowband amplification, whereas Josephson traveling-wave parametric amplifiers (JTWPAs) provide wider bandwidth and greater reuse potential, which is beneficial for parallel readout of multiple qubits (Castellanos-Beltran \& Lehnert, 2008; Macklin et al., 2015). Therefore, amplifier Q-IP should include parameters such as gain, bandwidth, saturation power, pump conditions, noise temperature, packaging interfaces, and cryogenic thermal load.

Standard subarrays provide reusable mesoscale structures, such as $2 \times 2$ or $3 \times 3$ qubit coupling units, readout clusters, or surface-code tiles, enabling large-scale systems to construct complex topologies modularly. The work by Murray et al. on frequency allocation in fixed-frequency processors, as well as Shi et al.'s subsequent improvements to frequency-planning optimization algorithms, both show that subarray Q-IP must be accompanied by executable frequency-allocation rules, collision-risk models, and crosstalk budgets, rather than only static layouts (Murray et al., 2021; Shi et al., 2024). The MQT-DASQA framework proposed by Kunasaikaran et al. further shows that application-oriented subarray Q-IP also needs to parameterize topology constraints, coupling graphs, and control resources together, so that architectural-level composition can be automated (Kunasaikaran et al., 2024). Its reuse value ultimately depends on whether frequency allocation, crosstalk suppression, readout reuse, packaging boundaries, and cross-module calibration can be stably modeled, and whether they can be independently regressed in prevalidated databases such as SQuADDS or in equivalent test-chip data (Shanto et al., 2024).

\subsection{Logical and Error-Correction Q-IP}

Logical Q-IP corresponds to high-level abstraction structures in fault-tolerant quantum computing, including surface-code patches, error-syndrome extraction circuits, lattice-surgery boundaries, logical-qubit routing architectures, and logical memory/entanglement modules (Fowler et al., 2012; Litinski, 2019). This class of Q-IP is not a purely software-level quantum gate library, but a hardware-logic composite module that jointly encapsulates physical qubit arrays, measurement qubits, readout scheduling, decoding interfaces, and control resources.

Within the surface-code framework, a logical qubit is composed of multiple physical qubits, and its fault-tolerance capability is determined by the code distance $d$. The logical error rate can be approximated as:

\[
P_L \approx A \left(\frac{p}{p_{th}}\right)^{\frac{d+1}{2}}
\]

Here, $p$ is the physical error rate, $p_{th}$ is the fault-tolerance threshold, and $A$ is a constant related to the specific noise model and circuit implementation. The applicability of this relationship presupposes that the physical error rate is below the threshold and that the error model, syndrome measurement, and decoding flow satisfy the corresponding assumptions. The experimental results of Barends et al. approaching the surface-code threshold in transmon systems show that the physical error-rate range promised by logical Q-IP must come from repeatable cryogenic calibration rather than idealized simulation (Barends et al., 2014). In practical engineering, issues such as leakage errors, correlated errors, readout crosstalk, real-time decoding latency, and control-line congestion must also be addressed. The hierarchical modular surface-code architecture proposed by Nickerson et al. and the hardware co-design work of the QuIRC Collaboration on lattice-surgery routing cards both show that logical Q-IP must also define module boundaries, interconnection interfaces, and control/decoding resource-allocation rules (Fowler et al., 2012; Nickerson et al., 2016; Gambetta et al., 2017; QuIRC Collaboration, 2023; National Academies, 2024).

The core role of logical Q-IP is to encapsulate complex error-correction structures into standard modules, allowing system designers to invoke logical qubits, logical memory blocks, or lattice-surgery interfaces without redesigning all underlying measurement circuits and error-correction scheduling logic in every project. However, such encapsulation must be accompanied by explicit physical assumptions, including the required physical error-rate range, code distance, measurement cycle, area overhead, readout parallelism, decoding interface, and boundary-connection rules. Without these metadata, logical Q-IP can serve only as a conceptual illustration and cannot become a licensable and reusable engineering asset.

\subsection{Licensing, Verification, and Responsibility}

Q-IP cores must be encapsulated as a combination of parameterized PCells, model cards, SPICE-Q files, test-chip data, and integration documentation, and must undergo complete characterization and verification under the PDK constraints of the target foundry to ensure their repeatability and performance consistency in real manufacturing environments (Cross et al., 2022). Drawing on classical SoC IP reuse methodology, Q-IP releases cannot deliver only design files; they must also deliver interface specifications, constraint conditions, verification cases, version records, known limitations, and integration guides (Keating \& Bricaud, 2002).

The verification flow usually includes process design-rule checks (DRC/LVS/DFM), joint electromagnetic and circuit-level simulation (SPICE-Q), noise and decoherence model verification, calibration against cryogenic experimental data, system-level error-propagation analysis, multi-PDK/multi-batch revalidation, and fab-out manufacturing-file checks. For logical Q-IP, syndrome measurement cycles, decoding interfaces, logical error-rate models, and boundary-connection rules must also be additionally verified. For readout/amplifier Q-IP, gain, bandwidth, noise temperature, pump conditions, and cryogenic thermal load must also be verified.

From the perspective of engineering governance, Q-IP simultaneously plays the dual role of a technical unit and a definition of responsibility boundaries. Drawing on classical hard-IP/soft-IP licensing practices, Q-IP licensing usually also needs to distinguish between "layout- and process-bound modules" and "model/interface-portable modules": the licensing scope of the former is usually limited to a specific PDK version and foundry node, whereas the latter may be reused across projects under the premise of re-sign-off and revalidation, but may not exceed the frequency window, packaging conditions, and noise assumptions declared in the model card (Keating \& Bricaud, 2002; Mitchell et al., 2019). Once an IP module is packaged and released, its performance metrics must hold within the constraint space defined by the PDK; if a designer uses the IP in a process, frequency window, packaging condition, or temperature regime outside the scope of the model card, revalidation is usually required. Licensing contracts should clearly define the division of responsibility among IP suppliers, fabless designers, and foundries: IP suppliers are responsible for model and interface declarations, foundries are responsible for process statistics and manufacturing consistency, and fabless designers are responsible for system-level integration and application-scenario validation. Third-party verification databases, such as SQuADDS, or foundry PCM data can serve as independent audit baselines, reducing the reuse risks associated with "single-team self-validation" (Shanto et al., 2024; Wilkinson et al., 2016).

Therefore, this mechanism is not only a matter of technical standards, but also the core governance structure for ensuring manufacturability and credibility in the fabless quantum ecosystem. A mature Q-IP market must establish version management, revalidation rules, data-access permissions, failure responsibilities, and licensing scopes; otherwise, Q-IP reuse may mistake single-project experience for cross-platform commitments and thereby increase system risk (Krantz et al., 2019; Fowler et al., 2012; Mitchell et al., 2019; Wilkinson et al., 2016).

\section{Commercialization Pathways and Industrial Coordination}

> Literature index: The discussion in this chapter on Fabless-Foundry industrial division of labor, modular production networks, IP reuse, and service contracts draws primarily on the classic literature on VLSI/Fabless and modular industries (Mead \& Conway, 1980; Baldwin \& Clark, 2000; Langlois \& Robertson, 1992; Sturgeon, 2002; Macher et al., 2008; Keating \& Bricaud, 2002), and combines it with literature on quantum-industry scaling, PDK/Q-EDA, GDSII-to-wafer, data governance, and quantum-technology market monitoring (CMC Microsystems, 2023; QPDK Documentation, 2025; EDA-Q Collaboration, 2025; GDSII-to-Wafer Collaboration, 2026; Wilkinson et al., 2016; McKinsey \& Company, 2025; National Academies, 2024).

This chapter discusses the industrial coordination problems that must be resolved as Fabless quantum-chip design moves from a technical framework toward a commercialization ecosystem. The commercialization pathway considered here is neither a simple outsourcing of manufacturing to foundries nor an assumption that quantum chips can immediately replicate the mature CMOS industrial chain. Rather, it analyzes how, under conditions in which PDKs, PCells, SPICE-Q, Q-EDA, Q-IP, cryogenic testing, and data feedback gradually mature, repeated investment can be reduced, foundational component development can be accelerated, barriers to entry can be lowered, and a governable coordination network can be formed among Foundries, tool vendors, IP suppliers, Fabless design firms, and application customers.

\subsection{Reducing Duplicated Industry Investment}

One of the core objectives of the quantum Fabless ecosystem is to reduce duplicated investment in repeatable stages of the ``design--manufacturing--validation'' loop. Under the traditional integrated device manufacturer (IDM) model, each quantum-chip company often needs to independently complete the full process from qubit design, layout implementation, micro/nanofabrication, and packaging interconnection to cryogenic experimental validation. This highly vertically integrated structure is conducive to rapid early-stage closed-loop iteration, but it also easily causes multiple teams to repeatedly consume engineering resources on similar problems such as Josephson-junction process windows, readout resonators, frequency allocation, coupler calibration, and test-structure design (Krantz et al., 2019; Kjaergaard et al., 2020; National Academies, 2024).

In a Fabless--Foundry structure, this problem can be mitigated through standardized PDKs (Process Design Kits), reusable PCell/Q-IP, and Q-EDA toolchains. Design companies do not need to re-explore the entire parameter space of foundational devices in every project. Instead, they can conduct system-level compositional design based on validated Qubit PCells, readout PCells, coupler PCells, and Logical PCells, and then complete tape-out, testing, and model feedback on Foundry platforms with micro/nanofabrication and cryogenic testing capabilities. The essence of this model is to migrate some repeatable quantum-chip development activities from ``exploratory experimental loops within individual projects'' to a standardized engineering process supported by ``Foundry-PDK-test structures-Q-EDA.'' Its effectiveness depends on PDK version stability, the credibility of model cards, testing throughput, and IP licensing boundaries, rather than being automatically guaranteed by the Fabless organizational form (CMC Microsystems, 2023; QPDK Documentation, 2025; GDSII-to-Wafer Collaboration, 2026).

From the perspective of resource optimization, this transformation can be formalized as a constrained optimization problem. Under the traditional model, total R\&D cost can be expressed as:

\[
C_{IDM} = \sum_{i=1}^{N} (C_{design,i} + C_{fab,i} + C_{test,i})
\]

where each firm $i$ must repeatedly bear the cost of the complete R\&D chain. Under the Fabless model, through PDK and Q-EDA sharing mechanisms, the total cost is restructured as:

\[
C_{Fabless} = C_{PDK} + C_{foundry} + \sum_{i=1}^{N} C_{design,i}^{reuse}
\]

where $C_{design,i}^{reuse}$ is not necessarily much smaller than $C_{design,i}$, but depends on the credibility of PCell models, Q-IP licensing prices, cryogenic testing schedules, and revalidation costs. A more cautious formulation is that the exploratory cost borne by an individual project decreases only when repeated device development, frequency tuning, coupling-strength calibration, and readout-chain validation can be absorbed by PDKs, test structures, and Q-IP modules (Krantz et al., 2019; Shanto et al., 2024).

More importantly, this system introduces a cross-firm data feedback mechanism. Foundries, such as the Suzhou production line, not only execute manufacturing tasks but also continuously optimize the parameter spaces of Qubit PCells and Logical PCells by collecting design and testing data from different Fabless companies, thereby gradually pushing the performance of foundational components toward the statistical boundaries attainable by the process. This process is closer to governed data feedback and empirical risk minimization than to unconstrained data aggregation:

\[
\theta^{*} = \arg\min_{\theta \in \Omega_{PDK}} \mathbb{E}_{\mathcal{D}_{multi}}[\mathcal{L}(\theta)]
\]

where $\mathcal{D}_{multi}$ denotes the joint data distribution from multiple design companies, but its effectiveness depends on data permissions, consistency of measurement conditions, model-card records, and FAIR metadata governance (Mitchell et al., 2019; Wilkinson et al., 2016). At the level of industrial structure, this mechanism can reduce the waste of resources caused by repeated trial and error, but it cannot completely replace project-level validation. Each generation of quantum chips still requires re-signoff for packaging, frequency windows, crosstalk, and system calibration.

For start-ups, this structural change is especially important. Firms can externalize part of the burden of foundational-device R\&D, process exploration, and cryogenic testing to the Foundry and PDK system, and perform circuit-level design and system-level optimization within the PDK-constrained space. This transformation helps lower barriers to entry and gradually shifts quantum-hardware development from a purely heavy-asset, research-driven model toward an engineering model of ``shared manufacturing infrastructure + differentiated system design.'' It should be emphasized that the specific magnitude of cost savings depends on PDK maturity, foundry yield, cryogenic testing capacity, and IP licensing prices; this paper does not provide deterministic quantification of savings that has not been validated by the industry (National Academies, 2024; CMC Microsystems, 2023).

From the perspective of macro-level industrial structure, this mechanism is expected to alleviate the current pattern of regionally isolated development, such as Beijing's emphasis on academia, Hefei's emphasis on system-oriented R\&D, Suzhou's emphasis on process technology, and Shenzhen's emphasis on applications, and to promote the formation of a distributed design network supported by unified interfaces and traceable data. However, regional division of labor does not mean that all regions should completely stop building overlapping capabilities. In the early stage, moderate redundant validation and multi-route exploration still help reduce technological uncertainty. More cautiously, the number of Fabless design actors may grow with the degree of PDK openness, the availability of Q-EDA, Foundry response speed, cryogenic testing capability, and the clarity of application demand, thereby providing conditions for quantum computing to transition from the research stage to the engineering and commercialization stage (Cross et al., 2022; McKinsey \& Company, 2025; National Academies, 2024).

\subsection{Accelerating Foundational Component Development}

In the quantum Fabless ecosystem, the development efficiency of foundational components (fundamental building blocks) directly determines the technological iteration speed of the entire industry. Foundational components include reusable quantum IP units such as Qubit PCells, Logical PCells, readout modules, and couplers. Under the traditional IDM model, these components usually depend on the internal R\&D process of a single institution, resulting in long development cycles and difficulty in reuse across teams (Krantz et al., 2019).

In the Fabless--Foundry architecture, through a unified PDK and Q-EDA toolchain, foundational component development can gradually shift from ``point optimization'' to ``multi-party collaborative optimization.'' Design companies can iterate different component configurations within the standardized parameter space $\Omega_{PDK}$ and reuse the test structures, model cards, and validated PCell results released by the Foundry. However, this does not mean that low-level physical modeling and cryogenic experimental validation are completely eliminated. For new frequency windows, packaging structures, coupler topologies, or logical patches, SPICE-Q simulation, PDK signoff, and cryogenic benchmarking still need to be completed again. This process can be abstracted as the following optimization problem:

\[
\theta^{*} = \arg\min_{\theta \in \Omega_{PDK}} \mathcal{L}_{component}(\theta)
\]

where $\theta$ denotes component design parameters, and $\mathcal{L}_{component}$ denotes the component-performance loss function, such as decoherence time, gate fidelity, or readout error rate.

In this system, foundry production lines such as those in Suzhou not only undertake manufacturing tasks but may also gradually evolve into ``data learning centers for foundational components.'' By aggregating experimental and tape-out data from multiple Fabless design companies, they can continuously optimize Qubit PCells and Logical PCells and thereby form a closed-loop feedback system. This process essentially constitutes a distributed empirical learning framework, whose form can be expressed as:

\[
\mathcal{D}_{t+1} = \mathcal{D}_{t} \cup \{(x_i, y_i)\}_{new}
\]

where $\mathcal{D}_{t}$ is the historical design and testing dataset, and newly added data are used to continuously update process models and component-parameter distributions.

This data-driven optimization mechanism enables foundational component development to gradually shift from traditional ``experience-and-trial-and-error driven'' development to ``statistical-learning driven'' development. For example, the frequency drift of Qubit PCells, coupler crosstalk, readout signal noise, Purcell limits, and sensitivity to packaging modes can all be systematically calibrated through cross-project data statistics and then used to update PDK statistical corners and model cards. However, these data must satisfy requirements for comparable measurement conditions, clear version records, and traceable metadata; otherwise, cross-project aggregation may misidentify batch differences, measurement-chain differences, or packaging differences as device-design regularities (Mitchell et al., 2019; Wilkinson et al., 2016).

At the same time, this mechanism can improve resource utilization efficiency. Because part of the work of component-performance optimization can be completed in shared data spaces and standardized test structures, foundries can more intensively optimize key process nodes, such as Josephson-junction uniformity, dielectric loss tangent $\tan\delta$, CPW/readout-resonator loss, and microwave interconnect loss, thereby improving overall yield and consistency. The improvement here remains conditional: if cryogenic testing capacity is insufficient, data permissions are restricted, or failure-mode records are incomplete, shared data will not automatically translate into improved component performance (McRae et al., 2020; Murray, 2021; Shanto et al., 2024).

From a system perspective, this model gives foundational component development conditional scale effects. As the number of Fabless design companies $N$ increases, data scale, device diversity, and model-calibration opportunities increase, but optimization efficiency does not automatically improve monotonically with $N$. Multi-project data become usable model improvements only when data formats, testing protocols, PDK versions, and failure-mode records remain consistent. A more reasonable relationship can be written as:

\[
\frac{d\mathcal{P}}{dt} = f(\mathcal{D}_{quality}, N_{validated}, \Omega_{PDK}, C_{test})
\]

where $\mathcal{P}$ denotes the component-performance boundary, $\mathcal{D}_{quality}$ denotes the scale of quality-controlled data, $N_{validated}$ denotes the number of projects that have completed comparable validation, and $C_{test}$ denotes the cryogenic testing-capacity constraint. Prevalidated databases such as SQuADDS show that the key to data-driven component development is not merely accumulating more layout samples, but establishing a closed loop among parameter--performance mapping, simulation workflows, and experimental validation (Shanto et al., 2024).

Therefore, production lines such as those in Suzhou are not merely manufacturing nodes in this ecosystem, but may also become data-feedback centers for the evolution of foundational components. By continuously absorbing multi-source design data and feeding it back into the PDK and Q-EDA systems, foundational component performance is expected to gradually approach the physical limits of the process, thereby accelerating the evolution of quantum chips from prototype systems to scalable engineering systems. However, this judgment must be premised on simultaneous improvements in testing throughput, data governance, and the degree of Foundry openness (Cross et al., 2022; Krantz et al., 2019; GDSII-to-Wafer Collaboration, 2026).

\subsection{Lowering Barriers to Entry}

One core objective of the Fabless quantum-chip ecosystem is to lower the barriers to entry for quantum-hardware R\&D once conditions mature, enabling start-ups and academic institutions to participate in certain forms of quantum-chip design and innovation without building complete cleanroom, micro/nanofabrication, and cryogenic testing systems themselves. Under the traditional IDM model, the R\&D of quantum logical qubits often requires long-term process iteration, cryogenic testing, micro/nanofabrication, packaging interconnection, and joint debugging with control systems, and may involve high tape-out and debugging costs. These requirements constitute the main barriers to industry entry (Krantz et al., 2019; Kjaergaard et al., 2020; National Academies, 2024; McKinsey \& Company, 2025).

In the Fabless--Foundry architecture, this structural barrier can be partially weakened by PDKs (Process Design Kits), Q-EDA toolchains, standardized test structures, and Q-IP modules. Designers no longer need to reconstruct qubits from low-level device physics in every project, but can invoke validated Qubit PCells, readout PCells, coupler PCells, or Logical PCells within the parameterized design space $\Omega_{PDK}$, thereby transforming part of system design into a constrained optimization problem:

\[
\theta^{*} = \arg\min_{\theta \in \Omega_{PDK}} \mathcal{L}_{system}(\theta)
\]

where $\theta$ denotes quantum-chip design parameters, and $\mathcal{L}_{system}$ denotes a system-level performance loss function, such as a composite metric of logical error rate, gate fidelity, and decoherence time.

The key significance of this transformation lies in the ``upward shift of abstraction level'': quantum-chip design is partially elevated from the device-physics layer to the system-architecture layer, enabling designers to explore topology, frequency, readout, and error-correction structures in a platform environment analogous to classical EDA. However, this abstraction cannot be understood as fully equivalent to classical SoC design, because Transmon frequency drift, material loss, packaging modes, and cryogenic calibration still feed back into system performance. What Fabless lowers is the threshold for entering low-level manufacturing and repeated device exploration, not the demand for quantum-hardware engineering knowledge (Kjaergaard et al., 2020; National Academies, 2024).

Under this model, start-ups and academic institutions can focus on application-oriented quantum processor design, such as Application-Specific Quantum Integrated Circuits (ASQICs) for quantum-chemistry simulation, combinatorial optimization problems, or machine-learning acceleration. Quantum chemistry has long been regarded as an important application direction for fault-tolerant quantum computing, and the resource estimate by Reiher et al. for the nitrogenase reaction mechanism shows that real application design must simultaneously consider algorithms, error-correction overhead, gate-set compilation, and hardware-architecture constraints (Reiher et al., 2017). Therefore, an ASQIC is not a simple mapping of an application algorithm onto a chip, but a task-specific hardware co-design process under the combined effects of PDK constraints, Q-IP reuse, and error budgets.

From the perspective of cost structure, the entry barrier can be reconstructed from the traditional model:

\[
C_{entry}^{IDM} = C_{cleanroom} + C_{cryogenic} + C_{process\_dev} + C_{iteration}
\]

to the Fabless model:

\[
C_{entry}^{Fabless} = C_{design} + C_{PDK\_access} + C_{cloud\_EDA}
\]

where high-capital-expenditure items such as $C_{cleanroom}$ and $C_{cryogenic}$ are partially externalized to foundries and infrastructure providers under the Fabless model. However, designers still need to bear the costs of system architecture, simulation validation, Q-IP licensing, cryogenic testing schedules, and application integration. Thus, Fabless lowers the threshold for repeated manufacturing and low-level process exploration, rather than eliminating the complexity of quantum-hardware design itself.

In addition, the Fabless model also introduces an effect of ``capability democratization.'' Because Q-IP modules are reusable and composable, different institutions can develop highly differentiated quantum systems on the basis of the same physical platform without repeatedly solving the problems of low-level physical implementation. This structure resembles the transition in the classical semiconductor industry from transistor design to standard-cell design (Weste \& Harris, 2011).

From the system level, this model may expand the number of innovation participants $N$, but the rate of technological progress is not determined solely by the number of participants. A more reasonable formulation is to express overall innovation capability as a function jointly constrained by interface maturity, testing capacity, data-feedback quality, Q-IP reusability, and application demand:

\[
R_{innovation}=f(N,S_{PDK},Q_{EDA},C_{test},D_{feedback},A_{market})
\]

where $S_{PDK}$ denotes PDK stability, $Q_{EDA}$ denotes toolchain maturity, $C_{test}$ denotes cryogenic testing capacity, $D_{feedback}$ denotes the quality of data that can be fed back, and $A_{market}$ denotes the clarity of application demand. Only when these constraints improve simultaneously can more participants be transformed into effective innovation, rather than producing fragmented tools, repeated tape-outs, and validation congestion.

Therefore, the significance of the Fabless quantum ecosystem lies in lowering some economic and technical thresholds and gradually changing the distributional structure of quantum-hardware innovation: from a centralized R\&D system dominated by a small number of large vertically integrated institutions to an open design network jointly involving Foundries, Q-EDA, Q-IP, Fabless design firms, and application teams. This transition still depends on standard interfaces, model credibility, and application pull, and should not be understood as an industrial restructuring that will occur automatically in the short term (Cross et al., 2022; Preskill, 2018; McKinsey \& Company, 2025).

\subsection{Market Structure and Coordination}

The core of the Fabless quantum-chip ecosystem is to construct a four-in-one coordination structure of ``factory--design--market--software platform.'' In this system, the Foundry provides reusable manufacturing capability, test structures, and process-data interfaces; Fabless companies are responsible for system-level design and application development; PDK and Q-EDA toolchains constitute the foundational software layer connecting design and manufacturing; and Q-IP suppliers accumulate licensable device, circuit, and logical modules. The goal of this division-of-labor model is to gradually decouple the quantum-chip industry from a fully vertically integrated structure into a governable horizontal collaboration network. Its efficiency gains depend on stable interfaces, clear responsibility boundaries, usable data feedback, and clear application demand, rather than arising automatically from organizational separation itself (Baldwin \& Clark, 2000; Sturgeon, 2002; Macher et al., 2008; Krantz et al., 2019).

Within this structure, ``softwareization'' is one of the key drivers. By uniformly encapsulating physical device design, quantum-circuit simulation, and layout generation as software workflows, quantum-chip design gradually shifts from being purely experiment-driven to being driven in parallel by computation and experimental feedback. Designers can complete a considerable part of system design in high-level abstraction spaces, but still need to confirm physical realizability through Foundry signoff, cryogenic testing, and model-card updates. Work on OpenQASM 3, QIR, KQCircuits, QPDK, and GDSII-to-wafer indicates that the softwareization of quantum hardware has already emerged as a direction, but a mature ecosystem depends on the interoperability of these interfaces with manufacturing data packages, PDK versions, and Q-IP licensing boundaries (Cross et al., 2022; QIR Alliance, 2021; KQCircuits Documentation, 2024; GDSII-to-Wafer Collaboration, 2026).

From a system perspective, this coordination structure can be modeled as a multilayer optimization problem:

\[
\min_{\theta_d, \theta_f} \mathcal{L}_{system}(\theta_d, \theta_f)
\]

where $\theta_d$ denotes design-space parameters on the Fabless side, and $\theta_f$ denotes manufacturing-process parameters on the Foundry side. This bilevel optimization structure is coupled through the PDK, enabling design and manufacturing to converge collaboratively within a unified constrained space.

\subsubsection{Expansion Effects of the Fabless Model}

Under the Fabless model, start-ups can design quantum chips based on standardized PDKs and Q-EDA tools, and reduce their direct dependence on low-level process development. This structural decoupling helps lower barriers to industry entry, but the growth in the number of design firms still depends on the degree of PDK openness, foundry capacity, cryogenic testing throughput, Q-IP licensing costs, and real application demand.

From the perspective of network effects, the scale growth of the Fabless ecosystem may have nonlinear characteristics, but it is inappropriate to directly assume that an increase in the number of design actors will necessarily produce superlinear innovation. Let the number of design companies be $N$; then the system innovation capability $R$ is more suitably expressed as a function jointly constrained by interface maturity, data availability, and application demand:

\[
R = f(N, S_{interface}, D_{feedback}, A_{market})
\]

where $S_{interface}$ denotes the stability of PDK/Q-EDA/Q-IP interfaces, $D_{feedback}$ denotes the quality of manufacturing and cryogenic testing data that can be fed back, and $A_{market}$ denotes the clarity of application demand. The theory of modular production networks shows that an increase in participants brings specialization and parallel innovation gains only when interface boundaries are clear and responsibility governance is sufficient (Baldwin \& Clark, 2000; Sturgeon, 2002).

In addition, the Fabless model promotes the maturation of IP reuse mechanisms. Q-IP modules, such as Qubit PCells and Logical PCells, can be shared among different design companies, transforming foundational component development from ``repeated construction'' into ``combinatorial innovation.'' However, reuse efficiency depends on whether the modules carry auditable models, cryogenic validation records, and re-signoff rules, rather than on whether the modules are nominally encapsulated.

\subsubsection{Mechanisms for Accelerating Industry Development}

In the Fabless coordination system, improvements in quantum-chip production efficiency come mainly from two sources: first, shortened development cycles brought by design-flow standardization; and second, process optimization driven by manufacturing feedback data.

Under the traditional model, each generation of quantum chips often requires a complete design--manufacturing--testing loop to be repeated. Under the Fabless model, if PDKs, PCells, and test structures are already mature, part of the foundational-device validation can be reused, and project-level iteration cycles are expected to shorten. This relationship can be expressed as:

\[
\mathbb{E}[T_{cycle}^{Fabless}] < \mathbb{E}[T_{cycle}^{IDM}]
\]

where $T_{cycle}$ denotes a single technology-iteration cycle. This inequality does not hold automatically, but depends on Foundry response speed, the maturity of signoff workflows, and cryogenic testing resources.

Meanwhile, in this system, foundries such as Suzhou production lines with micro/nanofabrication foundations should not only undertake one-off manufacturing tasks, but can also gradually become centers of data aggregation and process optimization. On the premise of compliant authorization and metadata consistency, by aggregating design, tape-out, and cryogenic testing data from multiple Fabless companies, the Foundry can continuously optimize the model parameters of Qubit PCells, readout PCells, coupler PCells, and Logical PCells, thereby improving yield prediction and version governance. However, this process requires clear data ownership, anonymization mechanisms, measurement protocols, and model-card update rules; otherwise, multi-party data sharing will introduce risks of commercial confidentiality leakage and model misuse (Mitchell et al., 2019; Wilkinson et al., 2016; QPDK Documentation, 2025).

This mechanism helps the quantum-chip industry move from ``regionally isolated development,'' such as Beijing's emphasis on research, Hefei's emphasis on system R\&D, Suzhou's emphasis on process technology, and Shenzhen's emphasis on applications, toward a coordinated network structure driven by unified standards. Public industry reports show that Chinese quantum-computing enterprises and industrial parks are evolving toward multi-regional collaboration networks, but standardized PDKs and cross-platform Q-EDA interfaces remain at an early stage (ChinaQuantum, 2025; Entangled Future, 2026; The Quantum Insider, 2026; McKinsey \& Company, 2025). In this network, different regions do not completely abandon overlapping capability building, but gradually form division of labor and collaboration around unified PDKs, testing protocols, and Q-IP interfaces while preserving necessary redundant validation.

From the perspective of macro-level industrial outcomes, this structural change is expected to increase the speed of market expansion and, in the medium to long term, form more application-oriented quantum-chip design actors, thereby promoting the transition of quantum computing from a research-oriented stage to engineering and commercialization expansion (Cross et al., 2022; Preskill, 2018).

\subsection{Business Models and Service Contracts}

In the Fabless quantum-chip ecosystem, the core business model may gradually shift from traditional ``one-off hardware delivery'' toward a ``PDK-driven, continuous-service coordination structure.'' In this system, a quantum chip is no longer merely a physical product, but a system-level engineering service jointly defined by Q-IP modules, the PDK-constrained space, Q-EDA design workflows, cryogenic testing services, and model updates. This transformation expands the industrial structure from single-product delivery to a service portfolio composed of platforms, tools, IP, manufacturing, and calibration (Keating \& Bricaud, 2002; Cross et al., 2022; McKinsey \& Company, 2025).

At the level of service contracts, cooperation between Fabless companies and Foundries usually includes five core categories: PDK access and process-use authorization, Q-IP module invocation and reuse rights, Q-EDA tool and manufacturing data-package services, cryogenic testing and model-feedback arrangements, and joint validation and delivery standards based on performance metrics. This contractual structure essentially extends traditional manufacturing contracts into an integrated coordination agreement for ``design--manufacturing--validation--calibration--model updating,'' and requires clear responsibility boundaries among Foundries, Fabless design firms, Q-IP suppliers, and application customers.

From the perspective of system modeling, the stability of functionalized quantum chips can be characterized through foundational component consistency. If a system consists of $N$ Q-IP modules, its overall performance fluctuation can be expressed as:

\[
\sigma_{system}^2 = \sum_{i=1}^{N} \sigma_{QIP,i}^2 + \sum_{i \neq j} \mathrm{Cov}(QIP_i, QIP_j)
\]

where the first term denotes the intrinsic noise contribution of each independent Q-IP module, and the second term denotes correlated errors among modules. The goal of standardized PDKs and unified Q-IP interface design is to identify and control covariance terms, such as correlated errors caused by common packaging modes, shared readout chains, frequency crowding, and control crosstalk. However, covariance does not automatically disappear because of interface standardization, and must still be validated through system-level simulation, test chips, and cryogenic calibration.

The key to this mechanism is the ``sufficient understanding and standardization of foundational components.'' Once Qubit PCells and Logical PCells are sufficiently characterized at the PDK level, their behavior no longer depends only on the results of one-off experimental tuning, but is constrained by statistical process models, model cards, and version governance. Thus, functionalized quantum chips built by different Fabless companies on the basis of the same foundational components may exhibit higher consistency and predictability, provided that their packaging, readout, control, and application workloads remain within the applicable range of the model.

At the level of business models, this structure supports a transition from one-off sales to a multilayer service system. PDK subscription services correspond to access rights to process nodes and version updates; Q-IP licensing and reuse fees correspond to validated device, circuit, and logical modules; cloud-based Q-EDA services correspond to simulation, design-rule checking, manufacturing data-package generation, and model management; and performance-based joint validation contracts write the responsibility boundaries among Foundries, Fabless design firms, and Q-IP suppliers into the delivery process. Such a service portfolio has governance logic similar to interface documentation, verification cases, version records, and known limitations in classical SoC IP reuse, but the quantum context must additionally incorporate cryogenic testing conditions, the applicable ranges of noise models, and data-feedback permissions (Keating \& Bricaud, 2002; Mitchell et al., 2019; Wilkinson et al., 2016).

This model means that the Foundry exists not only as a manufacturer, but also as a ``provider of standardized stability,'' improving the reliability of the entire ecosystem by continuously optimizing the parameter spaces of foundational components.

From the perspective of industrial outcomes, as foundational component stability improves, correlated errors are identified, and system-level revalidation is passed, the performance variance of functionalized quantum chips is expected to decline, thereby promoting quantum-computing applications from ``experimentally feasible'' toward ``engineering usable.'' This process can be formalized as:

\[
\mathbb{E}[P_{fail}] \downarrow \quad \text{as} \quad \sigma_{QIP} \downarrow
\]

where $P_{fail}$ denotes the system failure probability.

Therefore, a business model based on standardized Q-IP and PDK systems not only changes the delivery mode of quantum chips, but also provides a clearer engineering foundation for the stability, scalability, and responsibility governance of functionalized quantum chips. Its commercial implementation still depends on the simultaneous maturation of Foundry yield, cryogenic testing throughput, model credibility, Q-IP licensing costs, and application demand (Cross et al., 2022; Krantz et al., 2019; GDSII-to-Wafer Collaboration, 2026; McKinsey \& Company, 2025).

\section{Application Markets and Product Strategy}

\subsection{Application-Specific Quantum Integrated Circuits}

In the Fabless quantum-chip ecosystem, Application-Specific Quantum Integrated Circuits (ASQICs) constitute a potential product form oriented toward application markets. Unlike general-purpose quantum processors, ASQICs are architecturally optimized for specific computational tasks, such as quantum-chemistry simulation, combinatorial optimization, or quantum machine-learning acceleration, thereby maximizing task performance under limited qubit resources (Preskill, 2018). It should be carefully noted that ASQICs at present are more a design direction for future fault-tolerant or high-quality NISQ hardware than an already mature commercial product category. Their value depends on whether target applications can provide clear circuit structures, error budgets, and resource constraints.

From the application literature, quantum chemistry, combinatorial optimization, and variational algorithms are the three directions most suitable as early demonstration targets for ASQICs. The quantum resource estimate by Reiher et al. for complex chemical reaction mechanisms shows that real chemical applications must be specified in terms of the number of logical qubits, gate depth, error-correction overhead, and compilation constraints before they can guide hardware architecture in reverse (Reiher et al., 2017). QAOA, proposed by Farhi et al., indicates that combinatorial optimization applications naturally contain structured parameters such as problem graphs, cost Hamiltonians, mixing Hamiltonians, and circuit depth $p$, which can be mapped to requirements for coupling graphs, connectivity, gate fidelity, and readout scheduling (Farhi et al., 2014). The review of variational quantum algorithms by Cerezo et al. further points out that NISQ-stage applications are usually limited by noise, trainability, circuit depth, and classical optimizer stability. Therefore, task-specific chips should not pursue only more qubits, but should also optimize connectivity, crosstalk, measurement parallelism, and error-mitigation interfaces around target circuit families (Cerezo et al., 2021).

Fabless quantum-chip designers conduct system-level design through standardized PDKs and PCell libraries, enabling dedicated quantum chips to be iteratively constructed within a constrained process space, rather than being completed automatically by stable standard cells as in mature classical ASICs. Their design process can be abstracted as a constrained optimization problem:

\[
\theta^{*} = \arg\min_{\theta \in \Omega_{PDK}} \mathcal{L}_{task}(\theta)
\]

where $\mathcal{L}_{task}$ denotes the performance loss function of a specific application task, such as quantum-state fidelity, optimization convergence speed, chemical energy error, approximation ratio, sampling quality, or the effective circuit success rate after error mitigation. For quantum-chemistry chips, the constraints may focus on long coherence times, compilable two-qubit gate sequences, and fault-tolerant resource estimates. For combinatorial-optimization chips, the constraints may focus on problem-graph embedding, local connectivity, parameterized gate depth, and rapid repeated sampling. For variational-algorithm chips, the constraints include trainability, the risk of noise-induced barren plateaus, measurement throughput, and classical--quantum feedback latency (Farhi et al., 2014; Cerezo et al., 2021).

Therefore, ASQICs are essentially an ``application-driven quantum-hardware co-design paradigm.'' Their core advantage is not that they guarantee rapid commercialization, but that, through Q-IP reuse and the PDK-constrained space, they translate application workloads into verifiable hardware requirements, error budgets, and test-chip metrics.

\subsection{Calibration and Benchmarking Chips}

In the Fabless system, quantum foundries not only undertake production functions but are also responsible for providing standardized calibration chips and benchmarking chips to characterize process consistency and qubit-performance boundaries. Such chips usually include standardized Transmon arrays, CPW resonators, Josephson-junction monitoring structures, coupling structures, and readout chains, and are used to systematically evaluate key metrics such as decoherence time, resonator $Q$ factors, gate fidelity, material loss, and frequency drift (Krantz et al., 2019; McRae et al., 2020; McDonald et al., 2022).

At the physical level, the performance of superconducting quantum chips depends strongly on material and interface quality, such as Josephson-junction uniformity, dielectric loss tangent $\tan\delta$, and two-level systems (TLS). These factors directly determine qubit decoherence behavior:

\[
T_1^{-1} = \Gamma_{rad} + \Gamma_{TLS} + \Gamma_{Purcell}
\]

where the terms correspond respectively to radiation loss, material-defect loss, and the Purcell effect.

Through dedicated calibration chips, foundries can establish a unified performance-calibration system across batches and across Fabless companies, thereby providing a data foundation for PDK model updates. Here, ``benchmarking'' should not be equated only with a single qubit's $T_1/T_2$ or a single gate fidelity measurement, but should cover circuit width, depth, connectivity, readout parallelism, crosstalk, frequency collisions, leakage, and application-related workloads. The volumetric benchmark framework proposed by Blume-Kohout and Young emphasizes that quantum processor capability should be evaluated through a two-dimensional width--depth space. This idea can be used in the design of Fabless Foundry test chips: test structures should characterize not only the physical boundaries of devices, but also the system boundaries of executable circuit families (Blume-Kohout \& Young, 2020). In scenarios closer to product delivery, calibration chips should also include benchmark circuits related to ASQIC workloads, such as QAOA layers, variational ansatz fragments, readout multiplexing patterns, and short-range error-correction cycles, in order to test whether PDK models can predict error accumulation in real application circuits (Farhi et al., 2014; Cerezo et al., 2021).

\subsection{Education and Developer Hardware}

Education and developer hardware are important product categories in the Fabless quantum ecosystem for lowering cognitive barriers and cultivating a developer ecosystem. Through standardized quantum-chip modules, public or controlled-access simplified PDK interfaces, and cloud-based simulation environments, students and engineers can understand the relationships among qubit operations, frequency planning, layout constraints, readout chains, and error budgets without directly building complete cryogenic measurement and control systems.

This type of hardware is usually based on simplified Q-IP structures, such as small-scale Transmon arrays, basic readout systems, or virtualized PCell/PDK sandboxes, enabling users to observe the effects of quantum superposition, entanglement, decoherence, and measurement error on circuit results in a controlled environment. It should not be positioned as a ``low-cost substitute for research-grade hardware,'' but rather as a developer platform: it connects logical interfaces such as OpenQASM/QIR, Q-EDA layout generation, SPICE-Q models, and Foundry testing data in an educational manner (Cross et al., 2022; QIR Alliance, 2021; OpenQASM Working Group, 2022).

This structure essentially extends quantum hardware from a single research tool into a learnable, verifiable, and reproducible development platform, but its dissemination still depends on whether cloud access, pedagogical PDK examples, anonymized cryogenic testing data, and intellectual-property boundaries are clear. After IBM Quantum Experience opened access to cloud-based quantum processors in 2016, it quickly became a teaching and early-experimentation platform. Santos's educational introduction to IBM Quantum Experience and AbuGhanem's review of the evolution of IBM cloud quantum hardware both show that cloud access can expand the developer ecosystem, but also exposes real engineering constraints such as hardware noise, connectivity topology, and queue-resource limits (Santos, 2017; AbuGhanem, 2024). Therefore, education and developer hardware should not display only ideal quantum gates, but should incorporate noise, calibration drift, device queues, backend selection, and result-reproducibility experiments into teaching workflows.

\subsection{Cloud Platforms and Foundry-Coupled Services}

In the Fabless system, cloud platforms and quantum foundries form a tightly coupled digital manufacturing-service system. The core deliverable that a foundry provides to designers is the PDK (Process Design Kit), but the PDK should not be merely a static documentation package. It should include design rules (DRC/LVS/DFM), parameterized PCell libraries, SPICE-Q models, noise and material models, test-structure data, and fab-out manufacturing-file specifications that can be invoked by cloud-based Q-EDA. The GDSII-to-wafer work emphasizes that wafer-level superconducting quantum-chip manufacturing requires converting design files into executable manufacturing data packages, which means that cloud platforms must cover the full chain from logical description, layout generation, and rule checking to mask-data preparation (GDSII-to-Wafer Collaboration, 2026).

This system enables a considerable portion of quantum-chip design simulation, rule checking, and submission preparation to be completed in the cloud, but it still requires physical signoff, mask-data preparation, and the cryogenic testing loop on the Foundry side. Therefore, a more accurate formulation is not ``design is manufacturing,'' but a digital coordination paradigm of ``cloud design--process signoff--manufacturing feedback.'' Its value lies in reducing information-conversion costs and design-submission errors, rather than bypassing the physical constraints of manufacturing.

From a system perspective, the cloud-foundry coupling structure can be expressed as:

\[
\text{Flow} = f(\text{Design}_{cloud}, \text{PDK}, \text{Foundry})
\]

where the cloud is responsible for design iteration and simulation, and the foundry is responsible for physical implementation and data feedback, thereby forming a closed-loop optimization system (Cross et al., 2022). From the perspective of product strategy, such services should be divided into three levels. The first level is logical-circuit and application-workload services, supporting OpenQASM/QIR, VQA/QAOA templates, and benchmark circuits. The second level is physical-design services, supporting PCell invocation, SPICE-Q simulation, DRC/LVS/DFM checking, and fab-out data-package generation. The third level is manufacturing and testing feedback services, responsible for feeding cryogenic measurements, failure modes, and model-card updates back into the PDK. The development experience of IBM's quantum cloud platform shows that cloud access can significantly expand experimental and developer communities, but if device calibration information, operating histories, and benchmarking metrics are missing, customers will find it difficult to translate cloud-based experimental results into purchasable hardware capability (Santos, 2017; AbuGhanem, 2024; Blume-Kohout \& Young, 2020).

\subsection{Procurement and Customer Success}

In the Fabless quantum-chip industry, procurement and customer-success systems involve not only chip sales, but also functionalized quantum chips and continuous service-delivery models. Customers no longer purchase a single piece of hardware, but an overall solution that includes PDK access, Q-IP licensing, cloud simulation, benchmarking, system integration, calibration support, and version maintenance. Because quantum-hardware performance is highly sensitive to cryogenic environments, calibration cycles, readout chains, and application circuit families, procurement contracts must expand the ``deliverables'' from bare chips to operational capability packages: the chip itself, control/readout interfaces, model cards, passed benchmarks, reproducible experimental scripts, maintenance windows, and boundaries of failure responsibility (Mitchell et al., 2019; Wilkinson et al., 2016).

This model shifts the quantum-chip supply chain from ``product delivery'' to ``capability delivery,'' meaning that what the customer obtains is operational, calibratable, and maintainable quantum-computing capability rather than a single device. For early markets, the focus of customer success is not to promise general quantum advantage, but to decompose target applications into verifiable circuit families, error budgets, benchmarks, and maintenance protocols. This is consistent with Preskill's judgment regarding application uncertainty in the NISQ stage and with the idea of volumetric benchmarks, which characterize capability boundaries by circuit width and depth (Preskill, 2018; Blume-Kohout \& Young, 2020).

From the perspective of market structure, the growth of functionalized quantum chips can be viewed as an application-demand-driven process:

\[
G_{market} \propto \sum_{i} U_{application,i} \cdot A_{hardware,i}
\]

where $U_{application}$ denotes the intensity of application demand, and $A_{hardware}$ denotes hardware availability. However, this expression should not be interpreted as a deterministic formula for market growth, because $U_{application}$ itself depends on problem scale, the cost of classical alternatives, quantum error rates, runtime queue times, and integration costs; $A_{hardware}$ must also be defined by reproducible benchmarks, maintenance stability, and customer on-site/cloud operation records. For fields such as quantum chemistry, optimization, and machine learning, purchasers need to see not only potential advantages in papers, but also the correspondence among task scale, executable circuits, error-mitigation strategies, and total cost of ownership (Reiher et al., 2017; Farhi et al., 2014; Cerezo et al., 2021).

Therefore, through standardized PDK and Q-IP systems, the Fabless model may gradually transform the quantum-chip market from an experiment-driven structure into an application-driven structure. However, commercial implementation depends on the joint maturation of application value, hardware availability, service contracts, and customer-success mechanisms, rather than being determined solely by successful chip tape-out (Preskill, 2018; Krantz et al., 2019; McKinsey \& Company, 2025).

\section{Roadmap, Risks, and Conclusion}

\subsection{Near-Term Roadmap}

In the near-term stage (1--3 years), the core objectives of the Fabless quantum-chip ecosystem are to complete PDK standardization, build foundational Q-IP module libraries, and deploy the first usable Q-EDA toolchains. The key task at this stage is not to pursue large-scale expansion in qubit number, but to establish a minimal closed loop of ``design--simulation--tape-out--cryogenic testing--model feedback,'' enabling Fabless designers to complete an end-to-end quantum-chip development process without requiring vertically integrated capabilities (Kjaergaard et al., 2020; GDSII-to-Wafer Collaboration, 2026). The near-term roadmap should be centered on auditable milestones, such as PDK version freeze, test-structure coverage, SPICE-Q model error, DRC/LVS/DFM pass rates, cryogenic testing schedules, and model-card release, rather than using only qubit number as the progress metric.

Specifically, the near-term roadmap includes three parallel directions. The first is the release of PDK version 1.0, whose contents should cover foundational Transmon PCells, standard readout-resonator models, minimum manufacturability design rules (DRC/LVS/DFM), test structures, and model cards, rather than only layout-layer definitions. The second is the initial engineering of the SPICE-Q simulation environment, enabling designers to complete Hamiltonian extraction, noise budgeting, frequency-collision analysis, and readout-chain analysis at room temperature, and to align the results with cryogenic testing data. The third is joint validation tape-out between pilot foundry production lines, such as platforms in Suzhou with micro/nanofabrication foundations, and Fabless design companies, using test-chip data to calibrate PDK models in reverse. Google Quantum AI's experiment showing error suppression in surface-code logical qubits as code distance increases demonstrates that ``scaling'' in the roadmap cannot merely count the number of physical qubits, but must also validate whether the logical error rate decreases as expected with system expansion (Google Quantum AI, 2023). Subsequent below-threshold surface-code experiments further show that true near-term milestones should include real-time decoding latency, rare correlated error events, suppression factors of logical error rate with code distance, and break-even logical lifetime. These metrics better reflect whether the fault-tolerance pathway has entered a scalable regime than the simple presence of ``more physical qubits'' (Google Quantum AI and Collaborators, 2025).

This stage can be formalized as a system-maturity function:

\[
M(t) = \alpha P_{PDK}(t) + \beta Q_{EDA}(t) + \gamma I_{foundry}(t)
\]

where $M(t)$ denotes ecosystem maturity, and the three terms respectively correspond to the contribution weights of PDKs, EDA tools, and foundry capabilities.

\subsection{Medium-Term Scaling}

The objective of the medium-term stage (3--7 years) is to promote the coordinated maturation of Fabless quantum-chip design actors, Q-EDA tools, and Foundry service capabilities, and to gradually form a layered ecosystem structure analogous to that of the classical semiconductor industry. At this stage, the Q-IP market begins to mature, and PDKs become standard interfaces for cross-project reuse. Cross-foundry compatibility, however, must be gradually achieved through model recalibration, rule mapping, and revalidation, rather than being naturally available. It should be emphasized that medium-term scaling is not equivalent to directly entering million-scale fault-tolerant computing. The resource estimate by Gidney and Ekera for factoring RSA-2048 shows that, even under relatively optimistic assumptions for superconducting platforms, practical fault-tolerant algorithms may still require physical qubits at the ten-million scale, stringent surface-code cycle times, and complex space-time layout optimization (Gidney \& Ekera, 2021). Therefore, the medium-term goal is more reasonably positioned as the formation of ``reliable logical patches, reusable subarrays, stable Foundry processes, and assessable resource models,'' rather than the immediate realization of universal fault-tolerant quantum computing.

The scaling of the quantum-chip industry depends on three key mechanisms. First, the reuse rate of Q-IP modules must increase, enabling complex quantum systems to be assembled from standard submodules; however, the reuse rate must be counted together with the revalidation pass rate. Second, cross-foundry process consistency must improve, allowing performance distributions under different physical implementation pathways to gradually converge; however, migration across Foundries usually still requires model recalibration and test-chip validation. Third, cloud-based Q-EDA platforms must implement automatic optimization and design-space search, thereby reducing design complexity; however, automated results must enter physical signoff and the cryogenic testing loop. The National Academies' summary of the scaling challenges for quantum supercomputers also shows that medium-term bottlenecks usually arise from the combined constraints of control, wiring, refrigeration, error correction, and system integration, rather than from a single chip area or qubit number (National Academies, 2024).

From the perspective of system growth, this stage should not simply assume exponential growth. A more reasonable representation is to treat the number of Fabless design actors as a function of PDK openness, toolchain maturity, Foundry capacity, and application demand:

\[
N_{Fabless}(t) = f(P_{PDK}, Q_{EDA}, I_{foundry}, D_{market})
\]

where $P_{PDK}$ denotes the degree of PDK standardization and openness, $Q_{EDA}$ denotes toolchain maturity, $I_{foundry}$ denotes foundry and testing infrastructure capability, and $D_{market}$ denotes the intensity of application demand (Preskill, 2018; GDSII-to-Wafer Collaboration, 2026). This function should also be constrained by cryogenic testing capacity, packaging supply chains, control electronics, real-time decoding, and data governance. If these constraints do not expand synchronously, an increase in the number of Fabless actors may only produce more queuing, repeated validation, and model inconsistency, rather than automatically generating industrial scale effects (National Academies, 2024).

\subsection{Risk List}

Although the Fabless quantum-chip model has significant scaling potential, it faces risks across multiple dimensions. Physical-layer risks include fluctuations in decoherence time, material-interface defects, Josephson-junction process instability, frequency collisions, and packaging-mode coupling, all of which may lead to PDK model mismatch (McRae et al., 2020; McDonald et al., 2022; Levenson-Falk \& Shanto, 2025). Google Quantum AI's surface-code experiments further show that even when logical error rates begin to improve with increasing code distance, systems remain constrained by low-probability high-energy events, correlated errors, and incomplete error budgets. Therefore, the Fabless pathway must incorporate abnormal-event records and failure-mode feedback into PDK governance (Google Quantum AI, 2023).

System-level risks mainly arise from the accumulation of nonlinear coupling errors among Q-IP modules, including residual ZZ coupling, readout crosstalk, packaging modes, control-line thermal load, and decoding latency. If these errors are not additive, calibratable, or predictable at module boundaries, the assumptions of modular design will fail. Application-layer risks also cannot be ignored: many near-term applications depend on variational quantum algorithms and hybrid optimization workflows, and the results of McClean et al. on barren plateaus show that parameterized quantum circuits may exhibit gradient disappearance and training difficulty after scaling up. This means that ASQIC or NISQ application markets cannot evaluate only hardware availability, but must also assess algorithm trainability, measurement overhead, and error-mitigation costs (McClean et al., 2018; Cerezo et al., 2021). Ecosystem-layer risks include format fragmentation caused by standards competition, such as incompatible SPICE-Q extensions; model non-updatability caused by closed foundry data; and difficulty in defining reuse responsibility caused by unclear Q-IP licensing boundaries. Economic-layer risks arise from insufficient early market demand, excessive costs for cryogenic testing services, insufficient Foundry capacity utilization, overly long application-value validation cycles, and non-unified customer acceptance metrics.

Aggregate risk can be expressed as:

\[
R_{total} = R_{phys} + R_{sys} + R_{eco} + R_{market}
\]

where any single risk that is too high may block the formation of the ecosystem closed loop (Krantz et al., 2019). More precisely, the expression above is only a risk-classification framework rather than a directly additive physical quantity, because physical risks, system risks, ecosystem risks, and market risks are coupled. For example, insufficient cryogenic testing throughput simultaneously amplifies model mismatch, customer-delivery delays, and Q-IP revalidation costs; rare correlated error events may simultaneously affect logical error rates, model-card credibility, and roadmap judgment (Google Quantum AI and Collaborators, 2025). Therefore, risk governance should be centered on versioned model cards, failure-mode databases, cross-batch test structures, and contractual responsibility boundaries.

\subsection{Evaluation Metrics}

To quantify the development state of the Fabless quantum-chip ecosystem, it is necessary to establish a cross-layer evaluation-metric system covering three dimensions: the physical layer, the design layer, and the ecosystem layer.

Physical-layer metrics include qubit $T_1/T_2$ times, gate fidelity, readout error rates, frequency drift, material loss, Purcell limits, and cross-batch variance. Design-layer metrics include Q-IP reuse rate, revalidation pass rate, design convergence speed, SPICE-Q simulation error, DRC/LVS/DFM pass rates, and fab-out manufacturing-file completeness. System-layer metrics should further record the width, depth, logical error rate, readout parallelism, and decoding latency of executable circuits. The volumetric benchmark framework proposed by Blume-Kohout and Young shows that quantum processor capability cannot be represented only by point fidelity or qubit number, but should characterize the boundary of passable circuit families in the circuit width--depth space (Blume-Kohout \& Young, 2020). Ecosystem-layer metrics include the number of Fabless design actors, PDK usage frequency, test-chip data feedback speed, cross-PDK/cross-Foundry revalidation costs, and Q-IP licensing dispute rate.

An aggregate performance metric can be defined as a weighted function:

\[
S_{ecosystem} = w_1 F_{fidelity} + w_2 C_{design} + w_3 E_{scalability} + w_4 V_{validation}
\]

where the weights reflect the degree to which different indicators are relied upon at different stages of industrial development. $V_{validation}$ denotes validation and governance maturity, including model-card completeness, test-structure coverage, cryogenic data-feedback speed, Q-IP revalidation pass rate, and the reproducibility of customer benchmarks. For systems entering the fault-tolerance pathway, surface-code cycle time, real-time decoding latency, logical-error-rate suppression factors, rare correlated error frequency, and physical/logical resource overhead should also be recorded separately. Google 2023 and below-threshold experiments show that these metrics better reflect whether a system is approaching scalable fault-tolerant computation than does the sheer number of qubits (Google Quantum AI, 2023; Google Quantum AI and Collaborators, 2025).

The core function of this metric system is to shift quantum-hardware development from ``single-device performance optimization'' toward ``system-level ecosystem optimization.'' It should not replace specific product acceptance, but should serve as a shared evaluation language among Foundries, Fabless design firms, Q-EDA tool vendors, and customers, enabling roadmaps, risks, and investment decisions at different stages to be connected through the same set of auditable metrics.

\subsection{Conclusion}

The core contribution of Fabless quantum-chip design and commercialization pathways lies in gradually transforming quantum-hardware development from highly vertically integrated laboratory closed loops into an engineering system jointly supported by standardized interfaces, modular design, and softwareized toolchains. Through pillars such as PDKs, PCells, SPICE-Q, Q-EDA, and Q-IP, quantum-chip design can to some extent reduce repeated device development, process exploration, and low-level layout trial and error, while depositing part of the accumulated experience into reusable, verifiable, and traceable design assets (CMC Microsystems, 2023; QPDK Documentation, 2025; EDA-Q Collaboration, 2025; GDSII-to-Wafer Collaboration, 2026).

However, this paper does not understand the Fabless model as a simple replication of the classical semiconductor industry. The performance of superconducting quantum chips depends strongly on Josephson-junction statistical distributions, material-interface loss, packaging modes, frequency planning, cryogenic testing throughput, and system-calibration feedback. These factors mean that design and manufacturing can only achieve ``constrained decoupling,'' and cannot in the near term reach the high level of abstraction characteristic of mature CMOS digital chips (Koch et al., 2007; McRae et al., 2020; Levenson-Falk \& Shanto, 2025; National Academies, 2024). Therefore, the condition for the establishment of a Fabless quantum-chip ecosystem is not ``whether manufacturing is outsourced,'' but whether manufacturing capability can be stably expressed as PDKs, statistical models, test structures, signoff rules, and data-feedback interfaces.

From an industrial perspective, the long-term significance of this model is that quantum-computing hardware innovation no longer depends entirely on a small number of large vertically integrated actors, but may form a layered collaborative ecosystem jointly composed of Foundries, Fabless design firms, Q-EDA tool vendors, Q-IP suppliers, cryogenic testing platforms, and cloud service providers. For this transformation to occur, PDK maturity, Foundry yield, cryogenic testing capacity, Q-IP licensing mechanisms, and market demand must improve synchronously; it cannot be completed by a breakthrough in a single tool alone (Baldwin \& Clark, 2000; Macher et al., 2008; McKinsey \& Company, 2025).

Ultimately, the success of the Fabless model depends on one key condition: whether the stability and auditability of the design abstraction layer can be maintained under persistent physical uncertainty. Only when PCell model error, SPICE-Q simulation error, PDK version drift, cryogenic measurement data, manufacturing yield, logical error rates, and customer benchmarks can form a continuous closed loop can the quantum-chip industry move from the laboratory prototype stage into a scalable, divisible, and commercializable engineering stage. In other words, the endpoint of the Fabless quantum-chip pathway is not simply the appearance of more design companies, but the formation of an industrial institution capable of simultaneously encoding physical risk, system validation, application demand, and commercial responsibility into PDK, Q-IP, and Q-EDA workflows.
\section{Acknowledgment}

The authors acknowledge the assistance of AI in translation and content generation.

\section{Summary}

This paper systematically proposes a fabless (without in-house wafer fabrication) design and commercial production system for superconducting quantum chips. Its core idea is not simply to outsource manufacturing, but rather to transform quantum-chip design, on the basis of verifiable process, model, and data interfaces, from the vertically integrated IDM model into a Fabless-Foundry collaboration model in which design and manufacturing are progressively and conditionally decoupled. This system is jointly supported by five categories of infrastructure: process design kits (PDKs), parameterized device cells (PCells), SPICE-Q multiphysics simulation models, end-to-end Q-EDA workflows, and a market for reusable and licensable quantum intellectual property (Q-IP) (Mead \& Conway, 1980; Baldwin \& Clark, 2000; Macher et al., 2008; EDA-Q Collaboration, 2025).

At the technical level, a PDK defines the manufacturable design space, statistical process corners, boundaries of material loss, and signoff rules; a PCell library encapsulates commonly used components such as transmons, readout resonators, couplers, test structures, and logical patches as reusable objects; SPICE-Q is responsible for linking geometry, electromagnetics, Josephson nonlinearity, noise channels, and Hamiltonian parameters; and the Q-EDA workflow organizes design capture, layout generation, DRC/LVS/DFM, simulation-based verification, manufacturing-file generation, and the return flow of cryogenic test data into an auditable process. Recent work on QPDK, KQCircuits, EDA-Q, SQuADDS, and GDSII-to-Wafer indicates that this direction already has an early technical foundation, although mature industrial standards will still require further accumulation of model consistency, cross-batch statistical data, and foundry-level signoff procedures (CMC Microsystems, 2023; KQCircuits Documentation, 2024; QPDK Documentation, 2025; Shanto et al., 2024; GDSII-to-Wafer Collaboration, 2026).

At the industrial level, the value of the fabless quantum-chip model lies in reducing the costs of repeated device development, process exploration, and low-level layout trial and error, thereby enabling design organizations to focus more on architecture, Q-IP composition, error budgeting, application mapping, and system-level optimization. This value, however, is not unconditional. The manufacturing outcomes of quantum chips directly alter Hamiltonians, decoherence channels, frequency collisions, readout fidelity, and calibration complexity; therefore, design and manufacturing can be only partially decoupled, and only under the preconditions that models are trustworthy, PDKs are stable, test data can flow back, foundry yield can be measured, and Q-IP licensing mechanisms are clear (Koch et al., 2007; McRae et al., 2020; Levenson-Falk \& Shanto, 2025; National Academies, 2024).

This paper further points out that China's quantum-chip ecosystem has already formed a pattern of parallel exploration across multiple regions and multiple actors, but remains in a pre-standardization stage. Beijing, Hefei, Suzhou, Shenzhen, and other regions possess different advantages in scientific research, manufacturing, applications, and industrial support. To form a genuine fabless quantum ecosystem, the key is not for any single region to independently replicate a complete IDM capability, but to establish, across regions, shared Quantum PDKs, PCell libraries, Q-EDA interfaces, cryogenic test-data standards, and Q-IP transaction rules. Regions such as Suzhou, which have foundations in micro-nano fabrication and semiconductor support systems, may become potential foundry nodes; however, the viability of this role still depends on whether they can release stable process rules, statistical models, test-chip data, and externally auditable signoff flows.

Overall, the central proposition of fabless quantum-chip design and commercial production is to maintain the stability of the design-abstraction layer under persistent physical uncertainty. Only when PDKs, PCells, SPICE-Q, Q-EDA, Q-IP, cryogenic testing, and data governance form a closed loop can the quantum-chip industry move from laboratory prototypes and regional quasi-IDM models toward a reusable, divisible, and commercially viable engineering ecosystem. This process will not be accomplished overnight, but it provides a path that can be discussed, evaluated, and iteratively improved for fault-tolerant quantum computing to advance from research systems toward industrialized platforms.
\section{References}

1. Arute, F., Arya, K., Babbush, R., et al. (2019). Quantum supremacy using a programmable superconducting processor. \emph{Nature, 574}(7779), 505--510. https://doi.org/10.1038/s41586-019-1666-5; arXiv:1910.11333. https://arxiv.org/abs/1910.11333

2. Blais, A., Grimsmo, A. L., Girvin, S. M., \& Wallraff, A. (2021). Circuit quantum electrodynamics. \emph{Reviews of Modern Physics, 93}(2), 025005. https://doi.org/10.1103/RevModPhys.93.025005; arXiv:2005.12667. https://arxiv.org/abs/2005.12667

3. Breuer, H.-P., \& Petruccione, F. (2002). \emph{The theory of open quantum systems}. Oxford University Press.

4. ChinaQuantum. (2025). \emph{China Quantum Industry Brief Q4 2025}. Retrieved June 15, 2026, from https://www.chinaquantum.info/

5. CMC Microsystems. (2023). \emph{A process design kit for superconducting components} [White paper]. Retrieved June 15, 2026, from https://www.cmc.ca/wp-content/uploads/2023/01/WhitePaper\_Superconducting\_PDK\_2023.pdf

6. Cross, A. W., Bishop, L. S., Sheldon, S., Nation, P. D., \& Gambetta, J. M. (2022). OpenQASM 3: A broader and deeper quantum assembly language. \emph{ACM Transactions on Quantum Computing, 3}(3), Article 12, 1--50. https://doi.org/10.1145/3505636; arXiv:2104.14722. https://arxiv.org/abs/2104.14722

7. Devoret, M. H., \& Schoelkopf, R. J. (2013). Superconducting circuits for quantum information: An outlook. \emph{Science, 339}(6124), 1169--1174. https://doi.org/10.1126/science.1231930

8. ECNS. (2024, January 8). \emph{China's 3rd-generation superconductor quantum computer comes online}. Retrieved June 15, 2026, from https://www.ecns.cn/news/sci-tech/2024-01-08/detail-ihcwrpvv3803845.shtml

9. EDA-Q Collaboration. (2025). \emph{EDA-Q: Electronic design automation for superconducting quantum chip} [Preprint]. arXiv:2502.15386. https://arxiv.org/abs/2502.15386

10. Entangled Future. (2026). \emph{Chinese quantum computing companies}. Retrieved June 15, 2026, from https://entangledfuture.com/china/

11. Fowler, A. G., Mariantoni, M., Martinis, J. M., \& Cleland, A. N. (2012). Surface codes: Towards practical large-scale quantum computation. \emph{Physical Review A, 86}(3), 032324. https://doi.org/10.1103/PhysRevA.86.032324; arXiv:1208.0928. https://arxiv.org/abs/1208.0928

12. Gambetta, J. M., Chow, J. M., \& Steffen, M. (2017). Building logical qubits in a superconducting quantum computing system. \emph{npj Quantum Information, 3}, Article 2. https://doi.org/10.1038/s41534-016-0004-0; arXiv:1512.04129. https://arxiv.org/abs/1512.04129

13. GDSII-to-Wafer Collaboration. (2026). \emph{From GDSII to wafer: EDA design flow and data conversion for wafer-scale superconducting quantum chip fabrication} [Preprint]. arXiv:2604.11379. https://arxiv.org/abs/2604.11379

14. Hennessy, J. L., \& Patterson, D. A. (2019). \emph{Computer architecture: A quantitative approach} (6th ed.). Morgan Kaufmann.

15. IB Interview Questions. (2026). \emph{The semiconductor value chain: Fabless, foundries, and IDMs}. Retrieved June 15, 2026, from https://ibinterviewquestions.com/guides/tmt-investment-banking/semiconductor-value-chain

16. Kjaergaard, M., Schwartz, M. E., Braumueller, J., Krantz, P., Wang, J. I.-J., Gustavsson, S., \& Oliver, W. D. (2020). Superconducting qubits: Current state of play. \emph{Annual Review of Condensed Matter Physics, 11}, 369--395. https://doi.org/10.1146/annurev-conmatphys-031119-050605; arXiv:1905.13641. https://arxiv.org/abs/1905.13641

17. Koch, J., Yu, T. M., Gambetta, J., Houck, A. A., Schuster, D. I., Majer, J., Blais, A., Devoret, M. H., Girvin, S. M., \& Schoelkopf, R. J. (2007). Charge-insensitive qubit design derived from the Cooper pair box. \emph{Physical Review A, 76}(4), 042319. https://doi.org/10.1103/PhysRevA.76.042319; arXiv:cond-mat/0703002. https://arxiv.org/abs/cond-mat/0703002

18. KQCircuits Documentation. (2024). \emph{KQCircuits: KLayout Python library for integrated quantum circuit design} [Software documentation]. Retrieved June 15, 2026, from https://iqm-finland.github.io/KQCircuits/

19. Krantz, P., Kjaergaard, M., Yan, F., Orlando, T. P., Gustavsson, S., \& Oliver, W. D. (2019). A quantum engineer's guide to superconducting qubits. \emph{Applied Physics Reviews, 6}(2), 021318. https://doi.org/10.1063/1.5089550; arXiv:1904.06560. https://arxiv.org/abs/1904.06560

20. Litinski, D. (2019). A game of surface codes: Large-scale quantum computing with lattice surgery. \emph{Quantum, 3}, 128. https://doi.org/10.22331/q-2019-03-05-128

21. Macher, J. T., Mowery, D. C., \& Simcoe, T. S. (2008). e-Business and disintegration of the semiconductor industry value chain. \emph{Industry and Innovation, 15}(3), 155--181. https://doi.org/10.1080/13662710802033872

22. Mack, C. A. (2011). Fifty years of Moore's law. \emph{IEEE Transactions on Semiconductor Manufacturing, 24}(2), 202--207. https://doi.org/10.1109/TSM.2010.2096437

23. McDonald, M., et al. (2022). Laser-annealing Josephson junctions for yielding scaled-up superconducting quantum processors. \emph{npj Quantum Information, 8}, Article 60. https://doi.org/10.1038/s41534-021-00464-5

24. McRae, C. R. H., Wang, H., Gao, J., Vissers, M. R., Brecht, T., Dunsworth, A., Pappas, D. P., \& Mutus, J. Y. (2020). Materials loss measurements using superconducting microwave resonators. \emph{Review of Scientific Instruments, 91}(9), 091101. https://doi.org/10.1063/5.0017378; arXiv:2006.04718. https://arxiv.org/abs/2006.04718

25. Mead, C., \& Conway, L. (1980). \emph{Introduction to VLSI systems}. Addison-Wesley.

26. Minev, Z. K., Shanks, W. E., \& Gambetta, J. M. (2021). \emph{Qiskit Metal: An open-source framework for quantum device design and analysis} [Conference presentation]. IBM Research / APS March Meeting. Retrieved June 15, 2026, from https://research.ibm.com/publications/a-framework-for-quantum-device-designproject-qiskit-metal

27. Levenson-Falk, E. M., \& Shanto, S. A. (2025). A review of design concerns in superconducting quantum circuits. \emph{Materials for Quantum Technology, 5}, 022003. arXiv:2411.16967. https://arxiv.org/abs/2411.16967

28. Mueller, C., Cole, J. H., \& Lisenfeld, J. (2019). Towards understanding two-level-systems in amorphous solids: Insights from quantum circuits. \emph{Reports on Progress in Physics, 82}(12), 124501. https://doi.org/10.1088/1361-6633/ab3a7e; arXiv:1705.01108. https://arxiv.org/abs/1705.01108

29. Murray, C. E., Calusine, G., Melville, A., et al. (2021). \emph{Optimizing frequency allocation for fixed-frequency superconducting quantum processors} [Preprint]. arXiv:2112.01634. https://arxiv.org/abs/2112.01634

30. National Academies. (2024). \emph{How to build a quantum supercomputer: Scaling challenges and opportunities} [Preprint]. arXiv:2411.10406. https://arxiv.org/abs/2411.10406

31. Nickerson, N. H., Li, Y., \& Benjamin, S. C. (2016). Hierarchical surface code for network quantum computing with modules of arbitrary size. \emph{Nature Communications, 7}, Article 11185. https://doi.org/10.1038/ncomms11185; arXiv:1509.07796. https://arxiv.org/abs/1509.07796

32. Nigg, S. E., Paik, H., Vlastakis, B., Kirchmair, G., Shankar, S., Frunzio, L., Schoelkopf, R. J., \& Girvin, S. M. (2012). Black-box superconducting circuit quantization. \emph{Physical Review Letters, 108}(24), 240502. https://doi.org/10.1103/PhysRevLett.108.240502; arXiv:1204.0587. https://arxiv.org/abs/1204.0587

33. OpenQASM Working Group. (2022). \emph{OpenQASM 3 language specification and tooling} [Technical specification]. Retrieved June 15, 2026, from https://openqasm.com/

34. Murray, C. E. (2021). Material matters in superconducting qubits. \emph{Materials Science and Engineering: R: Reports, 146}, 100646. https://doi.org/10.1016/j.mser.2021.100646; arXiv:2106.05919. https://arxiv.org/abs/2106.05919

35. Preskill, J. (2018). Quantum computing in the NISQ era and beyond. \emph{Quantum, 2}, 79. https://doi.org/10.22331/q-2018-08-06-79; arXiv:1801.00862. https://arxiv.org/abs/1801.00862

36. QPDK Documentation. (2025). \emph{QPDK: Superconducting quantum process design kit} [Software documentation]. Retrieved June 15, 2026, from https://gdsfactory.github.io/quantum-rf-pdk/

37. QuIRC Collaboration. (2023). \emph{Co-designed superconducting architecture for lattice surgery of surface codes with Quantum Interface Routing Card} [Preprint]. arXiv:2312.01246. https://arxiv.org/abs/2312.01246

38. Shaydulin, R., Lotshaw, P. C., Larson, J., Ostrowski, J., \& Humble, T. S. (2023). Parameter transfer for quantum approximate optimization of weighted MaxCut. \emph{ACM Transactions on Quantum Computing, 4}(3), Article 19, 1--15. https://doi.org/10.1145/3606246

39. Shi, Y., et al. (2024). \emph{Efficient frequency allocation for superconducting quantum processors using improved optimization techniques} [Preprint]. arXiv:2410.20283. https://arxiv.org/abs/2410.20283

40. The Quantum Insider. (2026). \emph{10+ companies leading the quantum technologies race in China}. Retrieved June 15, 2026, from https://thequantuminsider.com/

41. Weste, N. H. E., \& Harris, D. M. (2011). \emph{CMOS VLSI design: A circuits and systems perspective} (4th ed.). Addison-Wesley.

42. Baldwin, C. Y., \& Clark, K. B. (2000). \emph{Design rules, volume 1: The power of modularity}. MIT Press.

43. Langlois, R. N., \& Robertson, P. L. (1992). Networks and innovation in a modular system: Lessons from the microcomputer and stereo component industries. \emph{Research Policy, 21}(4), 297--313. https://doi.org/10.1016/0048-7333(92)90030-8

44. Sturgeon, T. J. (2002). Modular production networks: A new American model of industrial organization. \emph{Industrial and Corporate Change, 11}(3), 451--496. https://doi.org/10.1093/icc/11.3.451

45. McKinsey \& Company. (2025). \emph{Quantum Technology Monitor 2025: The year of quantum: From concept to reality in 2025}. Retrieved June 16, 2026, from https://www.mckinsey.com/\textasciitilde{}/media/mckinsey/business\%20functions/mckinsey\%20digital/our\%20insights/the\%20year\%20of\%20quantum\%20from\%20concept\%20to\%20reality\%20in\%202025/quantum-monitor-2025.pdf

46. Place, A. P. M., Rodgers, L. V. H., Mundada, P., Smitham, B. M., Fitzpatrick, M., Leng, Z., Premkumar, A., Bryon, J., Vrajitoarea, A., Sussman, S., Cheng, G., Madhavan, T., Babla, H. K., Le, X. H., Gang, Y., Jaeck, B., Gyenis, A., Yao, N., Cava, R. J., de Leon, N. P., \& Houck, A. A. (2021). New material platform for superconducting transmon qubits with coherence times exceeding 0.3 milliseconds. \emph{Nature Communications, 12}, Article 1779. https://doi.org/10.1038/s41467-021-22030-5

47. Reed, M. D., Johnson, B. R., Houck, A. A., DiCarlo, L., Chow, J. M., Schuster, D. I., Frunzio, L., \& Schoelkopf, R. J. (2010). Fast reset and suppressing spontaneous emission of a superconducting qubit. \emph{Applied Physics Letters, 96}(20), 203110. https://doi.org/10.1063/1.3435463

48. Barends, R., Kelly, J., Megrant, A., Veitia, A., Sank, D., Jeffrey, E., White, T. C., Mutus, J., Fowler, A. G., Campbell, B., Chen, Y., Chen, Z., Chiaro, B., Dunsworth, A., Neill, C., O'Malley, P., Roushan, P., Vainsencher, A., Wenner, J., Korotkov, A. N., Cleland, A. N., \& Martinis, J. M. (2014). Superconducting quantum circuits at the surface code threshold for fault tolerance. \emph{Nature, 508}(7497), 500--503. https://doi.org/10.1038/nature13171

49. SkyWater PDK Authors. (2020). \emph{SkyWater Open Source PDK: Open source process design kit for SkyWater Technology Foundry's 130nm node} [Software documentation]. Retrieved June 16, 2026, from https://github.com/google/skywater-pdk

50. Mitchell, M., Wu, S., Zaldivar, A., Barnes, P., Vasserman, L., Hutchinson, B., Spitzer, E., Raji, I. D., \& Gebru, T. (2019). Model cards for model reporting. In \emph{Proceedings of the Conference on Fairness, Accountability, and Transparency} (pp. 220--229). ACM. https://doi.org/10.1145/3287560.3287596

51. Wilkinson, M. D., Dumontier, M., Aalbersberg, I. J., Appleton, G., Axton, M., Baak, A., Blomberg, N., Boiten, J.-W., da Silva Santos, L. B., Bourne, P. E., et al. (2016). The FAIR Guiding Principles for scientific data management and stewardship. \emph{Scientific Data, 3}, Article 160018. https://doi.org/10.1038/sdata.2016.18

52. Shanto, S., Kuo, A., Miyamoto, C., Zhang, H., Maurya, V., Vlachos, E., Hecht, M., Shum, C. W., \& Levenson-Falk, E. (2024). SQuADDS: A validated design database and simulation workflow for superconducting qubit design. \emph{Quantum, 8}, 1465. https://doi.org/10.22331/q-2024-09-09-1465

53. Kunasaikaran, J., Mato, K., \& Wille, R. (2024). A framework for the design and realization of alternative superconducting quantum architectures. In \emph{2024 IEEE 54th International Symposium on Multiple-Valued Logic (ISMVL)} (pp. 91--96). IEEE.

54. Geller, M. R., \& Fang, M. (2015). Tunable coupler for superconducting Xmon qubits: Perturbative nonlinear model. \emph{Physical Review A, 92}(1), 012320. https://doi.org/10.1103/PhysRevA.92.012320

55. Yan, F., Krantz, P., Sung, Y., Kjaergaard, M., Campbell, D. L., Orlando, T. P., Gustavsson, S., \& Oliver, W. D. (2018). Tunable coupling scheme for implementing high-fidelity two-qubit gates. \emph{Physical Review Applied, 10}(5), 054062. https://doi.org/10.1103/PhysRevApplied.10.054062

56. QIR Alliance. (2021). \emph{Quantum Intermediate Representation (QIR) specification} [Technical specification]. Retrieved June 16, 2026, from https://github.com/qir-alliance/qir-spec

57. IEEE P7130 Working Group. (2021). \emph{P7130: Standard for Quantum Technologies Definitions} [Standards project]. IEEE Standards Association. Retrieved June 16, 2026, from https://standards.ieee.org/ieee/7130/10680/

58. Keating, M., \& Bricaud, P. (2002). \emph{Reuse methodology manual for system-on-a-chip designs} (3rd ed.). Springer.

59. Castellanos-Beltran, M. A., \& Lehnert, K. W. (2008). Widely tunable parametric amplifier based on a SQUID array resonator. \emph{Applied Physics Letters, 91}, 083509. https://doi.org/10.1063/1.2773988

60. Macklin, C., O'Brien, K., Hover, D., Schwartz, M. E., Bolkhovsky, V., Zhang, X., Oliver, W. D., \& Siddiqi, I. (2015). A near-quantum-limited Josephson traveling-wave parametric amplifier. \emph{Science, 350}(6258), 307--310. https://doi.org/10.1126/science.aaa8525

61. Google Quantum AI. (2023). Suppressing quantum errors by scaling a surface code logical qubit. \emph{Nature, 614}, 676--681. https://doi.org/10.1038/s41586-022-05434-1; arXiv:2207.06431. https://arxiv.org/abs/2207.06431

62. Reiher, M., Wiebe, N., Svore, K. M., Wecker, D., \& Troyer, M. (2017). Elucidating reaction mechanisms on quantum computers. \emph{Proceedings of the National Academy of Sciences, 114}(29), 7555--7560. https://doi.org/10.1073/pnas.1619152114; arXiv:1605.03590. https://arxiv.org/abs/1605.03590

63. Blume-Kohout, R., \& Young, K. C. (2020). A volumetric framework for quantum computer benchmarks. \emph{Quantum, 4}, 362. https://doi.org/10.22331/q-2020-11-15-362; arXiv:1904.05546. https://arxiv.org/abs/1904.05546

64. Farhi, E., Goldstone, J., \& Gutmann, S. (2014). \emph{A quantum approximate optimization algorithm} [Preprint]. arXiv:1411.4028. https://arxiv.org/abs/1411.4028

65. Cerezo, M., Arrasmith, A., Babbush, R., Benjamin, S. C., Endo, S., Fujii, K., McClean, J. R., Mitarai, K., Yuan, X., Cincio, L., \& Coles, P. J. (2021). Variational quantum algorithms. \emph{Nature Reviews Physics, 3}, 625--644. https://doi.org/10.1038/s42254-021-00348-9; arXiv:2012.09265. https://arxiv.org/abs/2012.09265

66. Santos, A. C. (2017). The IBM Quantum Computer and the IBM Quantum Experience. \emph{Revista Brasileira de Ensino de Fisica, 39}(1), e1301. https://doi.org/10.1590/1806-9126-RBEF-2016-0155; arXiv:1610.06980. https://arxiv.org/abs/1610.06980

67. AbuGhanem, M. (2024). \emph{IBM Quantum Computers: Evolution, Performance, and Future Directions} [Preprint]. arXiv:2410.00916. https://arxiv.org/abs/2410.00916

68. Gidney, C., \& Ekera, M. (2021). How to factor 2048 bit RSA integers in 8 hours using 20 million noisy qubits. \emph{Quantum, 5}, 433. https://doi.org/10.22331/q-2021-04-15-433; arXiv:1905.09749. https://arxiv.org/abs/1905.09749

69. Google Quantum AI and Collaborators. (2025). Quantum error correction below the surface code threshold. \emph{Nature, 638}, 920--926. https://doi.org/10.1038/s41586-024-08449-y; arXiv:2408.13687. https://arxiv.org/abs/2408.13687

70. McClean, J. R., Boixo, S., Smelyanskiy, V. N., Babbush, R., \& Neven, H. (2018). Barren plateaus in quantum neural network training landscapes. \emph{Nature Communications, 9}, Article 4812. https://doi.org/10.1038/s41467-018-07090-4; arXiv:1803.11173. https://arxiv.org/abs/1803.11173

\end{document}